\documentclass{aa}  

\usepackage{graphicx, color}
\usepackage{txfonts}
\usepackage{comment}
\usepackage{booktabs}
\usepackage[allcolors=blue]{hyperref}
\newcommand{\orcidlink}[1]{\protect\href{https://orcid.org/#1}{\protect\includegraphics[width=8pt]{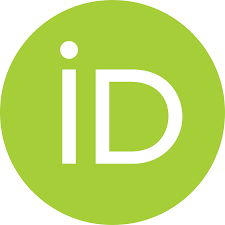}}}
\begin{document}

   \title{Evolving magnetic lives of Sun-like stars. }
\subtitle{I. Characterisation of the large-scale magnetic field with Zeeman-Doppler imaging}
   \titlerunning{Evolving magnetic lives of Sun-like stars}

   \author{S. Bellotti \inst{1,2}\orcidlink{0000-0002-2558-6920}
          \and
          T. L\"uftinger \inst{3}\orcidlink{0009-0000-4946-6942}
          \and
          S. Boro Saikia \inst{4}\orcidlink{0000-0002-3673-3746}
          \and
          C. P. Folsom \inst{5}\orcidlink{0000-0002-9023-7890}
          \and 
          P. Petit \inst{2}\orcidlink{0000-0001-7624-9222}
          \and
          J. Morin \inst{6}\orcidlink{0000-0001-5260-7179}
          \and
          M. G\"udel \inst{4}\orcidlink{0000-0001-9818-0588}
          \and
          J-F. Donati \inst{2}\orcidlink{0000-0001-5541-2887}
          \and
          E. Alecian\inst{7}\orcidlink{0000-0001-5260-7179}
          }
   \authorrunning{Bellotti et al.}
    
   \institute{
            Leiden Observatory, Leiden University,
            PO Box 9513, 2300 RA Leiden, The Netherlands\\
            \email{bellotti@strw.leidenuniv.nl}
        \and
            Institut de Recherche en Astrophysique et Plan\'etologie,
            Universit\'e de Toulouse, CNRS, IRAP/UMR 5277,
            14 avenue Edouard Belin, F-31400, Toulouse, France 
        \and
            Science Division, Directorate of Science, 
            European Space Research and Technology Centre (ESA/ESTEC),
            Keplerlaan 1, 2201 AZ, Noordwijk, The Netherlands
        \and
            University of Vienna, Department of Astrophysics, 
            T\"urkenschanzstrasse 17, A-1180 Vienna, Austria
        \and 
            Tartu Observatory, 
            University of Tartu, 
            Observatooriumi 1, Tõravere, 61602, Estonia
        \and
            Laboratoire Univers et Particules de Montpellier,
            Universit\'e de Montpellier, CNRS,
            F-34095, Montpellier, France
        \and
            Univ. Grenoble Alpes, CNRS, IPAG, 38000 Grenoble, France
             }
   \date{Received ; accepted }

% \abstract{}{}{}{}{} 
% 5 {} token are mandatory
 
  \abstract
   {Planets orbiting young, solar-type stars are embedded in a more energetic environment  than that of the solar neighbourhood. They experience harsher conditions due to enhanced stellar magnetic activity and wind shaping the secular evolution of a planetary atmosphere.}
   {This study is dedicated to the characterisation of the magnetic activity of eleven Sun-like stars, with ages between 0.2 and 6.1\,Gyr and rotation periods between 4.6 and 28.7\,d. Based on a sub-sample of six stars, we aim to study the large-scale magnetic field, which we then use to simulate the associated stellar wind and environment. Finally, we want to determine the conditions during the early evolution of planetary habitability.}
   {We analysed high-resolution spectropolarimetric data collected in 2018 and 2019 with Narval. We computed activity diagnostics from chromospheric lines such as \ion{Ca}{II} H\&K, H$\alpha$, and the \ion{Ca}{II} infrared triplet, as well as the longitudinal magnetic field from circularly polarised least-squares deconvolution profiles. For six stars exhibiting detectable circular polarisation signals, we reconstructed the large-scale magnetic field at the photospheric level by means of Zeeman-Doppler imaging (ZDI).}
   {In agreement with previous studies, we found a global decrease in the activity indices and longitudinal field with increasing age and rotation period. The large-scale magnetic field of the six sub-sample stars displays a strength between 1 and 25\,G and reveals substantial contributions from different components such as poloidal (40-90\%), toroidal (10-60\%), dipolar (30-80\%), and quadrupolar (10-40\%), with distinct levels of axisymmetry (6-84\%) and short-term variability of the order of months. Ultimately, this implies that exoplanets tend to experience a broad variety of stellar magnetic environments after their formation.}
   {} 


   \maketitle

%
%-------------------------------------------------------------------

\section{Introduction}

Stellar magnetic activity powers outflows and eruptive processes such as flares, coronal mass ejections, and energetic particle events. The collective action of these phenomena determines the space weather planets orbit within and plays a major role in the processing of planetary magnetospheres and upper atmospheres \citep[e.g.][]{Lammer2003,Lammer2012,Rugheimer2015}. Magnetic interactions in the outer stellar layers (chromosphere, transition region, and corona) produce energetic X-ray and extreme ultraviolet (EUV) radiation that can heat and ionise upper planetary atmospheres, triggering their evaporation and photochemistry \citep{CecchiPestellini2006,MurrayClay2009,Owen2012,Tsai2023,vanLooveren2025}. A similar effect is also induced by stellar cosmic rays \citep{RodgersLee2021,RodgersLee2021b,Raeside2025}. Ultimately, the evolution of stellar activity during the history of a star dictates any observed planetary atmosphere \citep{Penz2008,SanzForcada2011,Locci2019,Allan2019}. This is fundamental, for example, to explain the atmospheric evolution of Solar System planets \citep{Lammer2003,Wood2004,Kulikov2007,Airapetian2016},

In this context, studying Sun-like stars at different ages is relevant to understanding planetary atmospheric evolution at early stages of formation, as well as the consequences of such extreme conditions for the habitability of Earth-like planets \citep{See2014,Airapetian2020}. Observations of young stars indicate an enhanced magnetic activity and a harsher space environment compared to the current Sun \citep{Wood2006}, implying more frequent and more energetic phenomena \citep{Wood2004,Wood2006,Gudel2007}. For instance, X-ray and extreme ultraviolet (EUV) emission can be two or three orders of magnitude larger than in the case of our Sun \citep{Guinan2003,Ribas2005,Tu2015,Johnstone2021,Ketzer2023}. 

The nature of stellar activity is dictated by the intensity and structure of the magnetic field, which, in turn, depends on the rotation and internal structure of the star. Knowing the intensity of the magnetic fields on young planetary host stars, which can be several tens to hundreds of times stronger than that of the Sun \citep[e.g.][]{Folsom2018}, is a pivotal precondition to understanding whether planets can become habitable in the early evolution of a planetary system. Moreover, the habitability conditions are not solely determined by the star's temperature, but also by its magnetic field \citep{Vidotto2013,Cockell2016}, which can influence the survival of the planetary atmosphere \citep[see Fig.~5 in][for a recent example]{vanLooveren2025}.

The level of stellar magnetic activity is evaluated via proxies such as photometric variability, X-ray luminosity, and emission in the cores of chromospheric lines. The latter are typically \ion{Ca}{II} H\&K, H$\alpha,$ and \ion{Ca}{II} infrared triplet lines \citep{Wilson1968,Gizis2002,Busa2007}. A detailed characterisation of the stellar magnetic field can be carried out with spectropolarimetry, an observational technique that detects magnetically induced polarisation in spectral lines owing to the Zeeman effect \citep{Zeeman1897,Donati1997,Landstreet2009_1}. By collecting a time series of polarised spectra, it is possible to reconstruct the geometry of the large-scale magnetic field at the surface of the star by means of tomographic inversion, that is, Zeeman-Doppler imaging \citep[ZDI;][]{Semel1989,DonatiBrown1997}. The resulting maps of the magnetic field can then be used as boundary conditions to extrapolate the field lines outward and model the stellar wind via magnetohydrodynamical simulations \citep[e.g.][]{Vidotto2014,Nicholson2016,BoroSaikia2020,Folsom2020,Alvarado-Gomez2022b,Vidotto2023,Evensberget2021,Evensberget2022,Evensberget2023}.

The ZDI technique has been extensively applied to study the strength and configuration of the large-scale magnetic field of Sun-like stars of different ages and rotation periods \citep{DonatiCollier1997,Petit2008,Rosen2016,Folsom2016,Folsom2018,Willamo2022}. Distinct trends have emerged: for instance, slowly rotating Sun-like stars tend to exhibit poloidal-dominated field geometries, as shown by \citet{Petit2008} and \citet{Brown2022}, together with a decrease in chromospheric activity. Furthermore, the average magnetic field strength decreases with increasing age and rotation period \citep[e.g.][]{Vidotto2014,Folsom2018,Willamo2022}, which is in agreement with trends of the total, unsigned magnetic field measurements performed by theoretical modelling of Zeeman broadening \citep{Reiners2012,Kochukhov2020}. Repeated reconstructions of the large-scale magnetic field with ZDI have also revealed the presence of long-term trends and magnetic cycles for these stars \citep[e.g.][]{Fares2009,Rosen2016,BoroSaikia2016,Jeffers2017,AlvaradoGomez2018,Bellotti2025}.

In this work, we characterise the magnetic field and activity level for a sample of Sun-like stars whose ages span between 0.2 and 6.1\,Gyr and with their rotation periods between 4.6 and 28.7\,d. Thus, we are set to continue the exploration of magnetic activity properties across a wide stellar parameter space. For a sub-sample of six stars having detectable Zeeman signatures in circular polarisation and suitable temporal sampling of their rotation, we reconstructed the large-scale magnetic field geometry with ZDI. These six stars have ages between 0.2 and 1.6\,Gyr and rotation periods between 4.6 and 12.4\,d. In this age range, the relation between rotation period and age is not unique, thus, monitoring many stars in this age range is important. The magnetic field maps resulting from this study will be used in a subsequent work for stellar wind and environment modelling.

The paper is structured as follows. In Sect.~\ref{sec:observations}, we describe the sample of Sun-like stars examined in this study and their spectropolarimetric observations performed with the high-resolution spectropolarimeter Narval. In Sect.\ref{sec:indexes}, we outline the computation of canonical activity indices as well as the longitudinal magnetic field. In Sect.~\ref{sec:zdi}, we recall the principles and assumptions of ZDI. Finally, we discuss our results star-by-star and general trends in Sects.~\ref{sec:Results} and~\ref{sec:trends}, presenting our conclusions in Sect.~\ref{sec:conclusions}.

\section{Observations}\label{sec:observations}

\begin{figure}
    \includegraphics[width=\columnwidth]{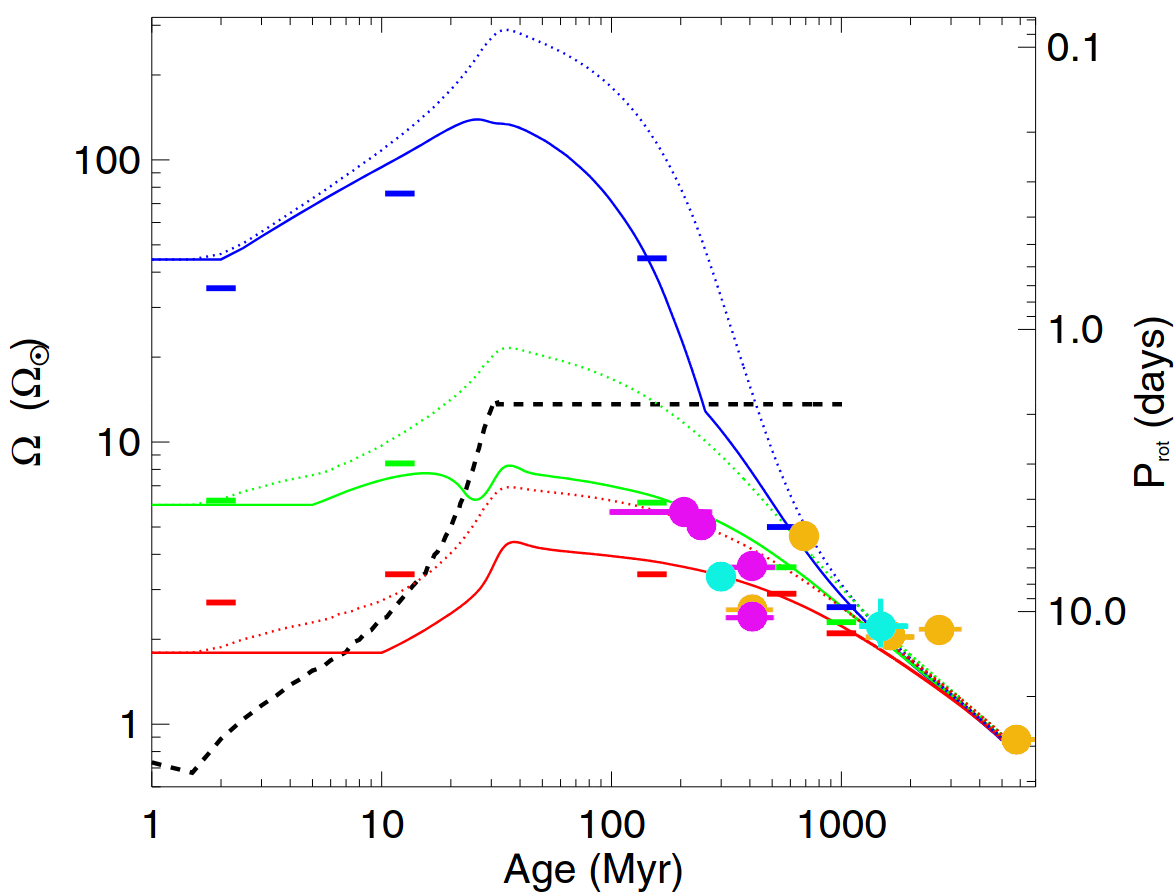}
    \caption{Rotational evolutionary tracks computed by \citet{Johnstone2015a,Johnstone2015b} and used in \citet{Tu2015}, with our sample stars overplotted. Magenta data points indicate stars with a ZDI map both in the literature and presented in this work, cyan data points indicate stars with a first ZDI reconstruction in this work, and yellow data points indicate stars without a ZDI map. Some of the error bars are smaller than the data point size. The curves represent the 10th (red), 50th (green), and 90th (blue) percentiles of the stellar rotational distribution, of the envelope (solid lines) and core (dotted lines). The small horizontal lines are the observational constraints for the corresponding percentiles. The dashed black line is a time-dependent saturation relation for the stellar dipolar field, the wind mass loss rate and the X-ray luminosity. For more information, see the work of \citet{Tu2015}.\label{fig:starsample}}
\end{figure}

The targets of our study are 11 G-type stars. According to the literature, they exhibit considerably distinct activity levels, along with likely differences in the magnetic field structure, wind properties, and high-energy output. We placed our stars on the rotation evolutionary tracks of \citet{Tu2015} to highlight this point, as illustrated in Fig.~\ref{fig:starsample}. The cited authors showed how the temporal evolution of stellar rotation, along with the associated X-ray and EUV luminosity, has a significant impact on the atmospheric evolution of exoplanets. Our stellar sample covers the three rotational evolution tracks (slow, medium, and fast), providing good coverage of the non-unique rotational history of Sun-like stars during the first few hundred million years. The Sun could have taken any of these three paths, so this coverage of the rotational evolution is necessary to remark on the magnetic history of the Sun.

Four stars in our sample (HD~1835, HD~82443, HD~189733, and HD~206860) have already been observed with spectropolarimetry, but these observations date back to different epochs. Therefore, our analyses also address the evolution of the magnetic field structure on timescales of many years, as reported previously for other Sun-like stars \citep{Donati2008,BoroSaikia2015,Rosen2016,Willamo2022,Bellotti2025}. 

The properties of our stars are summarised in Table~\ref{tab:star_properties}. The ages of the stars were mostly extracted from the work of \citet{Ramirez2012} and \citet{Linsky2020}. Furthermore, \citet{Ramirez2012} used either isochrones or gyrochronology to estimate the stellar age. The latter method is more accurate for young active stars. For our stars, the ages of HD~149026 and HD~219828 were obtained via isochrones, while for the remaining stars, ages were obtained from gyrochronology.

We analysed optical spectropolarimetric observations collected with Narval in 2018 and 2019\footnote{The data is available at \url{https://www.polarbase.ovgso.fr/}}. Narval is the spectropolarimeter on the 2~m T\'elescope Bernard Lyot (TBL) at the Pic du Midi Observatory in France \citep{Donati2003}, which operates between 370 and 1050\,nm at high resolution ($R = 65,000$). The observations were carried out in circular polarisation mode, providing both unpolarised (Stokes~$I$) and circularly polarised (Stokes~$V$) high-resolution spectra. The data were reduced with the \texttt{LIBRE-ESPRIT} pipeline \citep{Donati1997} and the continuum-normalised spectra were retrieved from PolarBase \citep{Petit2014}. We provide the full list of observations in Table~\ref{tab:log} and the total number of observations for each star in Table~\ref{tab:lsd_masks}.

The detection and characterisation of Zeeman signatures in circularly polarised light are performed by means of least-squares deconvolution \citep[LSD;][]{Donati1997,Kochukhov2010a} using the \textsc{lsdpy} code which is part of the Specpolflow software \citep{Folsom2025}\footnote{Available at \href{https://github.com/folsomcp/LSDpy}{https://github.com/folsomcp/LSDpy}}. This numerical technique produces high signal-to-noise ratio (S/N) line profiles (unpolarised and circularly polarised) from the combination of thousands of photospheric spectral lines included in a synthetic line list. The line lists were produced using the Vienna Atomic Line Database\footnote{\url{http://vald.astro.uu.se/}} \citep[VALD,][]{Ryabchikova2015}. They contain information of atomic lines with known Land\'e factor (indicated by g$_\mathrm{eff}$ and describing the magnetic sensitivity of a spectral line) and with a depth greater than 40\% the level of the unpolarised continuum \citep[following][]{Donati1997}. A summary of the line lists used and their properties is given in Table~\ref{tab:lsd_masks}. 

We recorded substantially lower S/N in Stokes~$V$ LSD profiles for two observations of HD\,206860 on July 26th 2018 and June 22nd 2019, as well as for HD\,43162 on December 19th 2018; hence,  these observations were excluded from the analysis. For HD\,206860, we also noticed that the Stokes~$V$ profile on July 28, 2018 featured several oscillations around the zero flux level and was also excluded. For three stars, namely, HD\,149026, HD\,190406, and HD\,219828, we did not report any clear detections of Zeeman signatures in Stokes~$V$ profiles. Indeed, for each star, the false-alarm probability \citep[FAP; see][for more details]{Donati1997} of the putative Zeeman signature is greater than $10^{-3}$. It was only in a couple of observations that we were able to see a marginal detection ($\mathrm{FAP}\sim10^{-3}-10^{-4}$). For this reason, these stars were excluded from the spectropolarimetric characterisation analyses, however, they were retained for the measurement of unpolarised magnetic activity indicators, as outlined in the next sections. In Appendix~\ref{app:stokesV}, we describe the polarisation signatures in Stokes~$V$ as well as spurious polarisation signatures in Stokes~$N$. In the next following, the observations have been phased with the following ephemeris,
\begin{align}
    \mathrm{HJD} = \mathrm{HJD}_0 + \mathrm{P}_\mathrm{rot}\cdot n_\mathrm{cyc},
    \label{eq:ephemeris}
\end{align}
where HJD$_\mathrm{0}$ is the heliocentric Julian Date reference (the first one of the time series for each star, see Table~\ref{tab:log}), P$_\mathrm{rot}$ is the rotation period of the star (see Table~\ref{tab:star_properties}), and $n_\mathrm{cyc}$ represents the rotation cycle.

\setlength{\tabcolsep}{4.1pt}
\begin{table*}[!t]
\caption{Properties of our stars.} 
\label{tab:star_properties}     
\centering                       
\begin{tabular}{l c c c l l l l c r r r}    
\toprule
Name & B & V & Dist & T$_\mathrm{eff}$ & $\log g$ & M & R & Age & P$_\mathrm{rot}$ & $v_\mathrm{eq}\sin i$ & Inc\\
& [mag] & [mag] & [pc] & [K] & [dex] & [M$_\odot$] & [R$_\odot$] & [Gyr] & [d] & [km\,s$^{-1}$] & [$^\circ$]\\
\midrule
HD\,1835   & 7.05$^{(1)}$ & 6.39$^{(1)}$ & 21.33$^{(8)}$ & 5837$^{(9)}$ & 4.47$^{(9)}$ & 0.980$^{(9)}$ & 0.958$^{(9)}$ & 0.40$\pm0.10^{(11)}$ & 7.68$^{(13)}$  & 7.0$^{(9)}$ & 65$^{(25)}$\\
HD\,28205  & 7.95$^{(2)}$ & 7.40$^{(2)}$ & 46.98$^{(8)}$ & 6234$^{(10)}$ & 4.34$^{(10)}$ & 1.210$^{(10)}$ & 1.228$^{(10)}$ & 0.63$\pm0.05^{(12)}$ & 5.87$^{(15)}$  & 10.6$^*$ & ... \\ 
HD\,30495  & 6.14$^{(3)}$ & 5.50$^{(3)}$ & 13.23$^{(8)}$ & 5759$^{(9)}$ & 4.39$^{(9)}$ & 0.860$^{(9)}$ & 0.983$^{(9)}$ & 0.40$\pm0.10^{(11)}$ & 11.36$\pm0.17^{(16)}$ & 4.1$^{(16)}$ & ... \\ 
HD\,43162  & 7.07$^{(1)}$ & 6.37$^{(1)}$ & 16.69$^{(8)}$ & 5617$^{(10)}$ & 4.54$^{(10)}$ & 0.990$^{(10)}$ & 0.889$^{(10)}$ & 0.32$\pm0.04^{(12)}$ & 8.50$_{-0.8}^{+1.3(17)}$  & 7.0$^\dagger$ & 70$^*$\\ 
HD\,82443  & 7.78$^{(3)}$ & 7.05$^{(3)}$ & 18.07$^{(8)}$ & 5287$^{(10)}$ & 4.57$^{(10)}$ & 0.910$^{(10)}$ & 0.819$^{(10)}$ & 0.25$\pm0.05^{(12)}$ & 5.34$\pm0.07^{(14)}$ & 6.5$^{(14)}$ & 60$^{(14)}$ \\ 
HD\,114710 & 4.84$^{(4)}$ & 4.25$^{(5)}$ & 9.19$^{(8)}$ & 6075$^{(9)}$ & 4.57$^{(9)}$ & 1.540$^{(9)}$ & 1.063$^{(9)}$ & 1.60$\pm0.40^{(11)}$ & 12.35$^{(18)}$ & 4.1$^{(19)}$ & 70$^*$ \\ 
HD\,149026$^\star$ & 8.75$^{(1)}$ & 8.14$^{(1)}$ & 76.22$^{(8)}$ & 6084$^{(10)}$ & 4.17$^{(10)}$ & 1.140$^{(10)}$ & 1.460$^{(10)}$ & 2.78$\pm0.50^{(11)}$ & $ <12.30\pm1.20$ & 6.0$^{(21)}$ & ... \\ 
HD\,189733$^\star$ & 8.58$^{(6)}$ & 7.65$^{(6)}$ & 19.78$^{(8)}$ & 5023$^{(10)}$ & 4.58$^{(10)}$ & 0.776$^{(10)}$ & 0.840$^{(10)}$ & 0.40$\pm0.10^{(11)}$ & 11.94$\pm0.16^{(22)}$ & 2.97$^{(22)}$ & 75$^{(22)}$\\ 
HD\,190406 & 6.39$^{(1)}$ & 5.79$^{(1)}$ & 17.77$^{(8)}$ & 5932$^{(9)}$ & 4.45$^{(9)}$ & 1.131$^{(9)}$ & 1.052$^{(9)}$ & 1.70$\pm0.40^{(11)}$ & 13.94$^{(18)}$ & 4.0$^{*}$ & ... \\ 
HD\,206860$^\star$ & 5.29$^{(7)}$ & 4.70$^{(7)}$ & 18.13$^{(8)}$ & 5974$^{(9)}$ & 4.47$^{(9)}$ & 1.085$^{(9)}$ & 1.002$^{(9)}$ & 0.20$\pm0.10^{(11)}$ & 4.55$\pm0.03^{(23)}$  & 10.1$^{(23)}$ & 75$^{(23)}$ \\ 
HD\,219828$^\star$ & 8.78$^{(1)}$ & 8.01$^{(1)}$ & 72.53$^{(8)}$ & 5807$^{(10)}$ & 4.02$^{(10)}$ & 1.040$^{(10)}$ & 1.651$^{(10)}$ & 6.08$^{+0.9(11)}_{-0.6}$ & 28.70$\pm0.60^{(24)}$ & 3.7$^{(24)}$ & ... \\ 
\bottomrule 
\end{tabular}
\tablefoot{The columns indicate: identifier of the star, B band magnitude, V band magnitude, distance, effective temperature, surface gravity, mass, radius, age, rotation period, projected equatorial velocity ($v_\mathrm{eq}\sin i$), and stellar inclination. References: (1) \citet{Hog2000}, (2) \citet{Joner2006}, (3) SIMBAD \citep{Wenger2000}, (4) \citet{Ducati2002}, (5) \citet{vanBelle2009}, (6) \citet{Koen2010}, (7) \citet{BoroSaikia2015}, (8) \citet{GaiaCollaboration2020}, (9) \citet{Valenti2005}, (10) \citet{Peagert2021}, (11) \citet{Ramirez2012}, (12) \citet{Linsky2020}, (13) \citet{Gudel2007}, (14) \citet{Folsom2016}, (15) \citet{Pizzolato2003}, (16) \citet{Egeland2015}, (17) \citet{Marsden2014}, (18) \citet{Wright2011}, (19) \citet{Gray1997}, (20) \citet{Messina1999}, (21) \citet{Bonomo2017}, (22) \citet{Fares2010}, (23) \citet{BoroSaikia2015}, (24) \citet{Santos2016}, (25) \citet{Rosen2016}. The symbol $\star$ in the name of a star represents a known exoplanet host. The symbol $*$ indicates values of $v_\mathrm{eq}\sin i$ or inclination that we derived in this work from geometrical considerations. We estimated $v_\mathrm{eq}\sin i$ as the ratio between the radius and rotation period of the star. For stars like HD\,28205 and HD\,190406, these values are only references and were not used for the reconstructions because the stars have either no detected Zeeman signatures or too few observations. The symbol $\dagger$ means that we derived the value by means of ZDI optimisation.}
\end{table*}
\setlength{\tabcolsep}{6pt}

\section{Activity indices}\label{sec:indexes}

To gauge the magnetic activity level of our stars, we computed time series of activity indices. We used chromospheric spectral lines falling in the optical domain covered by Narval, namely the \ion{Ca}{II} H\&K lines, H$\alpha$, and \ion{Ca}{II} infrared triplet lines. In Table~\ref{tab:index_results}, we list the main statistical features of the time series for all three activity indicators.

The $\log R'_\mathrm{HK}$ quantifies the amount of flux in the \ion{Ca}{II} H\&K lines relative to the nearby continuum, without colour dependence and photospheric contribution \citep{Middelkoop1982,Noyes1984,Rutten1984}. The recipe to compute the index consists of three steps: i) measuring the $S$ index, ii) calibrating it to the Mount Wilson scale, and iii) converting it to the $\log R'_\mathrm{HK}$ index. Following the definition of \citet{Vaughan1978}, the $S$ index is measured via 
\begin{equation}
    S = \frac{aF_H+bF_K}{cF_R+dF_V}+e,
\end{equation}
where $F_H$ and $F_K$ are the fluxes in two triangular band passes with FWHM = 1.09\,{\AA} centred on the cores of the H line (3968.470\,{\AA}) and K line (3933.661\,{\AA}), whereas $F_R$ and $F_V$ are the fluxes within two 20-{\AA} rectangular band passes centred at 3901 and 4001\,{\AA}, respectively. The set of coefficients $\{a,b,c,d,e\}$ is used to calibrate the $S$ index from a specific instrument scale to the Mount Wilson scale, and were estimated by \citet{Marsden2014} for Narval. We then used the formula by \citet{Rutten1984} to convert the $S$ index to the $\log R'_\mathrm{HK}$ index. 

The H$\alpha$ index is measured as
\begin{equation}\label{eq:halpha}
    \mathrm{H}\alpha = \frac{F_{\mathrm{H}\alpha}}{H_R+H_V},
\end{equation}
where $F_{\mathrm{H}\alpha}$ is the flux within a rectangular band pass of 3.60\,{\AA} centred on the H$\alpha$ line at 6562.85\,{\AA}, and $H_V$ and $H_R$ are the fluxes within two rectangular bandpasses of 2.2\,{\AA} centred on 6558.85\,{\AA} and 6567.30\,{\AA} \citep{Gizis2002}. Finally, we followed \citet{Petit2013} and \citet{Marsden2014} to measure the \ion{Ca}{II} infrared triplet index as
\begin{equation}\label{eq:irt}
    \mathrm{CaII IRT} = \frac{\mathrm{IR1}+\mathrm{IR2}+\mathrm{IR3}}{\mathrm{IR}_R+\mathrm{IR}_V},
\end{equation}
where IR1, IR2, and IR3 are the fluxes within a rectangular band pass of 2\,{\AA} centred on the \ion{Ca}{II} lines at 8498.023\,{\AA}, 8542.091\,{\AA}, and 8662.410\,{\AA}, respectively, and IR$_R$ and IR$_V$ are the fluxes within two rectangular bandpasses of 5\,{\AA} centred on 8704.9\,{\AA} and 8475.8\,{\AA}.

\subsection{Longitudinal magnetic field}\label{sec:Bl}

We computed the disc-integrated, line-of-sight projected component of the large-scale magnetic field following \citet{Rees1979}. Formally, 
\begin{equation}
\mathrm{B}_l\;[G] = \frac{-2.14\cdot10^{11}}{\lambda_0 \mathrm{g}_{\mathrm{eff}}c}\frac{\int vV(v)dv}{\int(I_c-I)dv} \,,
\label{eq:Bl}
\end{equation}
where $\lambda_0$ (in nm) and $\mathrm{g}_\mathrm{eff}$ are the normalisation wavelength and Land\'e factor of the LSD profiles, $I_c$ is the unpolarised continuum level, $v$ is the radial velocity associated to a point in the spectral line profile in the star's rest frame (in km\,s$^{-1}$), and $c$ the speed of light in vacuum (in km\,s$^{-1}$). For all our stars, we set the normalisation parameters to $\lambda_0=700$\,nm and $\mathrm{g}_\mathrm{eff}=1.24$. The velocity range over which the integration is carried out encompass the width of both Stokes~$I$ and $V$ LSD profiles. In Table~\ref{tab:lsd_masks}, we list the velocity range for each star.

\section{Magnetic imaging}\label{sec:zdi}

We applied ZDI to reconstruct the large-scale magnetic field topology for the stars in our study. The magnetic field vector is expressed as the sum of a poloidal and toroidal component, each described via spherical harmonics formalism \citep{Lehmann2022}. The algorithm inverts a time series of Stokes~$V$ LSD profiles into a magnetic field map while applying a maximum-entropy regularisation scheme \citep{Skilling1984}. In practice, ZDI fits the spherical harmonics coefficients $\alpha_{\ell,m}$, $\beta_{\ell,m}$, and $\gamma_{\ell,m}$ (with $\ell$ and $m$ being the degree and order of the mode, respectively) in an iterative fashion, until a target reduced $\chi^2$ is reached \citep[for more information see][]{Skilling1984,Semel1989,Donati1997}. The coefficients $\alpha_{\ell,m}$, $\beta_{\ell,m}$, and $\gamma_{\ell,m}$ are complex numbers used to describe the radial poloidal, tangential poloidal, and toroidal field components, respectively. The algorithm searches for the maximum-entropy solution, that is, the magnetic field configuration compatible with the data and with the lowest information content. 

We employed the \texttt{zdipy} code described in \citet{Folsom2018} and adopted the weak-field approximation, for which Stokes~$V$ is proportional to the derivative of Stokes~$I$ with respect to wavelength \citep[e.g.][]{Landi1992}. As outlined in \citet{Folsom2018}, the local unpolarised line profiles are modelled with a Voigt kernel. For each star, we performed a $\chi^2_r$ minimisation between the median of the observed Stokes~$I$ LSD profiles and its model over a grid of line depth, Gaussian width and Lorentzian width values. The optimal values we used in the ZDI reconstruction are listed in Table~\ref{tab:wfa_stokesI}. 

We set the limb darkening coefficient to 0.6964 \citep{Claret2011} and the maximum degree of spherical harmonic coefficients to $l_\mathrm{max}=10$ for every ZDI reconstruction. The latter is consistent with the projected equatorial velocity ($v_\mathrm{eq}\sin i$) for our stars, which is at most 10.6\,km\,s$^{-1}$. $v_\mathrm{eq}\sin(i)$ determines $l_\mathrm{max}$ proportionally and the amount of magnetic field complexity one can image \citep{Hussain2009}. In fact, faster rotation implies more separated Zeeman signatures in radial velocity space, thus limiting polarity cancellation. In our case, we homogeneously set $l_\mathrm{max}$ equal to 10, but a lower value could have been used without changing the results, as most of our stars are slow rotators and the magnetic energy is stored in the low-$l$ degrees.

In cases where the  the stellar inclination was not available in the literature, it was estimated from geometrical considerations. More precisely, it was computed by comparing the stellar radius provided in the literature with the projected radius, $R\sin i=\mathrm{P}_\mathrm{rot}v_\mathrm{eq}\sin i/50.59$, where $R\sin i$ is measured in solar radii, P$_\mathrm{rot}$ is the stellar rotation period measured in days, and the $v_\mathrm{eq}\sin i$ unit is km\,s$^{-1}$. When the estimated inclination was larger than $80^{\circ}$, we adopted a value of $70^{\circ}$ to conservatively prevent mirroring effects between the northern and southern hemispheres.

The \textsc{zdipy} code includes differential rotation as a function of colatitude ($\theta$) expressed as
\begin{equation}\label{eq:diff_rot}
\Omega(\theta) = \Omega_\mathrm{eq} - d\Omega\sin^2(\theta),
\end{equation}
with $\Omega_\mathrm{eq}=2\pi/\mathrm{P}_\mathrm{rot}$ is the rotational frequency at the equator and $d\Omega$ is the differential rotation rate in rad\,d$^{-1}$. Following the parameter optimisation outlined in \citet{Donati2000} and \citet{Petit2002}, we generated a grid of (P$_\mathrm{rot}$, $d\Omega$) pairs and searched for the values that minimised the $\chi^2$ distribution between observations and synthetic LSD profiles, at a fixed entropy level. The best-fit parameters were measured by fitting a 2D paraboloid to the $\chi^2$ distribution, while the error bars were obtained from a variation of $\Delta\chi^2 = 1$ away from the minimum \citep{Press1992,Petit2002}. 

Once the stellar input parameters were fixed, it was necessary to determine a target $\chi_r^2$ that the ZDI algorithm is meant to converge to. This value represents the optimal fit quality of the Stokes~$V$ LSD profiles and should be the lowest possible. At the same time, it should be large enough to prevent overfitting the noise in the observations, which would lead to spurious artefacts in the final ZDI map. As outlined in \citet{Alvarado-Gomez2015}, finding the target $\chi_r^2$ is achieved by running ZDI over a grid of $\chi_r^2$ values, each time recording the entropy at convergence. In this way, we can observe the change in the information content of the ZDI map with $\chi^2_r$ and identify the target $\chi^2_r$ as the value corresponding to the maximum of the  change rate in the entropy.

\setlength{\tabcolsep}{5.5pt}
\begin{table*}[!t]
\caption{Properties of the magnetic maps.} 
\label{tab:zdi_output}     
\centering                       
\begin{tabular}{l c c c c c c c c c c c c c }      
\toprule
Star & Epoch & d$\Omega$ & $\langle|$B$_V|\rangle$   & $|$B$_\mathrm{max}|$ & $\langle B^2\rangle$  & f$_\mathrm{pol}$ & f$_\mathrm{tor}$  & f$_\mathrm{dip}$   & f$_\mathrm{quad}$  & f$_\mathrm{oct}$   & f$_\mathrm{axi}$ & f$_\mathrm{axi,pol}$ & f$_\mathrm{axi,tor}$ \\
& & [rad\,s$^{-1}$] & [G] & [G] & [$\times10^2$\,G$^2$] & [\%] & [\%] & [\%] & [\%] & [\%] & [\%] & [\%] & [\%]\\
\midrule
HD\,1835 & 2018.77   & $0.018\pm0.008$ & 7.4  & 29.3 & 0.78 & 91.1 & 8.9  & 32.1 & 28.4 & 27.4 & 6.4  & 6.9  & 1.2\\
HD\,43162 & 2019.01  & $\ldots$        & 11.0 & 23.3 & 1.50 & 63.8 & 36.2 & 84.0 & 11.1 & 3.5  & 63.9 & 45.7 & 96.0 \\
HD\,43162 & 2019.22  & $\ldots$        & 16.4 & 36.1 & 3.33 & 36.9 & 63.1 & 62.5 & 25.8 & 7.9  & 83.9 & 61.8 & 96.8 \\
HD\,82443 & 2019.22  & $0.114\pm0.022$ & 23.6 & 65.1 & 7.05 & 63.9 & 36.1 & 34.3 & 46.8 & 9.2 & 36.1 & 4.6  & 92.0\\
HD\,114710 & 2019.23 & $\ldots$        & 1.3  & 3.4  & 0.02 & 88.6 & 11.4 & 59.9 & 25.4 & 9.8  & 31.4 & 25.0 & 80.9\\
HD\,189733 & 2018.89 & $0.110^\dagger$ & 18.8 & 56.5 & 5.50 & 52.1 & 47.9 & 67.7 & 16.1 & 8.2  & 43.6 & 1.5  & 89.5\\
HD\,206860 & 2018.57 & $0.109\pm0.004$ & 20.8 & 47.7 & 5.10 & 85.5 & 14.5 & 46.8 & 16.8 & 13.5 & 64.6 & 61.0 & 85.7 \\
HD\,206860 & 2019.55 & $0.051\pm0.042$ & 15.8 & 35.3 & 2.80 & 69.2 & 30.8 & 59.4 & 21.0 & 10.5 & 73.4 & 65.2 & 91.9 \\
\bottomrule                                
\end{tabular}
\tablefoot{The following quantities are listed: star's name, epoch of observations, differential rotation rate, mean unsigned magnetic strength, maximum unsigned magnetic strength, total reconstructed magnetic energy, poloidal and toroidal magnetic energy as a fraction of the total energy, dipolar, quadrupolar, and octupolar magnetic energy as a fraction of the poloidal energy, axisymmetric magnetic energy as a fraction of the total energy, poloidal axisymmetric energy as a fraction of the poloidal energy, and toroidal axisymmetric energy as a fraction of the toroidal energy. The symbol $\dagger$ indicates that the value from \citet{Fares2017} was used.}
\end{table*}
\setlength{\tabcolsep}{6pt}

\section{Results}\label{sec:Results}

In this section, we describe the computation of activity indices and longitudinal magnetic field for each star individually. The average measurements are reported in Table~\ref{tab:index_results}, while all the values are listed in Table~\ref{app:log}. We also describe the magnetic field reconstructions with ZDI and (when possible) we compare them with previous maps available in the literature. The ZDI maps are given in Fig.~\ref{fig:zdi_A} and Fig.~\ref{fig:zdi_B} for the radial, azimuthal, and meridional components of the magnetic field. The properties of the ZDI maps and the results of the differential rotation search are summarised in Table~\ref{tab:zdi_output}. The table reports only the stars for which a Zeeman signature in Stokes~$V$ was detectable and with a number of observations producing sufficient rotational phase coverage to perform the tomographic inversion reliably.

We find average magnetic field strengths ranging from 1 to 25\,G. The magnetic field topologies are predominantly poloidal ($>60\%$), but there are cases where the toroidal component is dominant (63\%) or significant ($>20\%$). Most stars exhibit complex magnetic fields, as the dipolar component accounts for at most 60\% of the poloidal energy, and the quadrupolar and octupolar modes store between 10-40\% and 3-15\% of the energy. The axisymmetry of the reconstructed topologies also varies significantly within our sample, since the fraction of total energy in the corresponding modes spans between 6\% and 84\%, with two out of six stars featuring relatively axisymmetric magnetic orientations (65-73\% and 64-84\%). These results are in agreement with the expected magnetic topology for fast-rotating stars with Sun-like properties and interior, that is, with a convective envelope surrounding a radiative core \citep{Petit2005,Donati2009,Folsom2016,Rosen2016,Folsom2018,Willamo2022,Bellotti2025}. 

\subsection{HD\,1835 (BE\,Cet)}

HD\,1835 is a G2.5 dwarf \citep{Keenan1989} at a distance of 21.3\,pc \citep{GaiaCollaboration2020}. It is located in the Hyades cluster \citep{Montes2001} and its age was estimated to be 400\,Myr \citep{Gudel2007,Ramirez2012}, and its rotation period to 7.68\,d \citep{Gudel2007}. 

We measured $\log R'_\mathrm{HK}$ values between $-4.414$ and $-4.338$, with a median of $-4.384$ and a standard deviation of 0.025. These values are consistent with those reported in \citet{BoroSaikia2018} and \citet{Brown2022}. The H$\alpha$ index varies between 0.321 and 0.328, with a median of 0.326 and standard deviation of 0.002, and the \ion{Ca}{II} IRT ranges 0.853-0.890 with a median value of $0.866\pm0.010$. 

The longitudinal field is between $-5.7$ and $2.0$\,G, with a median value of $-2.2\pm2.5$\,G (the median error bar is 1.0\,G). In comparison, the range reported by \citet{Rosen2016} and \citet{Willamo2022} from 2013 and 2017 HARPS-Pol observations is approximately $-5$ to $10$\,G. While consistent, our 2018 range of B$_l$ values is narrower, possibly indicating a weakening of the magnetic field over a timescale of a year. 

\begin{figure}[t]
    \includegraphics[width=\columnwidth]{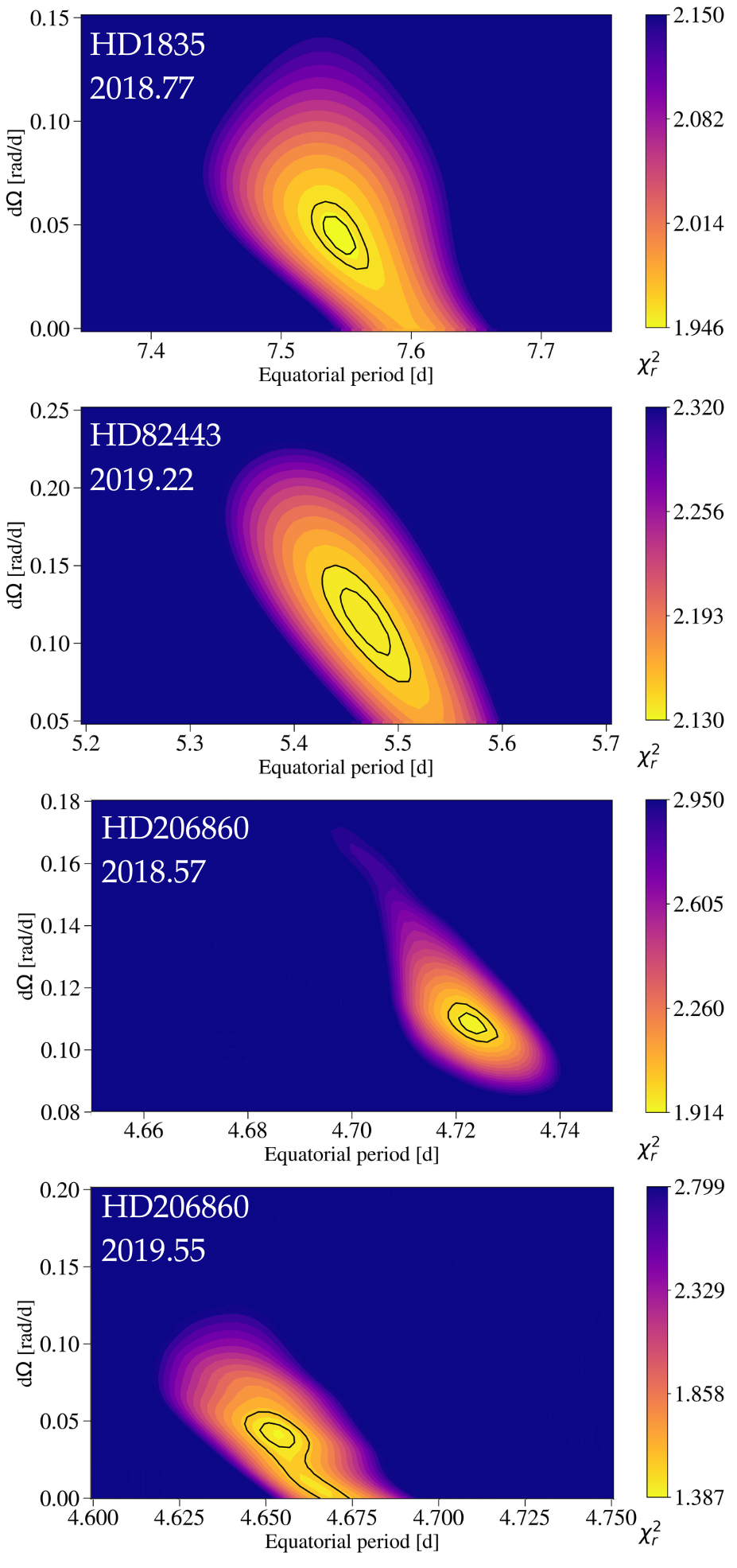}
    \caption{Joint search of differential rotation and equatorial rotation period for HD\,1835, HD\,82443, and HD\,206860. The panels illustrate the $\chi^2_r$ landscape over a grid of (P$_\mathrm{rot,eq}$,$d\Omega$) pairs, with the $1\sigma$ and $3\sigma$ contours. The best values are obtained by fitting a 2D paraboloid around the minimum, while their error bars are estimated from the projection of the $1\sigma$ contour on the respective axis \citep{Press1992}.}
    \label{fig:domega}
\end{figure}

We constrained differential rotation, as outlined in Sect.~\ref{sec:zdi}, finding that the optimised parameters are P$_\mathrm{rot}=7.57\pm0.02$\,d and $d\Omega=0.018\pm0.008$\,rad\,d$^{-1}$ (see Fig.~\ref{fig:domega}). Using Eq.~\ref{eq:diff_rot}, we estimated a rotation period at the pole of $7.74\pm0.08$\,d; thus, the rotation period of 7.68\,d reported in \citet{Gudel2007} is consistent with the expected equator-pole range of rotation rates. The initial $\chi^2_r$ was 6.5, which is associated to zero magnetic field map. The target $\chi_r^2$ of the reconstruction improved to 2.0 when solid body rotation was assumed, and down to 1.4 when differential rotation was included. The deviation of $\chi^2_r$ from 1.0 is symbolic of intrinsic variability that occurred during the time span of our observations and which is unaccounted for by the ZDI model. The time series for the modelled Stokes~$V$ profiles are given in Appendix~\ref{app:stokesV}

\begin{figure}[t]
    \includegraphics[width=\columnwidth]{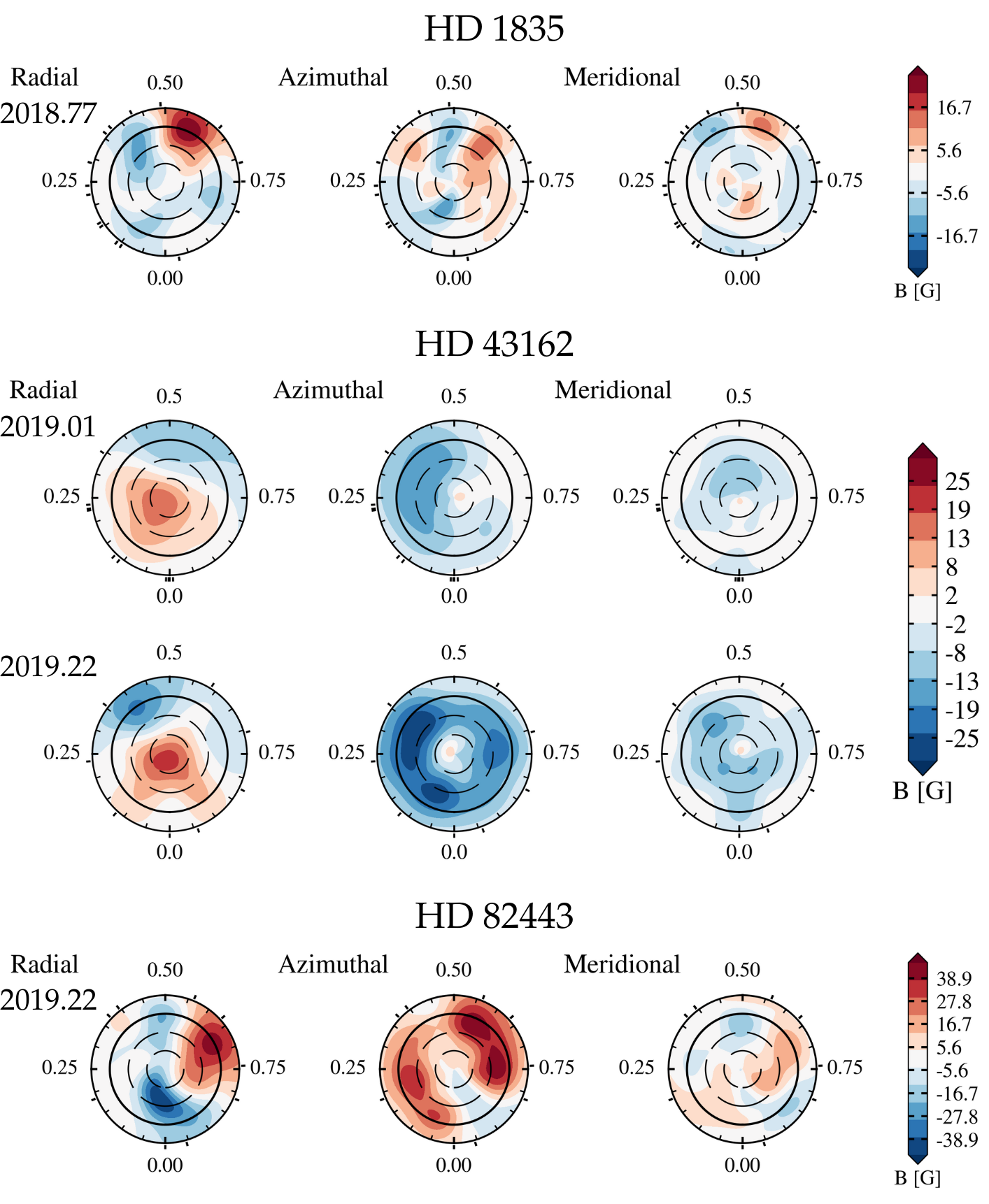}
    \caption{Reconstructed large-scale magnetic field map of HD\,1835, HD\,43162, and HD~82443 in flattened polar view. From the left, the radial, azimuthal, and meridional components of the magnetic field vector are illustrated. The radial ticks are located at the rotational phases when the observations were collected (see Eq.~\ref{eq:ephemeris}), while the concentric circles represent different stellar latitudes: -30\,$^{\circ}$, +30\,$^{\circ}$, and +60\,$^{\circ}$ (dashed lines), as well as the equator (solid line).}
    \label{fig:zdi_A}
\end{figure}

Our ZDI reconstruction of HD\,1835 is shown in Fig.~\ref{fig:zdi_A}, exhibiting low-latitude, almost equatorial magnetic structures. The field has a predominantly poloidal component (91\% of the total magnetic energy), of which 32\%, 28\%, and 37\% is in the dipolar, quadrupolar and octupolar modes. The field is largely non-axisymmetric, since the axisymmetric component (that is, $\ell\geq1$ and $m=0$) accounts for 6\% of the total energy. The reconstructed field shows similarities with the 2013 map of \citet{Rosen2016} and the 2017 map of \citet{Willamo2022}, both in terms of dominant component and axisymmetry. Compared to \citet{Rosen2016} reconstruction, our map manifests an increased quadrupolar component at the expense of the dipolar component, indicating a higher complexity. Compared to \citet{Willamo2022}, our ZDI reconstruction of the radial field features a more extended area covered by negative polarity.

The average magnetic field strength we recovered is 7\,G, which is approximately half the value from previous reconstructions. This goes in the same direction as the decreased range of variability of the longitudinal field. For HD\,1835, \citet{Egeland2017} reported two superimposed activity cycles with two main periodicities of 7.8\,yr and 20.8\,yr. In a similar manner, \citet{BoroSaikia2018} estimated periodicities of 9.06\,yr and 22\,yr. Although there are three large-scale magnetic field maps reconstructed for this star, they are more recent than the temporal baseline of $S$-index values used by \citet{Egeland2017} and \citet{BoroSaikia2018}, which complicates the comparison of the field topology and the phase of the $S$-index cycle.

\subsection{HD\,28205 (V993\,Tau)}

HD\,28205 is a G0.0 dwarf \citep{Linsky2020} at a distance of 46.98\,pc \citep{GaiaCollaboration2020}. It is located in the Hyades cluster \citep{Kraft1965} and its age was estimated to be 630\,Myr \citep{Linsky2020}, and its rotation period to 5.87\,d \citep{Pizzolato2003}. 

We measured $\log R'_\mathrm{HK}$ values between $-4.453$ and $-4.424$, with a median of $-4.438\pm0.009$ (the median error is 0.036 in comparison). These values are about 0.06 dex larger than the average index given in \citep{Brown2022}. The H$\alpha$ index is stable around $0.313\pm0.001$, and the \ion{Ca}{II} IRT varies between 0.843 and 0.849 with a median value of $0.848\pm0.002$. 

The longitudinal field was measured to be between $-2.9$ and $0.8$\,G, with a median value of $-1.0\pm1.3$\,G (the median error bar is 1.7\,G). The value is on the same order of magnitude as the unsigned, average longitudinal field reported by \citet{Brown2022}. Since we only had five observations for this star, mostly clustered around rotational phase 0.0, we did not perform temporal analyses or ZDI.

\subsection{HD\,30495 (IX\,Eri)}

HD\,30495 is a G2.5 dwarf \citep{Gray2006} at a distance of 13.24\,pc \citep{GaiaCollaboration2020}. It has an age of 400\,Myr \citep{Ramirez2012} and a rotation period of 11.36\,d \citep{Egeland2015}. 

We measured $\log R'_\mathrm{HK}$ values between $-4.481$ and $-4.495$, with a median of $-4.490\pm0.006$ (the median error is 0.030 in comparison). These values are compatible with \citet{BoroSaikia2018} and \citet{Brown2022}, but on the higher end of the range listed. This could be explained by the chromospheric activity cycle of the star, composed of two superimposed periodicities of 2 and 12\,yr \citep{Egeland2015,Brandenburg2017}. More precisely, extrapolating the trend shown in Fig.~1 of \citet{Egeland2015}, we note that the 2015 observations analysed by \citet{Brown2022} lie at a lower activity state than our 2018 observations. The H$\alpha$ index we computed is constant around $0.323\pm0.001$, and the \ion{Ca}{II} IRT is stable around $0.855\pm0.001$. 

The longitudinal field was measured to be between $0.4$ and $3.1$\,G, with a median value of $1.6\pm1.1$\,G (the median error bar is 1.1\,G). The range is compatible with the unsigned, average longitudinal field reported by \citet{Brown2022}. Since we only have three observations for this star, we did not perform any temporal analyses or ZDI.

\subsection{HD\,43162 (V352\,CMa)}

HD\,43162 A is a G6.5 dwarf \citep{Linsky2020}, and it is the primary component of a triple system at a distance of 16.69\,pc \citep{Chini2014,GaiaCollaboration2020}. It has an age of 320\,Myr \citep{Linsky2020} and a rotation period of 8.50\,d \citep{Marsden2014}.

We measured $\log R'_\mathrm{HK}$ values between $-4.398$ and $-4.315$, with a median of $-4.364\pm0.023$. While these values indicate a higher activity level than reported in \citet{Marsden2014} and \citet{Lehtinen2016} by about 0.06 dex, they are compatible with the measurements by \citet{BoroSaikia2018} and \citet{Brown2022}. The H$\alpha$ index varies between 0.336 and 0.342, with a median of $0.339\pm0.002$, and the \ion{Ca}{II} IRT ranges between 0.891 and 0.919 with a median value of $0.903\pm0.007$. The longitudinal field was measured to be between $-4.9$ and $7.2$\,G, with a median value of $2.4\pm3.3$\,G (the median error bar is 1.5\,G). This range encompasses the value reported in \citet{Marsden2014} as well as the unsigned average in \citet{Brown2022}. 

We carried out a ZDI reconstructions for both epochs of HD\,43162. The maps are shown in Fig.~\ref{fig:zdi_A} and the model Stokes~$V$ profiles in Fig.~\ref{fig:stokesV_A}. In this case, reliable Stokes~$V$ models of the LSD profiles could only be obtained with an optimisation of $v_\mathrm{eq}\sin i$ based on $\chi^2$ minimisation. The idea is similar to the grid search used to find differential rotation (see Sect.~\ref{sec:zdi}); however, in this case, only a grid over $v_\mathrm{eq}\sin i$ was used, while all the other parameters were kept fixed. We obtained a value of 7\,km\,s$^{-1}$, which is lower than the value of 9.6\,km\,s$^{-1}$ estimated by \citet{Valenti2005}, but consistent with previous measurements by \citet{Zejda2012}. We then searched for the optimised values of rotation period and differential rotation jointly, but the results were inconclusive; hence, we performed ZDI reconstructions assuming solid body rotation. When searching for the optimised value of rotation period alone, we found a $\chi^2$ minimum at the expected rotation period of 8.5\,d \citep{Lehtinen2020} for both epochs; hence, we used this value for the ZDI maps.

For the 2019.01 epoch, we set a target $\chi^2_r$ to 1.7 from an initial value of 6.2. We found an average field strength of 11\,G, with the topology being 64\% poloidal and 36\% toroidal. The dipolar mode dominates the poloidal component by storing 84\% of the magnetic energy, while the quadrupolar and octupolar account for 11\% and 4\%, respectively. The axisymmetric component is 64\%. For 2019.22, the target $\chi^2_r$ is 1.5, from an initial value of 11.2. The average field strength is 16\,G, and the field configuration is predominantly toroidal (63\%). The poloidal component is still largely dipolar (63\%), but the quadrupolar and octupolar components increased to 26\% and 8\%. 

No polarity reversal is seen between the two epochs. While we cannot exclude an evolution of the field on short timescales, the difference in dominant topology between the two epochs could also be due to the scarce phase coverage affecting the 2019.01 epoch. Furthermore, we observe that the Stokes~$V$ signatures are asymmetric: a larger positive lobe than the negative one (see Fig.~\ref{fig:stokesV_A}). The spurious polarisation signature in Stokes~$N$ is negligible compared to the Stokes~$V$ signature; hence, it does not affect its shape and symmetry. The asymmetry in Stokes~$V$ has been encountered before for solar-like stars such as $\xi$~Boo~A \citep{Petit2005,Morgenthaler2012} as well as more massive stars such as $\theta$~Leo and $\varepsilon$ UMa \citep{Blazere2016}. We note that this might be due to vertical gradients of photospheric velocity and magnetic field strength \citep{LopezAriste2002}. Interestingly, we only found this feature in the case of HD~43162.

\subsection{HD\,82443 (DX~Leo)}

HD\,82443 is a G9.0 dwarf \citep{Gray2003}, and it is the primary of a binary system with an M6 dwarf \citep{Lepine2007} at a distance of 18.07\,pc \citep{GaiaCollaboration2020}. It is a member of the Her-Lyr moving group \citep{Gaidos1998,Eisenbeiss2013}, with an age of 250\,Myr \citep{Folsom2016,Linsky2020} and a rotation period of P$_\mathrm{rot}=5.377$\,d \citep{Folsom2016}. 

We measured $\log R'_\mathrm{HK}$ values between $-4.170$ and $-4.115$, with a median of $-4.147\pm0.016$ (the median error bar is 0.011). These values are 0.05 dex larger than the average values reported in \citet{Folsom2016}, \citet{BoroSaikia2018}, and \citet{Brown2022}. The H$\alpha$ index varies between 0.375 and 0.383, with a median of $0.380\pm0.003$, and the \ion{Ca}{II} IRT ranges 0.974-0.995 with a median value of $0.986\pm0.006$. The values of both indices are $\sim0.015$ units higher than the values measured by \citep{Folsom2016}.

The longitudinal field was measured to be between $-10.9$ and $1.3$\,G, with a median value of $-4.1\pm8.4$\,G (the median error bar is 1.3\,G). In comparison, \citet{Folsom2016} measured the longitudinal field to be between $\pm20$\,G, and \citet{Brown2022} obtained an average B$_l$ of 11\,G in absolute value.

We applied ZDI and fit the Stokes~$V$ time series down to $\chi^2_r=2.2$ from an initial value of 35.1, when assuming solid body rotation. We then searched for differential rotation and constrained P$_\mathrm{rot}=5.460\pm0.026$\,d and $d\Omega=0.114\pm0.022$~rad\,d$^{-1}$ (see Fig.~\ref{fig:domega}). We caution that measurements of the differential rotation with less than 15 observations across the stellar rotation may be biased due to rotational phase gaps \citep{Petit2002}. The rotation period is consistent with the value constrained with ZDI by \citet{Folsom2016} within $1\sigma$. Our differential rotation rate is $\sim2\sigma$ larger than the lower limit estimated from photometric light curves by \citet{Messina1999}. Using Eq.~\ref{eq:diff_rot}, we estimated a rotation period at the pole of $6.070\pm0.130$\,d. By assuming our P$_\mathrm{rot}$, $d\Omega$ pair, the target $\chi^2_r$ of the ZDI reconstruction was improved to 1.7.

The magnetic field topology is shown in Fig.~\ref{fig:zdi_A} and the properties listed in Table~\ref{tab:zdi_output}. We obtained an average magnetic field strength of 23\,G, which is consistent with the reconstruction of \citet{Folsom2016}. The magnetic field topology is predominantly poloidal (64\%), with 34\%, 47\%, and 9\% of the magnetic energy stored in the dipolar, quadrupolar, and octupolar modes, respectively. The field is also non-axisymmetric (36\%). In comparison, the reconstruction of \citet{Folsom2016} using 2013 data exhibits a predominantly dipolar (71\%) and non-axisymmetric (8\%) topology, indicating an increase in complexity with our recent observations. Considering the toroidal component specifically, the axisymmetric energy fraction increases from 38\% in \citet{Folsom2016} to 92\% in our reconstruction. This evolution occurred over a timescale of approximately 5\,yr, which is similar on the same order of magnitude as the photometric activity cycle reported by \citet{Baliunas1995,Messina1999,Lehtinen2016}. If we assume that the dynamo processes operating in the stellar interior are Sun-like, we could attribute the evolution of the axisymmetric-toroidal fraction to a correlated, equatorward distribution of starspots.

\begin{figure}[t]
    \includegraphics[width=\columnwidth]{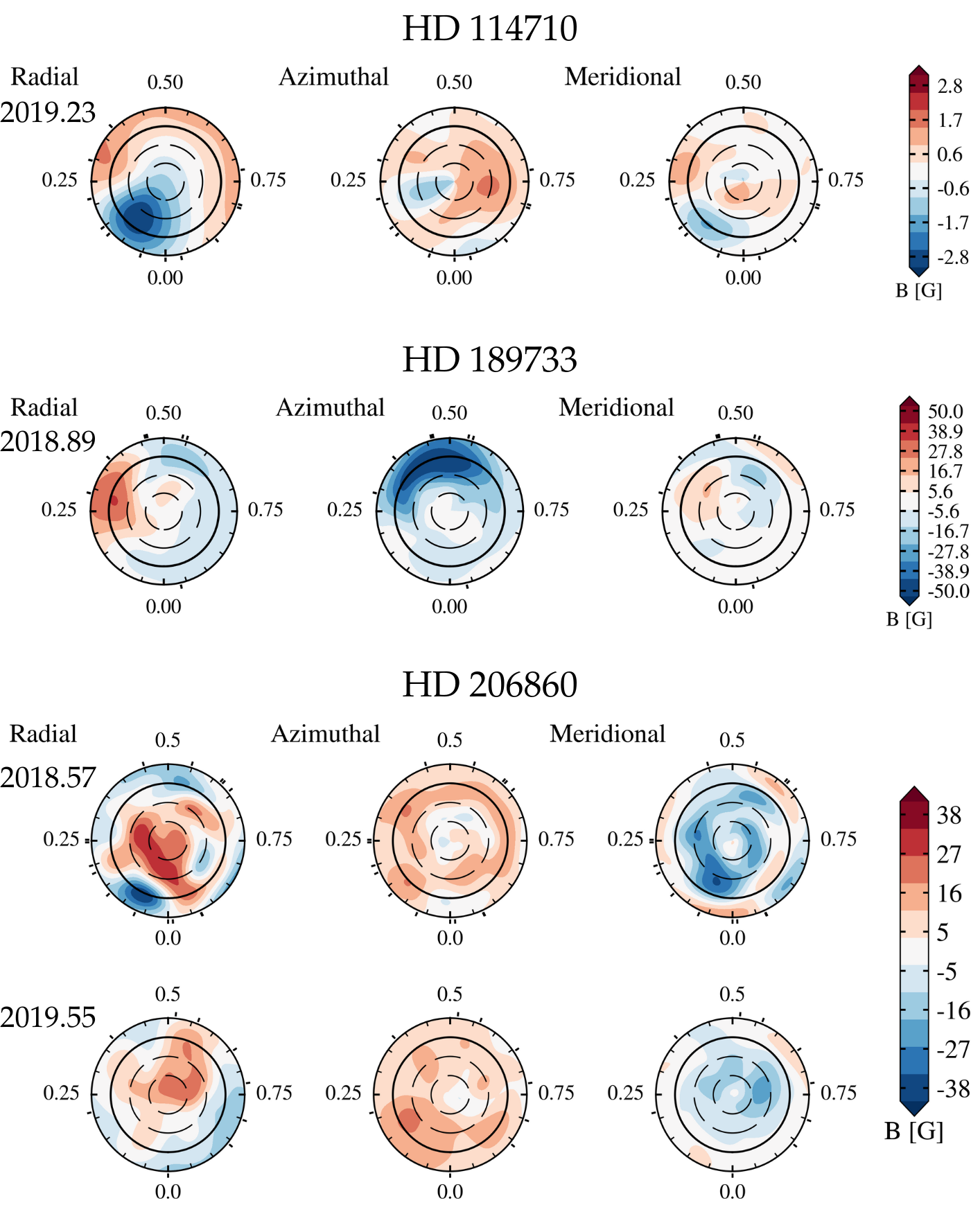}
    \caption{Reconstructed large-scale magnetic field map of HD\,114710, HD\,189733, and HD\,206860 for the two epochs. The format is the same as Fig.~\ref{fig:zdi_A}.}
    \label{fig:zdi_B}
\end{figure}

\subsection{HD\,114710 ($\beta$~Com)}

HD\,114710 is a G0.0 dwarf \citep{Keenan1989} at a distance of 9.19\,pc \citep{GaiaCollaboration2020} and part of a visual binary system \citep{Mason2001}. It has an age of 1.60\,Gyr \citep{Ramirez2012} and a rotation period of P$_\mathrm{rot}=12.35$\,d \citep{Wright2011}. 

We measured $\log R'_\mathrm{HK}$ values between $-4.633$ and $-4.591$, with a median of $-4.609\pm0.011$ (the median error bar is comparable to the standard deviation). These values are compatible with the measurements reported by \citet{BoroSaikia2018} and \citet{Brown2022}. The H$\alpha$ index varies between 0.312 and 0.315, with a median of $0.313\pm0.001$, and the \ion{Ca}{II} IRT ranges 0.793-0.813 with a median value of $0.798\pm0.006$. The longitudinal field was measured to be between $-2.7$ and $0.7$\,G, with a median value of $-0.3\pm1.0$\,G (where the median error bar is 0.5\,G). 

The Stokes~$V$ time series was fit down to $\chi^2_r=1.6$ from an initial value of 5.5 assuming solid body rotation, since the differential rotation search was inconclusive. The maps and the Stokes~$V$ fits are shown in Fig.~\ref{fig:zdi_B} and Fig.~\ref{fig:stokesV_A}, respectively. We caution that the amplitude of the spurious polarisation signature in Stokes~$N$ is of the same order of magnitude as the Zeeman signature in Stokes~$V$ for this star, but the shape and amplitude of Stokes~$V$ seems to be unaffected by the presence (or absence) of the Stokes~$N$ signature (see also Sect.~\ref{sec:observations}). The properties of the field topology are listed in Table~\ref{tab:zdi_output}. The magnetic field is predominantly poloidal (89\%), with 60\%, 25\%, and 10\% of the magnetic energy stored in the dipolar, quadrupolar, and octupolar modes, respectively. The field is also non-axisymmetric (31\% in the axisymmetric components) and exhibits an average field strength of 1.3\,G.

\subsection{HD\,149026 (HIP~808308)}

HD\,149026 is a G0.0 dwarf \citep{Sato2005} at a distance of 76.21\,pc \citep{GaiaCollaboration2020}. Its age was estimated to be 2.78\,Gyr \citep{Ramirez2012}. This star is known to host a Saturn-like planet with a mass of 0.38\,M$_\mathrm{Jup}$ on a 2.87\,d orbit \citep{Sato2005,Bonomo2017}. While there is no measurement of the rotation period for this star in the literature, we estimated a value of $12.3\pm1.2$\,d using the radius and the projected equatorial velocity (see~Table~\ref{tab:star_properties}).

From the three Narval observations, we measured $\log R'_\mathrm{HK}$ values between $-4.903$ and $-4.886$, with a median of $-4.900\pm0.010$. The value is consistent with \citet{BoroSaikia2018} and \citet{Brown2022}. The H$\alpha$ index is stable at 0.308, and the \ion{Ca}{II} IRT ranges 0.763-0.774 with a median value of $0.768\pm0.004$. 

Overall, this star is the least active of our sample. We did not detect any evident Zeeman signatures from the circularly polarised spectra of this star, but only one observation led to a marginal detection with a longitudinal magnetic field of $3.4\pm2.0$\,G. From observations with non-detections, we can still use Eq.~\ref{eq:Bl} to compute a $B_l$ value with error bar to determine the upper limits. The two non-detection observations led to $0.35\pm2.1$\,G and $2.2\pm2.2$\,G; thus, we placed a 3$\sigma$ upper limit of 6\,G to the longitudinal field detection (see Table~\ref{tab:index_results}).

\subsection{HD\,189733 (V452~Vul)}

HD\,189733 is a K2.0 dwarf \citep{Gray2003} at a distance of 19.78\,pc \citep{GaiaCollaboration2020}, and it is part of a visual binary system \citep{Mason2001}. The age is 400\,Myr \citep{Ramirez2012}, while the rotation period is 11.94\,d \citep{Fares2010}. The star hosts a hot Jupiter on a 2.22\,d orbit \citep{Bouchy2005b}.

We measured $\log R'_\mathrm{HK}$ values between $-4.485$ and $-4.449$, with a median of $-4.473\pm0.011$ (the median error bar is 0.026). Our range is compatible with the value computed by \citet{Brown2022} and 0.05\,dex larger than that reported by \citet{BoroSaikia2018}. The H$\alpha$ index varies between 0.362 and 0.364, with a median of $0.363\pm0.001$, and the \ion{Ca}{II} IRT ranges 0.862-0.880 with a median value of $0.867\pm0.005$.

The longitudinal field was measured to be between $-4.3$ and $1.8$\,G, with a median value of $-0.3\pm2.0$\,G (the median error bar is 1.1\,G). In comparison, \citet{Brown2022} measured an average B$_l$ of 4.5\,G in absolute value. 

We applied ZDI and fit the Stokes~$V$ time series down to $\chi^2_r=1.8$ from an initial value of 15.1, assuming solid body rotation. The maps are shown in Fig.~\ref{fig:zdi_B} and the Stokes~$V$ models in Fig.~\ref{fig:stokesV_A}. The search for differential rotation was inconclusive, owing to both a sparse time series and the poor phase coverage (mostly between 0.0 and 0.5). Thus, we attempted to carry out ZDI reconstructions with values of differential rotation rates obtained from previous ZDI reconstructions: $d\Omega=0.146\pm0.049$~rad\,d$^{-1}$\citep{Fares2010} and $d\Omega=0.110\pm0.050$~rad\,d$^{-1}$ from \citep{Fares2017}. In both cases, the value of the target $\chi^2_r$ was improved to $\sim$1.7. Compared to the ZDI map obtained assuming solid body rotation, the maps including $d\Omega$ differ by less than 0.5\,G in average field strength, less than 3\% in poloidal energy fraction, and less than 0.5\% for the axisymmetric energy fraction. The same was found for the dipolar and octupolar components, while the quadrupolar energy fraction decreased from 30 to 20\%, implying a simpler magnetic field configuration when differential rotation was accounted for. We decided to fix $d\Omega=0.110$~rad~d$^{-1}$ from \citet{Fares2017} since it was derived using the largest number of observations and for a data set in 2013, which is closest in time to our observations.

The properties of the magnetic topology are listed in Table~\ref{tab:zdi_output}. The field has an average field strength of 18\,G, and it has 52\% of the magnetic energy in the poloidal component, which is predominantly dipolar (68\%). The quadrupolar and octupolar components account for 16\% and 8\% of the energy, while the axisymmetric component is 44\%. Even though our map is reconstructed without an ideal phase coverage of the stellar rotation, our results are consistent with the previous reconstructions of \citet{Moutou2007,Fares2010,Fares2017}, with an equatorial magnetic spot of positive polarity in the radial component and an azimuthal component dominated by a negative polarity (see Fig.~3 in \citealt{Fares2017}). These authors studied the evolution of the magnetic topology of HD\,189733 from a nine-year spectropolarimetric monitoring. In addition, although a magnetic cycle could not be constrained, they corroborated the fast evolution of the topology over a few stellar rotations. This was manifested mainly in the variations in axisymmetry and poloidal-to-toroidal energy fraction, but the toroidal component was always the dominant one.

\subsection{HD\,190406 (GJ\,779)}

HD\,190406 is a G0.0 dwarf \citep{Gray2006} at a distance of 17.77\,pc \citep{GaiaCollaboration2020}, and it is part of a visual binary system \citep{Mason2001}. The estimated age is 1.70\,Gyr \citep{Ramirez2012} and the rotation period is P$_\mathrm{rot}=13.94$\,d \citep{Wright2011}.

We measured $\log R'_\mathrm{HK}$ values between $-4.719$ and $-4.663$, with a median of $-4.697\pm0.019$, which is consistent with the values reported by \citet{BoroSaikia2018} and \citet{Brown2022}. The H$\alpha$ index varies between 0.312 and 0.314, and the \ion{Ca}{II} IRT ranges 0.789-0.800 with a median value of $0.795\pm0.003$. We did not detect any Stokes~$V$ signatures in the eight Narval observations of this star. The average longitudinal field is 0.9\,G and the average uncertainty is 1.0\,G; thus, we set a 3$\sigma$ upper limit of 3\,G (see Table~\ref{tab:index_results}).

\subsection{HD\,206860 (HN~Peg)}

HD\,206860 is a G0.0 dwarf \citep{Gray2001} at a distance of 18.13\,pc \citep{GaiaCollaboration2020}, and it is part of a visual binary system \citep{Mason2001}. It is a member of the Her-Lyr moving group \citep{Eisenbeiss2013}, with an age of 200\,Myr \citep{Gudel2007,Ramirez2012} and a rotation period of P$_\mathrm{rot}=4.55$\,d \citep{BoroSaikia2015}. The star hosts a 16-M$_\mathrm{Jup}$ planet at a distance of 773\,au \citep{Luhman2007}.

We measured $\log R'_\mathrm{HK}$ values between $-4.356$ and $-4.294$, with a median of $-4.334\pm0.017$ (the median error bar is similar to the standard deviation). These values are at most 0.07 dex larger than the values reported in \citet{Marsden2014}, \citet{BoroSaikia2015}, \citet{BoroSaikia2018}, and \citet{Brown2022}. The H$\alpha$ index varies between 0.326 and 0.331, with a median of $0.327\pm0.001$, which is 0.012 and 0.005 units larger than the values of \citet{Marsden2014} and \citet{BoroSaikia2015}, respectively. The \ion{Ca}{II} IRT ranges 0.895-0.929 with a median value of $0.910\pm0.008$, which is comparable to the range of values from both \citet{Marsden2014} and \citet{BoroSaikia2015}.

The longitudinal field was measured to be between $2.5$ and $10.5$\,G, with a median value of $5.2\pm3.4$\,G (the median error bar is 1.5\,G). In comparison, \citet{BoroSaikia2015} measured longitudinal field values between $\pm14$\,G, and \citet{Brown2022} obtained an average B$_l$ of 6.4\,G in absolute value terms. 

The ZDI reconstructions were performed for 2018 and 2019 epochs of HD\,206860. The maps are shown in Fig.~\ref{fig:zdi_B} and the Stokes~$V$ models in Fig.~\ref{fig:stokesV_A}. Assuming solid body rotation, the target $\chi^2_r$ is 2.0 and 1.6 starting from 11.7 and 13.0 for the 2018 and 2019 epochs, respectively. We then constrained differential rotation via $\chi^2$ minimisation (see Fig.~\ref{fig:domega}). In 2018, we obtained P$_\mathrm{rot}=4.723\pm0.003$\,d and $d\Omega=0.109\pm0.004$\,rad\,d$^{-1}$ (implying a rotation period at the pole of $5.145\pm0.017$\,d), while in 2019 we obtained P$_\mathrm{rot}=4.650\pm0.017$\,d and $d\Omega=0.051\pm0.042$\,rad\,d$^{-1}$ (corresponding to $4.832\pm0.157$\,d at the pole). Using these values, the target $\chi^2_r$ was improved to 1.3 and 1.1 for the two epochs. The values of $d\Omega$ differ by roughly a factor of two, but they are reasonably consistent within uncertainties. The error bar on the differential rotation rate extracted from the 2019 epoch is larger owing to the incomplete phase coverage of the observations used in the analysis.

We note that the differential rotation rates we constrained for 2018 and 2019 are a factor of 2 and 4 smaller, respectively, than those measured by \citet{BoroSaikia2015}. If  we were to perform the ZDI reconstruction using $d\Omega=0.22$\,rad\,d$^{-1}$ from \citet{BoroSaikia2015}, the $\chi^2_r$ improvement would not be as substantial (of about 0.1-0.2) as when using our derived values. This may suggest an evolution of the differential shear at the surface of the star, which was also recorded for other stars \citep[AB~Dor, LQ~Hya, and HR~1099][]{Donati2003b}, and for the Sun in correlation with the magnetic cycle \citep{Beljan2022}.

The properties of the field topology are listed in Table~\ref{tab:zdi_output}. In 2018, the magnetic field is predominantly poloidal (86\%), with 47\%, 17\%, and 14\% of the magnetic energy stored in the dipolar, quadrupolar, and octupolar modes, respectively, while the axisymmetric component accounts for 65\% of the total magnetic energy. The situation is similar for 2019, the main difference being the decrease in the poloidal component to 69\% and the corresponding increase in the toroidal component. The poloidal component also becomes more dipolar (60\%). In terms of average magnetic field strength, the 2018 epoch featured a value of 21\,G, while in 2019 it decreased to 16\,G.

In general, the maps of HD\,206860 are consistent with both the ZDI reconstructions done by  \citet{Rosen2016}, with Narval data between 2007-2011 and HARPS-Pol in 2013, and by \citet{BoroSaikia2015}, with the Narval 2007-2011 time series. Indeed, they also feature a primarily poloidal, dipolar, and axisymmetric configuration, with the positive polarity predominantly located  in the northern hemisphere. The maps also share similarities in terms of complexity, with magnetic patches extending down to low latitudes. More specifically, our reconstructions from 2018 and 2019 resemble the previous ZDI maps based on the 2009 data.

\subsection{HD\,219828 (HIP\,115100)}

HD\,219828 is a G0.0 dwarf \citep{Moore1950} at a distance of 72.56\,pc \citep{GaiaCollaboration2020}. The estimated age is 6.08\,Gyr \citep{Ramirez2012} and the rotation period is P$_\mathrm{rot}=28.70$\,d \citep{Santos2016}. This is the most distant, oldest, and slowest rotator in our sample. The star hosts a hot Neptune (mass of 0.06\,M$_\mathrm{Jup}$) and a long-period planet (mass of 15\,M$_\mathrm{Jup}$) at a distance of 0.05 and 5.8\,au, respectively \citep{Santos2016}.

We measured $\log R'_\mathrm{HK}$ values between $-5.000$ and $-4.966$, with a median of $-4.986\pm0.011$, which is 0.1 dex lower than what was reported by \citet{BoroSaikia2018} . The H$\alpha$ index varies between 0.316 and 0.318, and the \ion{Ca}{II} IRT ranges 0.747-0.783 with a median value of $0.758\pm0.010$. We did not clearly detect Stokes~$V$ signatures in the ten Narval observations of this star. For one of them, there is a marginal detection and its associated longitudinal magnetic field is $3.6\pm2.0$\,G. The average longitudinal field is 1\,G and the average uncertainty is 1.8\,G, so we set a 3$\sigma$ upper limit of 5\,G.

\section{Activity trends}\label{sec:trends}

As summarised in Table~\ref{tab:star_properties}, the stars in our sample have ages between 0.2 and 6.1\,Gyr, while the rotation period ranges between 4.6 and 28.7\,d. Although the main purpose of this work is to provide magnetic field maps for modelling the stellar wind and environment in a subsequent paper, it is interesting to contextualise our measurements of activity indicators, longitudinal field, and magnetic field properties with previous studies. In particular, we  discuss the trends with respect to stellar age and rotation period below.

\begin{figure}
    \includegraphics[width=\columnwidth]{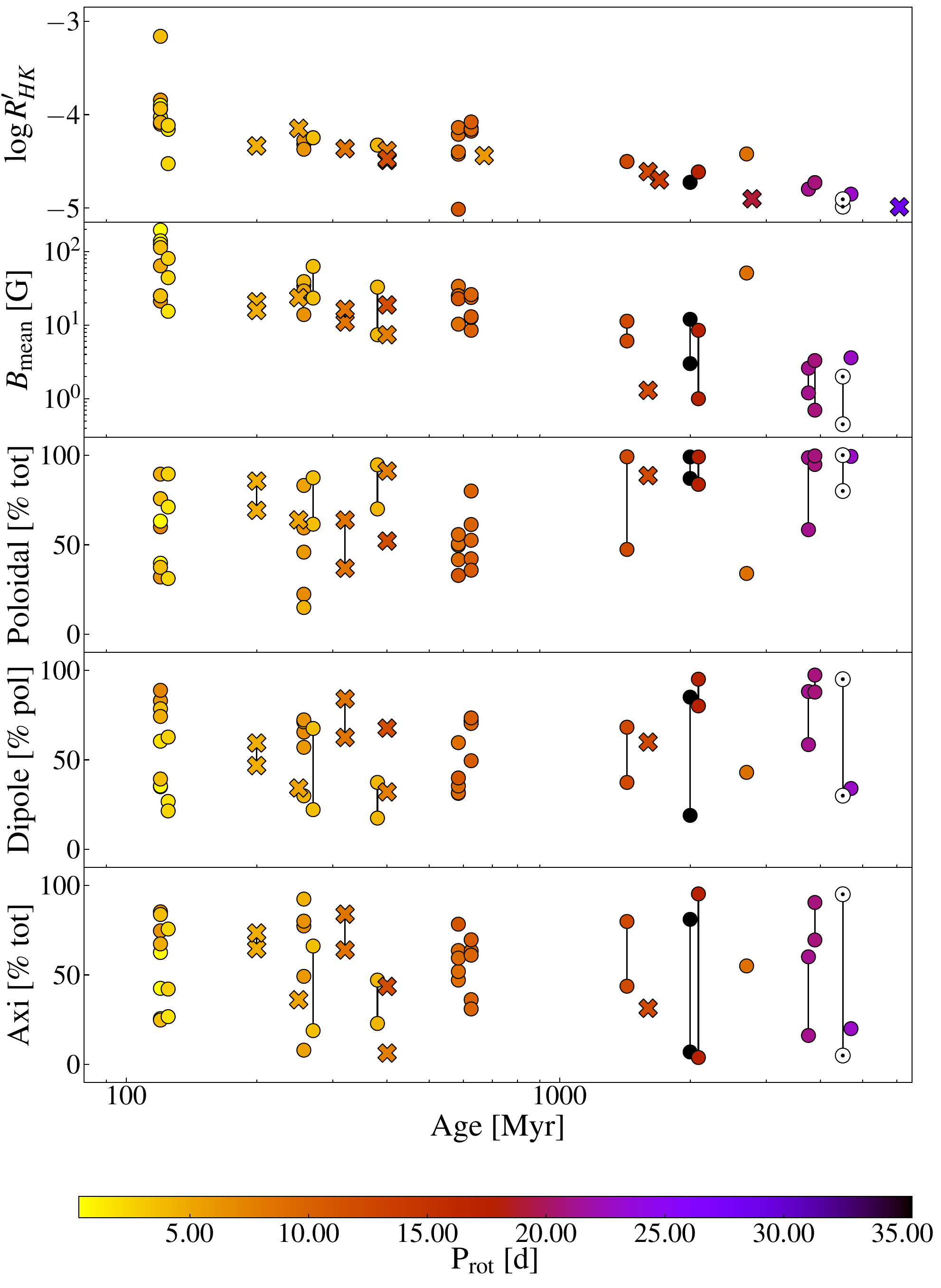}
    \caption{Trends of activity indices and reconstructed magnetic field geometry with age. From the top: $\log R'_\mathrm{HK}$, average magnetic field strength, poloidal fraction of the total energy, dipolar fraction of the poloidal energy, and axisymmetric fraction of the total energy. Properties measured for the same star but at different epochs are connected by a solid line. Our measurements are represented by `X' and measurements by \citet[][and references therein]{Folsom2016,Folsom2018} and \citet{Bellotti2025} are indicated with circles. The solar values  from the work of \citet{Egeland2017} and \citet{Vidotto2018} are also included.
        \label{fig:topology_age}}
\end{figure}

Studies of activity indices around cool stars on the main sequence by \citet{Pace2013} \citet{Marsden2014}, and \citet{Brown2022} showed a decrease in stellar activity with increasing stellar age and rotation period. We noted a consistent behaviour (see Fig.~\ref{fig:topology_age} and Table~\ref{tab:index_results}), where larger average values of activity indicators and longitudinal field towards younger and faster rotating stars. Quantitatively, we observe for example that our median longitudinal field decreases from 5\,G to less than 1\,G over from 0.1 to 4\,Gyr (or from 5\,d to 15\,d rotation period) and our $\log R'_\mathrm{HK}$ decreases from $-4.2$ to $-5.0$.

Turning to the large-scale magnetic field geometry, \citet{Vidotto2014b,Folsom2016,Rosen2016,Folsom2018} reported trends across different ages and rotation periods for Sun-like stars similar to those examined in this work. They reported a clear trend of decreasing average magnetic field with increasing age and rotation period, which we consistently observed for our stars as well. The average field strength of our six stars exhibits a decrease from 25 to 10\,G between 0.2 and 0.4\,Gyr. The oldest star in this sub-sample (at 1.6\,Gyr) has an average magnetic field strength of 1\,G. We show this trend in Fig.~\ref{fig:topology_age}, where we complement our measurements with those of \citet{Petit2008}, \citet{Folsom2016,Folsom2018}, and \citet{Bellotti2025}. 

Figure~\ref{fig:topology_age} also shows the variation with age and rotation period of the magnetic energy fraction in the poloidal, dipolar, and axisymmetric components. The addition of our measurements did not reveal clear trends in dipolar nor axisymmetric energy fraction, as already reported in \citet{Folsom2018}. For the poloidal energy fraction, we expect the magnetic field of slow rotators and old stars to have a predominantly poloidal geometry, while faster-rotating and younger stars to have mixed geometries \citep{Petit2008,Folsom2016}. Our measurements are consistent with these expectations.

Overall, although the age estimate is not used in a ZDI reconstruction, inhomogeneous or inaccurate age measurements can increase scatter. Moreover, the search of these trends is complicated by temporal variations in the magnetic field in the form of intrinsic variability (short-term) or magnetic cycles (long-term), as revealed also for other Sun-like stars by spectropolarimetry \citep[e.g.][]{BoroSaikia2016,Mengel2016,Fares2017,Jeffers2022,Bellotti2025}. For two of the stars examined here, HD~43162 and HD~206860, we reconstructed two maps corresponding to two distinct observational epochs. The magnetic properties reconstructed with ZDI are different at these epochs, which increases the scatter as illustrated in Fig.~\ref{fig:topology_age}. A more thorough search for trends of the magnetic geometry over age and characterisation of this scatter would require multiple ZDI maps for the same star over different years to include such a temporal variability \citep{Jeffers2023}. 

\section{Conclusions}\label{sec:conclusions}

In this study, we  analysed the spectropolarimetric time series of eleven Sun-like stars collected with Narval between 2018 and 2019. Our aim has been to characterise the magnetic properties of these stars for subsequent modelling of their stellar environments, which is topical in light of space-based missions dedicated to planetary atmospheric characterisation (e.g. $JWST$ currently and $Ariel$ in the future), as well as for direct imaging (e.g. LIFE \citealp{Quanz2018,Quanz2022} and the Habitable World Observatory \citealp{2021pdaa.book.....N}). Indeed, knowledge of the stellar magnetic environment that exoplanets are embedded in feeds back to, for instance, the interpretation of atmospheric signatures due to star-planet interactions \citep[e.g.][]{Carolan2021,Gupta2023}, as well as  habitability assessments \citep[e.g.][]{Vidotto2013,Lingam2019,vanLooveren2024}.

The stellar ages and rotation periods of our stars span different regions of the parameter space, from 0.2 to 6.1\,Gyr and from 4.6 to 28.7\,d, respectively, allowing us to sample stellar activity and magnetic field properties at different stages on the main sequence. As a first-order characterisation of the stars' magnetic activity, we computed chromospheric activity indices such as $\log R'_\mathrm{HK}$, H$\alpha$, and the \ion{Ca}{II} infrared triplet from unpolarised spectra as well as the longitudinal magnetic field B$_l$ from circularly polarised spectra. Overall, we noted trends expected from previous work: in particular, stellar activity decreasing with increasing age and rotation period \citep[e.g.][]{Marsden2014,Brown2022}. 

For the six stars with a detectable circular polarisation signature and large number of observations, we reconstructed the large-scale magnetic field configuration via ZDI. This was the first magnetic field reconstruction for two stars: HD\,43162 and HD\,114710. The properties of the large-scale magnetic field are in good agreement in terms of field strength and complexity with stars of similar properties. We found average magnetic field strengths ranging from 1 to 25\,G, where the lowest values belong to old or slow rotating stars, consistently with what is expected for this stellar parameter space \citep[see e.g.][]{Rosen2015,Folsom2016,Folsom2018}. Most stars exhibit complex magnetic fields that are predominantly poloidal and with a significant ($>$10\%) toroidal component and with distinct degrees of axisymmetry. 

In agreement with \citet{See2015} and \citet{Folsom2018}, we found that the poloidal-dominated topologies have various levels of axisymmetry, whereas the toroidal-dominated topologies are mostly axisymmetric. We also observed a decreasing trend of the average field strength with increasing age and rotation period. We did not record evident trends among the magnetic field properties, such as the poloidal fraction or axisymmetric fraction with respect to stellar age and rotation period, which could be due to temporal variations in the field topology, namely, in the magnetic cycles.

Our results stress the importance of spectropolarimetric studies to investigate the magnetic field properties of exoplanet-hosting stars, given the variety of properties of the large-scale magnetic field for Sun-like stars over time. This likely translates into a variety of stellar wind properties and offers a range of diverse magnetic environments in which exoplanets could be embedded. Ultimately, this study comprises a systematic attempt to characterise the evolution of environmental conditions and habitability of exoplanets from  the perspective of stellar magnetism.

\section{Data availability}

All analysed spectropolarimetric observations collected with Narval are available in Polarbase. They can be found at \url{https://www.polarbase.ovgso.fr/}

\begin{acknowledgements}

S.B. acknowledges funding by the Dutch Research Council (NWO) under the project "Exo-space weather and contemporaneous signatures of star-planet interactions" (with project number OCENW.M.22.215 of the research programme "Open Competition Domain Science- M"). S.B. also acknowledges funding from the SCI-S department of the European Space Agency (ESA), under the Science Faculty Research fund E/0429-03. S.B.S. acknowledges funding from the Austrian Science Fund (FWF) Lise Meitner project M2829-N. C.P.F. acknowledges funding from the European Union's Horizon Europe research and innovation program under grant agreement No. 101079231 (EXOHOST), and from the United Kingdom Research and Innovation Horizon Europe Guarantee Scheme (grant No. 10051045).

This work used the Dutch national e-infrastructure with the support of the SURF Cooperative using grant nos. EINF-2218 and EINF-5173. Based on observations obtained at the Canada-France-Hawaii Telescope (CFHT) which is operated by the National Research Council of Canada, the Institut National des Sciences de l'Univers of the Centre National de la Recherche Scientique of France, and the University of Hawaii. This work has made use of the VALD database, operated at Uppsala University, the Institute of Astronomy RAS in Moscow, and the University of Vienna; Astropy, a community-developed core Python package for Astronomy \citep{Astropy2013,Astropy2018}; NumPy \citep{VanderWalt2011}; Matplotlib: Visualization with Python \citep{Hunter2007}; SciPy \citep{Virtanen2020} and PyAstronomy \citep{Czesla2019}.

\end{acknowledgements}


\bibliographystyle{aa}
\bibliography{biblio}

\begin{thebibliography}{160}
\expandafter\ifx\csname natexlab\endcsname\relax\def\natexlab#1{#1}\fi

\bibitem[{{Airapetian} {et~al.}(2020){Airapetian}, {Barnes}, {Cohen},
  {Collinson}, {Danchi}, {Dong}, {Del Genio}, {France}, {Garcia-Sage},
  {Glocer}, {Gopalswamy}, {Grenfell}, {Gronoff}, {G{\"u}del}, {Herbst},
  {Henning}, {Jackman}, {Jin}, {Johnstone}, {Kaltenegger}, {Kay}, {Kobayashi},
  {Kuang}, {Li}, {Lynch}, {L{\"u}ftinger}, {Luhmann}, {Maehara}, {Mlynczak},
  {Notsu}, {Osten}, {Ramirez}, {Rugheimer}, {Scheucher}, {Schlieder},
  {Shibata}, {Sousa-Silva}, {Stamenkovi{\'c}}, {Strangeway}, {Usmanov},
  {Vergados}, {Verkhoglyadova}, {Vidotto}, {Voytek}, {Way}, {Zank}, \&
  {Yamashiki}}]{Airapetian2020}
{Airapetian}, V.~S., {Barnes}, R., {Cohen}, O., {et~al.} 2020, International
  Journal of Astrobiology, 19, 136

\bibitem[{{Airapetian} \& {Usmanov}(2016)}]{Airapetian2016}
{Airapetian}, V.~S. \& {Usmanov}, A.~V. 2016, \apjl, 817, L24

\bibitem[{{Allan} \& {Vidotto}(2019)}]{Allan2019}
{Allan}, A. \& {Vidotto}, A.~A. 2019, \mnras, 490, 3760

\bibitem[{{Alvarado-G{\'o}mez} {et~al.}(2022){Alvarado-G{\'o}mez}, {Cohen},
  {Drake}, {Fraschetti}, {Poppenhaeger}, {Garraffo}, {Chebly}, {Ilin},
  {Harbach}, \& {Kochukhov}}]{Alvarado-Gomez2022b}
{Alvarado-G{\'o}mez}, J.~D., {Cohen}, O., {Drake}, J.~J., {et~al.} 2022, \apj,
  928, 147

\bibitem[{{Alvarado-G{\'o}mez} {et~al.}(2018){Alvarado-G{\'o}mez}, {Hussain},
  {Drake}, {Donati}, {Sanz-Forcada}, {Stelzer}, {Cohen}, {Amazo-G{\'o}mez},
  {Grunhut}, {Garraffo}, {Moschou}, {Silvester}, \&
  {Oksala}}]{AlvaradoGomez2018}
{Alvarado-G{\'o}mez}, J.~D., {Hussain}, G. A.~J., {Drake}, J.~J., {et~al.}
  2018, \mnras, 473, 4326

\bibitem[{{Alvarado-G{\'o}mez} {et~al.}(2015){Alvarado-G{\'o}mez}, {Hussain},
  {Grunhut}, {Fares}, {Donati}, {Alecian}, {Kochukhov}, {Oksala}, {Morin},
  {Redfield}, {Cohen}, {Drake}, {Jardine}, {Matt}, {Petit}, \&
  {Walter}}]{Alvarado-Gomez2015}
{Alvarado-G{\'o}mez}, J.~D., {Hussain}, G.~A.~J., {Grunhut}, J., {et~al.} 2015,
  \aap, 582, A38

\bibitem[{{Astropy Collaboration} {et~al.}(2018){Astropy Collaboration},
  {Price-Whelan}, {Sip{\H{o}}cz}, {G{\"u}nther}, {Lim}, {Crawford}, {Conseil},
  {Shupe}, {Craig}, {Dencheva}, {Ginsburg}, {VanderPlas}, {Bradley},
  {P{\'e}rez-Su{\'a}rez}, {de Val-Borro}, {Aldcroft}, {Cruz}, {Robitaille},
  {Tollerud}, {Ardelean}, {Babej}, {Bach}, {Bachetti}, {Bakanov}, {Bamford},
  {Barentsen}, {Barmby}, {Baumbach}, {Berry}, {Biscani}, {Boquien}, {Bostroem},
  {Bouma}, {Brammer}, {Bray}, {Breytenbach}, {Buddelmeijer}, {Burke},
  {Calderone}, {Cano Rodr{\'\i}guez}, {Cara}, {Cardoso}, {Cheedella}, {Copin},
  {Corrales}, {Crichton}, {D'Avella}, {Deil}, {Depagne}, {Dietrich}, {Donath},
  {Droettboom}, {Earl}, {Erben}, {Fabbro}, {Ferreira}, {Finethy}, {Fox},
  {Garrison}, {Gibbons}, {Goldstein}, {Gommers}, {Greco}, {Greenfield},
  {Groener}, {Grollier}, {Hagen}, {Hirst}, {Homeier}, {Horton}, {Hosseinzadeh},
  {Hu}, {Hunkeler}, {Ivezi{\'c}}, {Jain}, {Jenness}, {Kanarek}, {Kendrew},
  {Kern}, {Kerzendorf}, {Khvalko}, {King}, {Kirkby}, {Kulkarni}, {Kumar},
  {Lee}, {Lenz}, {Littlefair}, {Ma}, {Macleod}, {Mastropietro}, {McCully},
  {Montagnac}, {Morris}, {Mueller}, {Mumford}, {Muna}, {Murphy}, {Nelson},
  {Nguyen}, {Ninan}, {N{\"o}the}, {Ogaz}, {Oh}, {Parejko}, {Parley}, {Pascual},
  {Patil}, {Patil}, {Plunkett}, {Prochaska}, {Rastogi}, {Reddy Janga},
  {Sabater}, {Sakurikar}, {Seifert}, {Sherbert}, {Sherwood-Taylor}, {Shih},
  {Sick}, {Silbiger}, {Singanamalla}, {Singer}, {Sladen}, {Sooley},
  {Sornarajah}, {Streicher}, {Teuben}, {Thomas}, {Tremblay}, {Turner},
  {Terr{\'o}n}, {van Kerkwijk}, {de la Vega}, {Watkins}, {Weaver}, {Whitmore},
  {Woillez}, {Zabalza}, \& {Astropy Contributors}}]{Astropy2018}
{Astropy Collaboration}, {Price-Whelan}, A.~M., {Sip{\H{o}}cz}, B.~M., {et~al.}
  2018, AJ, 156, 123

\bibitem[{{Astropy Collaboration} {et~al.}(2013){Astropy Collaboration},
  {Robitaille}, {Tollerud}, {Greenfield}, {Droettboom}, {Bray}, {Aldcroft},
  {Davis}, {Ginsburg}, {Price-Whelan}, {Kerzendorf}, {Conley}, {Crighton},
  {Barbary}, {Muna}, {Ferguson}, {Grollier}, {Parikh}, {Nair}, {Unther},
  {Deil}, {Woillez}, {Conseil}, {Kramer}, {Turner}, {Singer}, {Fox}, {Weaver},
  {Zabalza}, {Edwards}, {Azalee Bostroem}, {Burke}, {Casey}, {Crawford},
  {Dencheva}, {Ely}, {Jenness}, {Labrie}, {Lim}, {Pierfederici}, {Pontzen},
  {Ptak}, {Refsdal}, {Servillat}, \& {Streicher}}]{Astropy2013}
{Astropy Collaboration}, {Robitaille}, T.~P., {Tollerud}, E.~J., {et~al.} 2013,
  A\&A, 558, A33

\bibitem[{{Bagnulo} {et~al.}(2009){Bagnulo}, {Landolfi}, {Landstreet}, {Landi
  Degl'Innocenti}, {Fossati}, \& {Sterzik}}]{Bagnulo2009}
{Bagnulo}, S., {Landolfi}, M., {Landstreet}, J.~D., {et~al.} 2009, PASP, 121,
  993

\bibitem[{{Baliunas} {et~al.}(1995){Baliunas}, {Donahue}, {Soon}, {Horne},
  {Frazer}, {Woodard-Eklund}, {Bradford}, {Rao}, {Wilson}, {Zhang}, {Bennett},
  {Briggs}, {Carroll}, {Duncan}, {Figueroa}, {Lanning}, {Misch}, {Mueller},
  {Noyes}, {Poppe}, {Porter}, {Robinson}, {Russell}, {Shelton}, {Soyumer},
  {Vaughan}, \& {Whitney}}]{Baliunas1995}
{Baliunas}, S.~L., {Donahue}, R.~A., {Soon}, W.~H., {et~al.} 1995, ApJ, 438,
  269

\bibitem[{{Bellotti} {et~al.}(2023){Bellotti}, {Fares}, {Vidotto}, {Morin},
  {Petit}, {Hussain}, {Bourrier}, {Donati}, {Moutou}, \&
  {H{\'e}brard}}]{Bellotti2023a}
{Bellotti}, S., {Fares}, R., {Vidotto}, A.~A., {et~al.} 2023, \aap, 676, A139

\bibitem[{{Bellotti} {et~al.}(2025){Bellotti}, {Petit}, {Jeffers}, {Marsden},
  {Morin}, {Vidotto}, {Folsom}, {See}, \& {do Nascimento}}]{Bellotti2025}
{Bellotti}, S., {Petit}, P., {Jeffers}, S.~V., {et~al.} 2025, \aap, 693, A269

\bibitem[{{Bellotti} {et~al.}(2022){Bellotti}, {Petit}, {Morin}, {Hussain},
  {Folsom}, {Carmona}, {Delfosse}, \& {Moutou}}]{Bellotti2022}
{Bellotti}, S., {Petit}, P., {Morin}, J., {et~al.} 2022, \aap, 657, A107

\bibitem[{{Blaz{\`e}re} {et~al.}(2016){Blaz{\`e}re}, {Petit}, {Ligni{\`e}res},
  {Auri{\`e}re}, {Ballot}, {B{\"o}hm}, {Folsom}, {Gaurat}, {Jouve}, {Lopez
  Ariste}, {Neiner}, \& {Wade}}]{Blazere2016}
{Blaz{\`e}re}, A., {Petit}, P., {Ligni{\`e}res}, F., {et~al.} 2016, \aap, 586,
  A97

\bibitem[{{Bonomo} {et~al.}(2017){Bonomo}, {Desidera}, {Benatti}, {Borsa},
  {Crespi}, {Damasso}, {Lanza}, {Sozzetti}, {Lodato}, {Marzari}, {Boccato},
  {Claudi}, {Cosentino}, {Covino}, {Gratton}, {Maggio}, {Micela}, {Molinari},
  {Pagano}, {Piotto}, {Poretti}, {Smareglia}, {Affer}, {Biazzo}, {Bignamini},
  {Esposito}, {Giacobbe}, {H{\'e}brard}, {Malavolta}, {Maldonado}, {Mancini},
  {Martinez Fiorenzano}, {Masiero}, {Nascimbeni}, {Pedani}, {Rainer}, \&
  {Scandariato}}]{Bonomo2017}
{Bonomo}, A.~S., {Desidera}, S., {Benatti}, S., {et~al.} 2017, \aap, 602, A107

\bibitem[{{Boro Saikia} {et~al.}(2016){Boro Saikia}, {Jeffers}, {Morin},
  {Petit}, {Folsom}, {Marsden}, {Donati}, {Cameron}, {Hall}, {Perdelwitz},
  {Reiners}, \& {Vidotto}}]{BoroSaikia2016}
{Boro Saikia}, S., {Jeffers}, S.~V., {Morin}, J., {et~al.} 2016, A\&A, 594, A29

\bibitem[{{Boro Saikia} {et~al.}(2015){Boro Saikia}, {Jeffers}, {Petit},
  {Marsden}, {Morin}, \& {Folsom}}]{BoroSaikia2015}
{Boro Saikia}, S., {Jeffers}, S.~V., {Petit}, P., {et~al.} 2015, \aap, 573, A17

\bibitem[{{Boro Saikia} {et~al.}(2020){Boro Saikia}, {Jin}, {Johnstone},
  {L{\"u}ftinger}, {G{\"u}del}, {Airapetian}, {Kislyakova}, \&
  {Folsom}}]{BoroSaikia2020}
{Boro Saikia}, S., {Jin}, M., {Johnstone}, C.~P., {et~al.} 2020, \aap, 635,
  A178

\bibitem[{{Boro Saikia} {et~al.}(2018){Boro Saikia}, {Lueftinger}, {Jeffers},
  {Folsom}, {See}, {Petit}, {Marsden}, {Vidotto}, {Morin}, {Reiners}, {Guedel},
  \& {BCool Collaboration}}]{BoroSaikia2018}
{Boro Saikia}, S., {Lueftinger}, T., {Jeffers}, S.~V., {et~al.} 2018, \aap,
  620, L11

\bibitem[{{Bouchy} {et~al.}(2005){Bouchy}, {Udry}, {Mayor}, {Moutou}, {Pont},
  {Iribarne}, {da Silva}, {Ilovaisky}, {Queloz}, {Santos}, {S{\'e}gransan}, \&
  {Zucker}}]{Bouchy2005b}
{Bouchy}, F., {Udry}, S., {Mayor}, M., {et~al.} 2005, \aap, 444, L15

\bibitem[{{Brandenburg} {et~al.}(2017){Brandenburg}, {Mathur}, \&
  {Metcalfe}}]{Brandenburg2017}
{Brandenburg}, A., {Mathur}, S., \& {Metcalfe}, T.~S. 2017, \apj, 845, 79

\bibitem[{{Brown} {et~al.}(2022){Brown}, {Jeffers}, {Marsden}, {Morin}, {Boro
  Saikia}, {Petit}, {Jardine}, {See}, {Vidotto}, {Mengel}, {Dahlkemper}, \&
  {the BCool Collaboration}}]{Brown2022}
{Brown}, E.~L., {Jeffers}, S.~V., {Marsden}, S.~C., {et~al.} 2022, \mnras, 514,
  4300

\bibitem[{{Bus{\`a}} {et~al.}(2007){Bus{\`a}}, {Aznar Cuadrado}, {Terranegra},
  {Andretta}, \& {Gomez}}]{Busa2007}
{Bus{\`a}}, I., {Aznar Cuadrado}, R., {Terranegra}, L., {Andretta}, V., \&
  {Gomez}, M.~T. 2007, \aap, 466, 1089

\bibitem[{{Carolan} {et~al.}(2021){Carolan}, {Vidotto}, {Villarreal D'Angelo},
  \& {Hazra}}]{Carolan2021}
{Carolan}, S., {Vidotto}, A.~A., {Villarreal D'Angelo}, C., \& {Hazra}, G.
  2021, \mnras, 500, 3382

\bibitem[{{Cecchi-Pestellini} {et~al.}(2006){Cecchi-Pestellini}, {Ciaravella},
  \& {Micela}}]{CecchiPestellini2006}
{Cecchi-Pestellini}, C., {Ciaravella}, A., \& {Micela}, G. 2006, \aap, 458, L13

\bibitem[{{Chini} {et~al.}(2014){Chini}, {Fuhrmann}, {Barr}, {Pozo},
  {Westhues}, \& {Hodapp}}]{Chini2014}
{Chini}, R., {Fuhrmann}, K., {Barr}, A., {et~al.} 2014, \mnras, 437, 879

\bibitem[{{Claret} \& {Bloemen}(2011)}]{Claret2011}
{Claret}, A. \& {Bloemen}, S. 2011, A\&A, 529, A75

\bibitem[{{Cockell} {et~al.}(2016){Cockell}, {Bush}, {Bryce}, {Direito},
  {Fox-Powell}, {Harrison}, {Lammer}, {Landenmark}, {Martin-Torres},
  {Nicholson}, {Noack}, {O'Malley-James}, {Payler}, {Rushby}, {Samuels},
  {Schwendner}, {Wadsworth}, \& {Zorzano}}]{Cockell2016}
{Cockell}, C.~S., {Bush}, T., {Bryce}, C., {et~al.} 2016, Astrobiology, 16, 89

\bibitem[{{Czesla} {et~al.}(2019){Czesla}, {Schr{\"o}ter}, {Schneider},
  {Huber}, {Pfeifer}, {Andreasen}, \& {Zechmeister}}]{Czesla2019}
{Czesla}, S., {Schr{\"o}ter}, S., {Schneider}, C.~P., {et~al.} 2019, {PyA:
  Python astronomy-related packages}

\bibitem[{{Donati}(2003)}]{Donati2003}
{Donati}, J.~F. 2003, in Astronomical Society of the Pacific Conference Series,
  Vol. 307, Solar Polarization, ed. J.~{Trujillo-Bueno} \& J.~{Sanchez
  Almeida}, 41

\bibitem[{{Donati} \& {Brown}(1997)}]{DonatiBrown1997}
{Donati}, J.~F. \& {Brown}, S.~F. 1997, A\&A, 326, 1135

\bibitem[{{Donati} \& {Collier Cameron}(1997)}]{DonatiCollier1997}
{Donati}, J.~F. \& {Collier Cameron}, A. 1997, MNRAS, 291, 1

\bibitem[{{Donati} {et~al.}(2003){Donati}, {Collier Cameron}, \&
  {Petit}}]{Donati2003b}
{Donati}, J.~F., {Collier Cameron}, A., \& {Petit}, P. 2003, \mnras, 345, 1187

\bibitem[{{Donati} \& {Landstreet}(2009)}]{Donati2009}
{Donati}, J.~F. \& {Landstreet}, J.~D. 2009, Annual Review of Astronomy \&
  Astrophysics, 47, 333

\bibitem[{{Donati} {et~al.}(2000){Donati}, {Mengel}, {Carter}, {Marsden},
  {Collier Cameron}, \& {Wichmann}}]{Donati2000}
{Donati}, J.~F., {Mengel}, M., {Carter}, B.~D., {et~al.} 2000, MNRAS, 316, 699

\bibitem[{{Donati} {et~al.}(2008){Donati}, {Morin}, {Petit}, {Delfosse},
  {Forveille}, {Auri{\`e}re}, {Cabanac}, {Dintrans}, {Fares}, {Gastine},
  {Jardine}, {Ligni{\`e}res}, {Paletou}, {Ramirez Velez}, \&
  {Th{\'e}ado}}]{Donati2008}
{Donati}, J.~F., {Morin}, J., {Petit}, P., {et~al.} 2008, MNRAS, 390, 545

\bibitem[{{Donati} {et~al.}(1997){Donati}, {Semel}, {Carter}, {Rees}, \&
  {Collier Cameron}}]{Donati1997}
{Donati}, J.~F., {Semel}, M., {Carter}, B.~D., {Rees}, D.~E., \& {Collier
  Cameron}, A. 1997, MNRAS, 291, 658

\bibitem[{{Ducati}(2002)}]{Ducati2002}
{Ducati}, J.~R. 2002, VizieR Online Data Catalog

\bibitem[{{Egeland}(2017)}]{Egeland2017}
{Egeland}, R. 2017, PhD thesis, Montana State University, Bozeman

\bibitem[{{Egeland} {et~al.}(2015){Egeland}, {Metcalfe}, {Hall}, \&
  {Henry}}]{Egeland2015}
{Egeland}, R., {Metcalfe}, T.~S., {Hall}, J.~C., \& {Henry}, G.~W. 2015, \apj,
  812, 12

\bibitem[{{Eisenbeiss} {et~al.}(2013){Eisenbeiss}, {Ammler-von Eiff}, {Roell},
  {Mugrauer}, {Adam}, {Neuh{\"a}user}, {Schmidt}, \&
  {Bedalov}}]{Eisenbeiss2013}
{Eisenbeiss}, T., {Ammler-von Eiff}, M., {Roell}, T., {et~al.} 2013, \aap, 556,
  A53

\bibitem[{{Evensberget} {et~al.}(2021){Evensberget}, {Carter}, {Marsden},
  {Brookshaw}, \& {Folsom}}]{Evensberget2021}
{Evensberget}, D., {Carter}, B.~D., {Marsden}, S.~C., {Brookshaw}, L., \&
  {Folsom}, C.~P. 2021, \mnras, 506, 2309

\bibitem[{{Evensberget} {et~al.}(2022){Evensberget}, {Carter}, {Marsden},
  {Brookshaw}, {Folsom}, \& {Salmeron}}]{Evensberget2022}
{Evensberget}, D., {Carter}, B.~D., {Marsden}, S.~C., {et~al.} 2022, \mnras,
  510, 5226

\bibitem[{{Evensberget} {et~al.}(2023){Evensberget}, {Marsden}, {Carter},
  {Salmeron}, {Vidotto}, {Folsom}, {Kavanagh}, {Pineda}, {Driessen}, \&
  {Strickert}}]{Evensberget2023}
{Evensberget}, D., {Marsden}, S.~C., {Carter}, B.~D., {et~al.} 2023, \mnras,
  524, 2042

\bibitem[{{Fares} {et~al.}(2017){Fares}, {Bourrier}, {Vidotto}, {Moutou},
  {Jardine}, {Zarka}, {Helling}, {Lecavelier des Etangs}, {Llama}, {Louden},
  {Wheatley}, \& {Ehrenreich}}]{Fares2017}
{Fares}, R., {Bourrier}, V., {Vidotto}, A.~A., {et~al.} 2017, \mnras, 471, 1246

\bibitem[{{Fares} {et~al.}(2009){Fares}, {Donati}, {Moutou}, {Bohlender},
  {Catala}, {Deleuil}, {Shkolnik}, {Collier Cameron}, {Jardine}, \&
  {Walker}}]{Fares2009}
{Fares}, R., {Donati}, J.~F., {Moutou}, C., {et~al.} 2009, MNRAS, 398, 1383

\bibitem[{{Fares} {et~al.}(2010){Fares}, {Donati}, {Moutou}, {Jardine},
  {Grie{\ss}meier}, {Zarka}, {Shkolnik}, {Bohlender}, {Catala}, \& {Collier
  Cameron}}]{Fares2010}
{Fares}, R., {Donati}, J.~F., {Moutou}, C., {et~al.} 2010, \mnras, 406, 409

\bibitem[{{Folsom} {et~al.}(2018){Folsom}, {Bouvier}, {Petit}, {L{\`e}bre},
  {Amard}, {Palacios}, {Morin}, {Donati}, \& {Vidotto}}]{Folsom2018}
{Folsom}, C.~P., {Bouvier}, J., {Petit}, P., {et~al.} 2018, MNRAS, 474, 4956

\bibitem[{{Folsom} {et~al.}(2025){Folsom}, {Erba}, {Petit}, {Seadrow},
  {Stanley}, {Natan}, {Zaire}, {Oksala}, {Villadiego Forero}, {Moore}, \&
  {Catalan Olais}}]{Folsom2025}
{Folsom}, C.~P., {Erba}, C., {Petit}, V., {et~al.} 2025, arXiv e-prints,
  arXiv:2505.18476

\bibitem[{{Folsom} {et~al.}(2020){Folsom}, {{\'O} Fionnag{\'a}in}, {Fossati},
  {Vidotto}, {Moutou}, {Petit}, {Dragomir}, \& {Donati}}]{Folsom2020}
{Folsom}, C.~P., {{\'O} Fionnag{\'a}in}, D., {Fossati}, L., {et~al.} 2020,
  \aap, 633, A48

\bibitem[{{Folsom} {et~al.}(2016){Folsom}, {Petit}, {Bouvier}, {L{\`e}bre},
  {Amard}, {Palacios}, {Morin}, {Donati}, {Jeffers}, {Marsden}, \&
  {Vidotto}}]{Folsom2016}
{Folsom}, C.~P., {Petit}, P., {Bouvier}, J., {et~al.} 2016, MNRAS, 457, 580

\bibitem[{{Gaia Collaboration}(2020)}]{GaiaCollaboration2020}
{Gaia Collaboration}. 2020, VizieR Online Data Catalog, I/350

\bibitem[{{Gaidos}(1998)}]{Gaidos1998}
{Gaidos}, E.~J. 1998, \pasp, 110, 1259

\bibitem[{{Gizis} {et~al.}(2002){Gizis}, {Reid}, \& {Hawley}}]{Gizis2002}
{Gizis}, J.~E., {Reid}, I.~N., \& {Hawley}, S.~L. 2002, AJ, 123, 3356

\bibitem[{{Gray} \& {Baliunas}(1997)}]{Gray1997}
{Gray}, D.~F. \& {Baliunas}, S.~L. 1997, \apj, 475, 303

\bibitem[{{Gray} {et~al.}(2006){Gray}, {Corbally}, {Garrison}, {McFadden},
  {Bubar}, {McGahee}, {O'Donoghue}, \& {Knox}}]{Gray2006}
{Gray}, R.~O., {Corbally}, C.~J., {Garrison}, R.~F., {et~al.} 2006, \aj, 132,
  161

\bibitem[{{Gray} {et~al.}(2003){Gray}, {Corbally}, {Garrison}, {McFadden}, \&
  {Robinson}}]{Gray2003}
{Gray}, R.~O., {Corbally}, C.~J., {Garrison}, R.~F., {McFadden}, M.~T., \&
  {Robinson}, P.~E. 2003, \aj, 126, 2048

\bibitem[{{Gray} {et~al.}(2001){Gray}, {Napier}, \& {Winkler}}]{Gray2001}
{Gray}, R.~O., {Napier}, M.~G., \& {Winkler}, L.~I. 2001, \aj, 121, 2148

\bibitem[{{G{\"u}del}(2007)}]{Gudel2007}
{G{\"u}del}, M. 2007, Living Reviews in Solar Physics, 4, 3

\bibitem[{{Guinan} {et~al.}(2003){Guinan}, {Ribas}, \& {Harper}}]{Guinan2003}
{Guinan}, E.~F., {Ribas}, I., \& {Harper}, G.~M. 2003, \apj, 594, 561

\bibitem[{{Gupta} {et~al.}(2023){Gupta}, {Basak}, \& {Nandy}}]{Gupta2023}
{Gupta}, S., {Basak}, A., \& {Nandy}, D. 2023, \apj, 953, 70

\bibitem[{{H{\o}g} {et~al.}(2000){H{\o}g}, {Fabricius}, {Makarov}, {Urban},
  {Corbin}, {Wycoff}, {Bastian}, {Schwekendiek}, \& {Wicenec}}]{Hog2000}
{H{\o}g}, E., {Fabricius}, C., {Makarov}, V.~V., {et~al.} 2000, \aap, 355, L27

\bibitem[{{Hunter}(2007)}]{Hunter2007}
{Hunter}, J.~D. 2007, Computing in Science and Engineering, 9, 90

\bibitem[{{Hussain} {et~al.}(2009){Hussain}, {Collier Cameron}, {Jardine},
  {Dunstone}, {Ramirez Velez}, {Stempels}, {Donati}, {Semel}, {Aulanier},
  {Harries}, {Bouvier}, {Dougados}, {Ferreira}, {Carter}, \&
  {Lawson}}]{Hussain2009}
{Hussain}, G.~A.~J., {Collier Cameron}, A., {Jardine}, M.~M., {et~al.} 2009,
  MNRAS, 398, 189

\bibitem[{{Jeffers} {et~al.}(2017){Jeffers}, {Boro Saikia}, {Barnes}, {Petit},
  {Marsden}, {Jardine}, {Vidotto}, \& {BCool Collaboration}}]{Jeffers2017}
{Jeffers}, S.~V., {Boro Saikia}, S., {Barnes}, J.~R., {et~al.} 2017, MNRAS,
  471, L96

\bibitem[{{Jeffers} {et~al.}(2022){Jeffers}, {Cameron}, {Marsden}, {Boro
  Saikia}, {Folsom}, {Jardine}, {Morin}, {Petit}, {See}, {Vidotto}, {Wolter},
  \& {Mittag}}]{Jeffers2022}
{Jeffers}, S.~V., {Cameron}, R.~H., {Marsden}, S.~C., {et~al.} 2022, A\&A, 661,
  A152

\bibitem[{{Jeffers} {et~al.}(2023){Jeffers}, {Kiefer}, \&
  {Metcalfe}}]{Jeffers2023}
{Jeffers}, S.~V., {Kiefer}, R., \& {Metcalfe}, T.~S. 2023, \ssr, 219, 54

\bibitem[{{Johnstone} {et~al.}(2021){Johnstone}, {Bartel}, \&
  {G{\"u}del}}]{Johnstone2021}
{Johnstone}, C.~P., {Bartel}, M., \& {G{\"u}del}, M. 2021, \aap, 649, A96

\bibitem[{{Johnstone} {et~al.}(2015{\natexlab{a}}){Johnstone}, {G{\"u}del},
  {Brott}, \& {L{\"u}ftinger}}]{Johnstone2015b}
{Johnstone}, C.~P., {G{\"u}del}, M., {Brott}, I., \& {L{\"u}ftinger}, T.
  2015{\natexlab{a}}, \aap, 577, A28

\bibitem[{{Johnstone} {et~al.}(2015{\natexlab{b}}){Johnstone}, {G{\"u}del},
  {L{\"u}ftinger}, {Toth}, \& {Brott}}]{Johnstone2015a}
{Johnstone}, C.~P., {G{\"u}del}, M., {L{\"u}ftinger}, T., {Toth}, G., \&
  {Brott}, I. 2015{\natexlab{b}}, \aap, 577, A27

\bibitem[{{Joner} {et~al.}(2006){Joner}, {Taylor}, {Laney}, \& {van
  Wyk}}]{Joner2006}
{Joner}, M.~D., {Taylor}, B.~J., {Laney}, C.~D., \& {van Wyk}, F. 2006, \aj,
  132, 111

\bibitem[{{Keenan} \& {McNeil}(1989)}]{Keenan1989}
{Keenan}, P.~C. \& {McNeil}, R.~C. 1989, \apjs, 71, 245

\bibitem[{{Ketzer} \& {Poppenhaeger}(2023)}]{Ketzer2023}
{Ketzer}, L. \& {Poppenhaeger}, K. 2023, \mnras, 518, 1683

\bibitem[{{Kochukhov} {et~al.}(2020){Kochukhov}, {Hackman}, {Lehtinen}, \&
  {Wehrhahn}}]{Kochukhov2020}
{Kochukhov}, O., {Hackman}, T., {Lehtinen}, J.~J., \& {Wehrhahn}, A. 2020,
  \aap, 635, A142

\bibitem[{{Kochukhov} {et~al.}(2010){Kochukhov}, {Makaganiuk}, \&
  {Piskunov}}]{Kochukhov2010a}
{Kochukhov}, O., {Makaganiuk}, V., \& {Piskunov}, N. 2010, A\&A, 524, A5

\bibitem[{{Koen} {et~al.}(2010){Koen}, {Kilkenny}, {van Wyk}, \&
  {Marang}}]{Koen2010}
{Koen}, C., {Kilkenny}, D., {van Wyk}, F., \& {Marang}, F. 2010, \mnras, 403,
  1949

\bibitem[{{Kraft}(1965)}]{Kraft1965}
{Kraft}, R.~P. 1965, \apj, 142, 681

\bibitem[{{Kulikov} {et~al.}(2007){Kulikov}, {Lammer}, {Lichtenegger}, {Penz},
  {Breuer}, {Spohn}, {Lundin}, \& {Biernat}}]{Kulikov2007}
{Kulikov}, Y.~N., {Lammer}, H., {Lichtenegger}, H. I.~M., {et~al.} 2007, \ssr,
  129, 207

\bibitem[{{Lammer} {et~al.}(2012){Lammer}, {G{\"u}del}, {Kulikov}, {Ribas},
  {Zaqarashvili}, {Khodachenko}, {Kislyakova}, {Gr{\"o}ller}, {Odert},
  {Leitzinger}, {Fichtinger}, {Krauss}, {Hausleitner}, {Holmstr{\"o}m},
  {Sanz-Forcada}, {Lichtenegger}, {Hanslmeier}, {Shematovich}, {Bisikalo},
  {Rauer}, \& {Fridlund}}]{Lammer2012}
{Lammer}, H., {G{\"u}del}, M., {Kulikov}, Y., {et~al.} 2012, Earth, Planets and
  Space, 64, 179

\bibitem[{{Lammer} {et~al.}(2003){Lammer}, {Selsis}, {Ribas}, {Guinan},
  {Bauer}, \& {Weiss}}]{Lammer2003}
{Lammer}, H., {Selsis}, F., {Ribas}, I., {et~al.} 2003, ApJl, 598, L121

\bibitem[{{Landi Degl'Innocenti}(1992)}]{Landi1992}
{Landi Degl'Innocenti}, E. 1992, {Magnetic field measurements.}, ed.
  F.~{Sanchez}, M.~{Collados}, \& M.~{Vazquez}, 71

\bibitem[{{Landstreet}(2009)}]{Landstreet2009_1}
{Landstreet}, J.~D. 2009, in EAS Publications Series, Vol.~39, EAS Publications
  Series, ed. C.~{Neiner} \& J.~P. {Zahn}, 1--20

\bibitem[{{Lehmann} \& {Donati}(2022)}]{Lehmann2022}
{Lehmann}, L.~T. \& {Donati}, J.~F. 2022, MNRAS, 514, 2333

\bibitem[{{Lehtinen} {et~al.}(2016){Lehtinen}, {Jetsu}, {Hackman}, {Kajatkari},
  \& {Henry}}]{Lehtinen2016}
{Lehtinen}, J., {Jetsu}, L., {Hackman}, T., {Kajatkari}, P., \& {Henry}, G.~W.
  2016, \aap, 588, A38

\bibitem[{{Lehtinen} {et~al.}(2020){Lehtinen}, {Spada}, {K{\"a}pyl{\"a}},
  {Olspert}, \& {K{\"a}pyl{\"a}}}]{Lehtinen2020}
{Lehtinen}, J.~J., {Spada}, F., {K{\"a}pyl{\"a}}, M.~J., {Olspert}, N., \&
  {K{\"a}pyl{\"a}}, P.~J. 2020, Nature Astronomy, 4, 658

\bibitem[{{L{\'e}pine} \& {Bongiorno}(2007)}]{Lepine2007}
{L{\'e}pine}, S. \& {Bongiorno}, B. 2007, \aj, 133, 889

\bibitem[{{Lingam} \& {Loeb}(2019)}]{Lingam2019}
{Lingam}, M. \& {Loeb}, A. 2019, International Journal of Astrobiology, 18, 527

\bibitem[{{Linsky} {et~al.}(2020){Linsky}, {Wood}, {Youngblood}, {Brown},
  {Froning}, {France}, {Buccino}, {Cranmer}, {Mauas}, {Miguel}, {Pineda},
  {Rugheimer}, {Vieytes}, {Wheatley}, \& {Wilson}}]{Linsky2020}
{Linsky}, J.~L., {Wood}, B.~E., {Youngblood}, A., {et~al.} 2020, \apj, 902, 3

\bibitem[{{Locci} {et~al.}(2019){Locci}, {Cecchi-Pestellini}, \&
  {Micela}}]{Locci2019}
{Locci}, D., {Cecchi-Pestellini}, C., \& {Micela}, G. 2019, \aap, 624, A101

\bibitem[{{L{\'o}pez Ariste}(2002)}]{LopezAriste2002}
{L{\'o}pez Ariste}, A. 2002, \apj, 564, 379

\bibitem[{{Luhman} {et~al.}(2007){Luhman}, {Patten}, {Marengo}, {Schuster},
  {Hora}, {Ellis}, {Stauffer}, {Sonnett}, {Winston}, {Gutermuth}, {Megeath},
  {Backman}, {Henry}, {Werner}, \& {Fazio}}]{Luhman2007}
{Luhman}, K.~L., {Patten}, B.~M., {Marengo}, M., {et~al.} 2007, \apj, 654, 570

\bibitem[{{Marsden} {et~al.}(2014){Marsden}, {Petit}, {Jeffers}, {Morin},
  {Fares}, {Reiners}, {do Nascimento}, {Auri{\`e}re}, {Bouvier}, {Carter},
  {Catala}, {Dintrans}, {Donati}, {Gastine}, {Jardine}, {Konstantinova-Antova},
  {Lanoux}, {Ligni{\`e}res}, {Morgenthaler}, {Ram{\`\i}rez-V{\`e}lez},
  {Th{\'e}ado}, {Van Grootel}, \& {BCool Collaboration}}]{Marsden2014}
{Marsden}, S.~C., {Petit}, P., {Jeffers}, S.~V., {et~al.} 2014, MNRAS, 444,
  3517

\bibitem[{{Mason} {et~al.}(2001){Mason}, {Wycoff}, {Hartkopf}, {Douglass}, \&
  {Worley}}]{Mason2001}
{Mason}, B.~D., {Wycoff}, G.~L., {Hartkopf}, W.~I., {Douglass}, G.~G., \&
  {Worley}, C.~E. 2001, \aj, 122, 3466

\bibitem[{{Mathias} {et~al.}(2018){Mathias}, {Auri{\`e}re}, {L{\'o}pez Ariste},
  {Petit}, {Tessore}, {Josselin}, {L{\`e}bre}, {Morin}, {Wade}, {Herpin},
  {Chiavassa}, {Montarg{\`e}s}, {Konstantinova-Antova}, {Kervella}, {Perrin},
  {Donati}, \& {Grunhut}}]{Mathias2018}
{Mathias}, P., {Auri{\`e}re}, M., {L{\'o}pez Ariste}, A., {et~al.} 2018, \aap,
  615, A116

\bibitem[{{Mengel} {et~al.}(2016){Mengel}, {Fares}, {Marsden}, {Carter},
  {Jeffers}, {Petit}, {Donati}, {Folsom}, \& {BCool
  Collaboration}}]{Mengel2016}
{Mengel}, M.~W., {Fares}, R., {Marsden}, S.~C., {et~al.} 2016, \mnras, 459,
  4325

\bibitem[{{Messina} {et~al.}(1999){Messina}, {Guinan}, {Lanza}, \&
  {Ambruster}}]{Messina1999}
{Messina}, S., {Guinan}, E.~F., {Lanza}, A.~F., \& {Ambruster}, C. 1999, \aap,
  347, 249

\bibitem[{{Middelkoop}(1982)}]{Middelkoop1982}
{Middelkoop}, F. 1982, \aap, 107, 31

\bibitem[{{Montes} {et~al.}(2001){Montes}, {L{\'o}pez-Santiago}, {G{\'a}lvez},
  {Fern{\'a}ndez-Figueroa}, {De Castro}, \& {Cornide}}]{Montes2001}
{Montes}, D., {L{\'o}pez-Santiago}, J., {G{\'a}lvez}, M.~C., {et~al.} 2001,
  \mnras, 328, 45

\bibitem[{{Moore} \& {Paddock}(1950)}]{Moore1950}
{Moore}, J.~H. \& {Paddock}, G.~F. 1950, \apj, 112, 48

\bibitem[{{Morgenthaler} {et~al.}(2012){Morgenthaler}, {Petit}, {Saar},
  {Solanki}, {Morin}, {Marsden}, {Auri{\`e}re}, {Dintrans}, {Fares}, {Gastine},
  {Lanoux}, {Ligni{\`e}res}, {Paletou}, {Ram{\'\i}rez V{\'e}lez}, {Th{\'e}ado},
  \& {Van Grootel}}]{Morgenthaler2012}
{Morgenthaler}, A., {Petit}, P., {Saar}, S., {et~al.} 2012, A\&A, 540, A138

\bibitem[{{Moutou} {et~al.}(2007){Moutou}, {Donati}, {Savalle}, {Hussain},
  {Alecian}, {Bouchy}, {Catala}, {Collier Cameron}, {Udry}, \&
  {Vidal-Madjar}}]{Moutou2007}
{Moutou}, C., {Donati}, J.~F., {Savalle}, R., {et~al.} 2007, A\&A, 473, 651

\bibitem[{{Murray-Clay} {et~al.}(2009){Murray-Clay}, {Chiang}, \&
  {Murray}}]{MurrayClay2009}
{Murray-Clay}, R.~A., {Chiang}, E.~I., \& {Murray}, N. 2009, \apj, 693, 23

\bibitem[{{National Academies of Sciences and Engineering,
  Medicine}(2021)}]{2021pdaa.book.....N}
{National Academies of Sciences and Engineering, Medicine}. 2021, {Pathways to
  Discovery in Astronomy and Astrophysics for the 2020s}

\bibitem[{{Nicholson} {et~al.}(2016){Nicholson}, {Vidotto}, {Mengel},
  {Brookshaw}, {Carter}, {Petit}, {Marsden}, {Jeffers}, {Fares}, \& {BCool
  Collaboration}}]{Nicholson2016}
{Nicholson}, B.~A., {Vidotto}, A.~A., {Mengel}, M., {et~al.} 2016, \mnras, 459,
  1907

\bibitem[{{Noyes} {et~al.}(1984){Noyes}, {Hartmann}, {Baliunas}, {Duncan}, \&
  {Vaughan}}]{Noyes1984}
{Noyes}, R.~W., {Hartmann}, L.~W., {Baliunas}, S.~L., {Duncan}, D.~K., \&
  {Vaughan}, A.~H. 1984, ApJ, 279, 763

\bibitem[{{Owen} \& {Jackson}(2012)}]{Owen2012}
{Owen}, J.~E. \& {Jackson}, A.~P. 2012, MNRAS, 425, 2931

\bibitem[{{Pace}(2013)}]{Pace2013}
{Pace}, G. 2013, \aap, 551, L8

\bibitem[{{Paegert} {et~al.}(2021){Paegert}, {Stassun}, {Collins}, {Pepper},
  {Torres}, {Jenkins}, {Twicken}, \& {Latham}}]{Peagert2021}
{Paegert}, M., {Stassun}, K.~G., {Collins}, K.~A., {et~al.} 2021, arXiv
  e-prints, arXiv:2108.04778

\bibitem[{{Penz} \& {Micela}(2008)}]{Penz2008}
{Penz}, T. \& {Micela}, G. 2008, A\&A, 479, 579

\bibitem[{{Petit} {et~al.}(2013){Petit}, {Auri{\`e}re}, {Konstantinova-Antova},
  {Morgenthaler}, {Perrin}, {Roudier}, \& {Donati}}]{Petit2013}
{Petit}, P., {Auri{\`e}re}, M., {Konstantinova-Antova}, R., {et~al.} 2013, in
  Lecture Notes in Physics, Berlin Springer Verlag, ed. J.-P. {Rozelot} \&
  C.~E.~. {Neiner}, Vol. 857, 231

\bibitem[{{Petit} {et~al.}(2008){Petit}, {Dintrans}, {Solanki}, {Donati},
  {Auri{\`e}re}, {Ligni{\`e}res}, {Morin}, {Paletou}, {Ramirez Velez},
  {Catala}, \& {Fares}}]{Petit2008}
{Petit}, P., {Dintrans}, B., {Solanki}, S.~K., {et~al.} 2008, \mnras, 388, 80

\bibitem[{{Petit} {et~al.}(2005){Petit}, {Donati}, {Auri{\`e}re}, {Landstreet},
  {Ligni{\`e}res}, {Marsden}, {Mouillet}, {Paletou}, {Toqu{\'e}}, \&
  {Wade}}]{Petit2005}
{Petit}, P., {Donati}, J.~F., {Auri{\`e}re}, M., {et~al.} 2005, \mnras, 361,
  837

\bibitem[{{Petit} {et~al.}(2002){Petit}, {Donati}, \& {Collier
  Cameron}}]{Petit2002}
{Petit}, P., {Donati}, J.~F., \& {Collier Cameron}, A. 2002, MNRAS, 334, 374

\bibitem[{{Petit} {et~al.}(2014){Petit}, {Louge}, {Th{\'e}ado}, {Paletou},
  {Manset}, {Morin}, {Marsden}, \& {Jeffers}}]{Petit2014}
{Petit}, P., {Louge}, T., {Th{\'e}ado}, S., {et~al.} 2014, PASP, 126, 469

\bibitem[{{Pizzolato} {et~al.}(2003){Pizzolato}, {Maggio}, {Micela},
  {Sciortino}, \& {Ventura}}]{Pizzolato2003}
{Pizzolato}, N., {Maggio}, A., {Micela}, G., {Sciortino}, S., \& {Ventura}, P.
  2003, \aap, 397, 147

\bibitem[{{Poljan{\v{c}}i{\'c} Beljan} {et~al.}(2022){Poljan{\v{c}}i{\'c}
  Beljan}, {Jurdana-{\v{S}}epi{\'c}}, {Jurki{\'c}}, {Braj{\v{s}}a},
  {Skoki{\'c}}, {Sudar}, {Ru{\v{z}}djak}, {Hr{\v{z}}ina}, {P{\"o}tzi},
  {Hanslmeier}, \& {Veronig}}]{Beljan2022}
{Poljan{\v{c}}i{\'c} Beljan}, I., {Jurdana-{\v{S}}epi{\'c}}, R., {Jurki{\'c}},
  T., {et~al.} 2022, \aap, 663, A24

\bibitem[{{Press} {et~al.}(1992){Press}, {Teukolsky}, {Vetterling}, \&
  {Flannery}}]{Press1992}
{Press}, W.~H., {Teukolsky}, S.~A., {Vetterling}, W.~T., \& {Flannery}, B.~P.
  1992, {Numerical recipes in FORTRAN. The art of scientific computing}

\bibitem[{{Quanz} {et~al.}(2018){Quanz}, {Kammerer}, {Defr{\`e}re}, {Absil},
  {Glauser}, \& {Kitzmann}}]{Quanz2018}
{Quanz}, S.~P., {Kammerer}, J., {Defr{\`e}re}, D., {et~al.} 2018, in Society of
  Photo-Optical Instrumentation Engineers (SPIE) Conference Series, Vol. 10701,
  Optical and Infrared Interferometry and Imaging VI, ed. M.~J.
  {Creech-Eakman}, P.~G. {Tuthill}, \& A.~{M{\'e}rand}, 107011I

\bibitem[{{Quanz} {et~al.}(2022){Quanz}, {Ottiger}, {Fontanet}, {Kammerer},
  {Menti}, {Dannert}, {Gheorghe}, {Absil}, {Airapetian}, {Alei}, {Allart},
  {Angerhausen}, {Blumenthal}, {Buchhave}, {Cabrera},
  {Carri{\'o}n-Gonz{\'a}lez}, {Chauvin}, {Danchi}, {Dandumont}, {Defr{\'e}re},
  {Dorn}, {Ehrenreich}, {Ertel}, {Fridlund}, {Garc{\'\i}a Mu{\~n}oz},
  {Gasc{\'o}n}, {Girard}, {Glauser}, {Grenfell}, {Guidi}, {Hagelberg},
  {Helled}, {Ireland}, {Janson}, {Kopparapu}, {Korth}, {Kozakis}, {Kraus},
  {L{\'e}ger}, {Leedj{\"a}rv}, {Lichtenberg}, {Lillo-Box}, {Linz}, {Liseau},
  {Loicq}, {Mahendra}, {Malbet}, {Mathew}, {Mennesson}, {Meyer}, {Mishra},
  {Molaverdikhani}, {Noack}, {Oza}, {Pall{\'e}}, {Parviainen}, {Quirrenbach},
  {Rauer}, {Ribas}, {Rice}, {Romagnolo}, {Rugheimer}, {Schwieterman},
  {Serabyn}, {Sharma}, {Stassun}, {Szul{\'a}gyi}, {Wang}, {Wunderlich},
  {Wyatt}, \& {LIFE Collaboration}}]{Quanz2022}
{Quanz}, S.~P., {Ottiger}, M., {Fontanet}, E., {et~al.} 2022, \aap, 664, A21

\bibitem[{{Raeside} {et~al.}(2025){Raeside}, {Rodgers-Lee}, \&
  {Rimmer}}]{Raeside2025}
{Raeside}, S.~R., {Rodgers-Lee}, D., \& {Rimmer}, P.~B. 2025, \aap, 697, A26

\bibitem[{{Ram{\'\i}rez} {et~al.}(2012){Ram{\'\i}rez}, {Fish}, {Lambert}, \&
  {Allende Prieto}}]{Ramirez2012}
{Ram{\'\i}rez}, I., {Fish}, J.~R., {Lambert}, D.~L., \& {Allende Prieto}, C.
  2012, \apj, 756, 46

\bibitem[{{Rees} \& {Semel}(1979)}]{Rees1979}
{Rees}, D.~E. \& {Semel}, M.~D. 1979, \aap, 74, 1

\bibitem[{{Reiners}(2012)}]{Reiners2012}
{Reiners}, A. 2012, Living Reviews in Solar Physics, 9, 1

\bibitem[{{Ribas} {et~al.}(2005){Ribas}, {Guinan}, {G{\"u}del}, \&
  {Audard}}]{Ribas2005}
{Ribas}, I., {Guinan}, E.~F., {G{\"u}del}, M., \& {Audard}, M. 2005, \apj, 622,
  680

\bibitem[{{Rodgers-Lee} {et~al.}(2021{\natexlab{a}}){Rodgers-Lee}, {Taylor},
  {Vidotto}, \& {Downes}}]{RodgersLee2021}
{Rodgers-Lee}, D., {Taylor}, A.~M., {Vidotto}, A.~A., \& {Downes}, T.~P.
  2021{\natexlab{a}}, \mnras, 504, 1519

\bibitem[{{Rodgers-Lee} {et~al.}(2021{\natexlab{b}}){Rodgers-Lee}, {Vidotto},
  \& {Mesquita}}]{RodgersLee2021b}
{Rodgers-Lee}, D., {Vidotto}, A.~A., \& {Mesquita}, A.~L. 2021{\natexlab{b}},
  \mnras, 508, 4696

\bibitem[{{Ros{\'e}n} {et~al.}(2016){Ros{\'e}n}, {Kochukhov}, {Hackman}, \&
  {Lehtinen}}]{Rosen2016}
{Ros{\'e}n}, L., {Kochukhov}, O., {Hackman}, T., \& {Lehtinen}, J. 2016, \aap,
  593, A35

\bibitem[{{Ros{\'e}n} {et~al.}(2015){Ros{\'e}n}, {Kochukhov}, \&
  {Wade}}]{Rosen2015}
{Ros{\'e}n}, L., {Kochukhov}, O., \& {Wade}, G.~A. 2015, ApJ, 805, 169

\bibitem[{{Rugheimer} {et~al.}(2015){Rugheimer}, {Kaltenegger}, {Segura},
  {Linsky}, \& {Mohanty}}]{Rugheimer2015}
{Rugheimer}, S., {Kaltenegger}, L., {Segura}, A., {Linsky}, J., \& {Mohanty},
  S. 2015, ApJ, 809, 57

\bibitem[{{Rutten}(1984)}]{Rutten1984}
{Rutten}, R.~G.~M. 1984, \aap, 130, 353

\bibitem[{{Ryabchikova} {et~al.}(2015){Ryabchikova}, {Piskunov}, {Kurucz},
  {Stempels}, {Heiter}, {Pakhomov}, \& {Barklem}}]{Ryabchikova2015}
{Ryabchikova}, T., {Piskunov}, N., {Kurucz}, R.~L., {et~al.} 2015, Phys. Scr.,
  90, 054005

\bibitem[{{Santos} {et~al.}(2016){Santos}, {Santerne}, {Faria}, {Rey},
  {Correia}, {Laskar}, {Udry}, {Adibekyan}, {Bouchy}, {Delgado-Mena}, {Melo},
  {Dumusque}, {H{\'e}brard}, {Lovis}, {Mayor}, {Montalto}, {Mortier}, {Pepe},
  {Figueira}, {Sahlmann}, {S{\'e}gransan}, \& {Sousa}}]{Santos2016}
{Santos}, N.~C., {Santerne}, A., {Faria}, J.~P., {et~al.} 2016, \aap, 592, A13

\bibitem[{{Sanz-Forcada} {et~al.}(2011){Sanz-Forcada}, {Micela}, {Ribas},
  {Pollock}, {Eiroa}, {Velasco}, {Solano}, \&
  {Garc{\'\i}a-{\'A}lvarez}}]{SanzForcada2011}
{Sanz-Forcada}, J., {Micela}, G., {Ribas}, I., {et~al.} 2011, \aap, 532, A6

\bibitem[{{Sato} {et~al.}(2005){Sato}, {Fischer}, {Henry}, {Laughlin},
  {Butler}, {Marcy}, {Vogt}, {Bodenheimer}, {Ida}, {Toyota}, {Wolf}, {Valenti},
  {Boyd}, {Johnson}, {Wright}, {Ammons}, {Robinson}, {Strader}, {McCarthy},
  {Tah}, \& {Minniti}}]{Sato2005}
{Sato}, B., {Fischer}, D.~A., {Henry}, G.~W., {et~al.} 2005, \apj, 633, 465

\bibitem[{{See} {et~al.}(2015){See}, {Jardine}, {Vidotto}, {Donati}, {Folsom},
  {Boro Saikia}, {Bouvier}, {Fares}, {Gregory}, {Hussain}, {Jeffers},
  {Marsden}, {Morin}, {Moutou}, {do Nascimento}, {Petit}, {Ros{\'e}n}, \&
  {Waite}}]{See2015}
{See}, V., {Jardine}, M., {Vidotto}, A.~A., {et~al.} 2015, MNRAS, 453, 4301

\bibitem[{{See} {et~al.}(2014){See}, {Jardine}, {Vidotto}, {Petit}, {Marsden},
  {Jeffers}, \& {do Nascimento}}]{See2014}
{See}, V., {Jardine}, M., {Vidotto}, A.~A., {et~al.} 2014, A\&A, 570, A99

\bibitem[{{Semel}(1989)}]{Semel1989}
{Semel}, M. 1989, A\&A, 225, 456

\bibitem[{{Skilling} \& {Bryan}(1984)}]{Skilling1984}
{Skilling}, J. \& {Bryan}, R.~K. 1984, MNRAS, 211, 111

\bibitem[{{Tsai} {et~al.}(2023){Tsai}, {Lee}, {Powell}, {Gao}, {Zhang},
  {Moses}, {H{\'e}brard}, {Venot}, {Parmentier}, {Jordan}, {Hu}, {Alam},
  {Alderson}, {Batalha}, {Bean}, {Benneke}, {Bierson}, {Brady}, {Carone},
  {Carter}, {Chubb}, {Inglis}, {Leconte}, {Line}, {L{\'o}pez-Morales},
  {Miguel}, {Molaverdikhani}, {Rustamkulov}, {Sing}, {Stevenson}, {Wakeford},
  {Yang}, {Aggarwal}, {Baeyens}, {Barat}, {de Val-Borro}, {Daylan}, {Fortney},
  {France}, {Goyal}, {Grant}, {Kirk}, {Kreidberg}, {Louca}, {Moran},
  {Mukherjee}, {Nasedkin}, {Ohno}, {Rackham}, {Redfield}, {Taylor}, {Tremblin},
  {Visscher}, {Wallack}, {Welbanks}, {Youngblood}, {Ahrer}, {Batalha}, {Behr},
  {Berta-Thompson}, {Blecic}, {Casewell}, {Crossfield}, {Crouzet}, {Cubillos},
  {Decin}, {D{\'e}sert}, {Feinstein}, {Gibson}, {Harrington}, {Heng},
  {Henning}, {Kempton}, {Krick}, {Lagage}, {Lendl}, {Lothringer}, {Mansfield},
  {Mayne}, {Mikal-Evans}, {Palle}, {Schlawin}, {Shorttle}, {Wheatley}, \&
  {Yurchenko}}]{Tsai2023}
{Tsai}, S.-M., {Lee}, E. K.~H., {Powell}, D., {et~al.} 2023, \nat, 617, 483

\bibitem[{{Tu} {et~al.}(2015){Tu}, {Johnstone}, {G{\"u}del}, \&
  {Lammer}}]{Tu2015}
{Tu}, L., {Johnstone}, C.~P., {G{\"u}del}, M., \& {Lammer}, H. 2015, \aap, 577,
  L3

\bibitem[{{Valenti} \& {Fischer}(2005)}]{Valenti2005}
{Valenti}, J.~A. \& {Fischer}, D.~A. 2005, \apjs, 159, 141

\bibitem[{{van Belle} \& {von Braun}(2009)}]{vanBelle2009}
{van Belle}, G.~T. \& {von Braun}, K. 2009, \apj, 694, 1085

\bibitem[{{van der Walt} {et~al.}(2011){van der Walt}, {Colbert}, \&
  {Varoquaux}}]{VanderWalt2011}
{van der Walt}, S., {Colbert}, S.~C., \& {Varoquaux}, G. 2011, Computing in
  Science and Engineering, 13, 22

\bibitem[{{Van Looveren} {et~al.}(2025){Van Looveren}, {Boro Saikia},
  {Herbort}, {Schleich}, {G{\"u}del}, {Johnstone}, \&
  {Kislyakova}}]{vanLooveren2025}
{Van Looveren}, G., {Boro Saikia}, S., {Herbort}, O., {et~al.} 2025, \aap, 694,
  A310

\bibitem[{{Van Looveren} {et~al.}(2024){Van Looveren}, {G{\"u}del}, {Boro
  Saikia}, \& {Kislyakova}}]{vanLooveren2024}
{Van Looveren}, G., {G{\"u}del}, M., {Boro Saikia}, S., \& {Kislyakova}, K.
  2024, \aap, 683, A153

\bibitem[{{Vaughan} {et~al.}(1978){Vaughan}, {Preston}, \&
  {Wilson}}]{Vaughan1978}
{Vaughan}, A.~H., {Preston}, G.~W., \& {Wilson}, O.~C. 1978, PASP, 90, 267

\bibitem[{{Vidotto} {et~al.}(2023){Vidotto}, {Bourrier}, {Fares}, {Bellotti},
  {Donati}, {Petit}, {Hussain}, \& {Morin}}]{Vidotto2023}
{Vidotto}, A.~A., {Bourrier}, V., {Fares}, R., {et~al.} 2023, \aap, 678, A152

\bibitem[{{Vidotto} {et~al.}(2014{\natexlab{a}}){Vidotto}, {Gregory},
  {Jardine}, {Donati}, {Petit}, {Morin}, {Folsom}, {Bouvier}, {Cameron},
  {Hussain}, {Marsden}, {Waite}, {Fares}, {Jeffers}, \& {do
  Nascimento}}]{Vidotto2014b}
{Vidotto}, A.~A., {Gregory}, S.~G., {Jardine}, M., {et~al.} 2014{\natexlab{a}},
  \mnras, 441, 2361

\bibitem[{{Vidotto} {et~al.}(2013){Vidotto}, {Jardine}, {Morin}, {Donati},
  {Lang}, \& {Russell}}]{Vidotto2013}
{Vidotto}, A.~A., {Jardine}, M., {Morin}, J., {et~al.} 2013, A\&A, 557, A67

\bibitem[{{Vidotto} {et~al.}(2014{\natexlab{b}}){Vidotto}, {Jardine}, {Morin},
  {Donati}, {Opher}, \& {Gombosi}}]{Vidotto2014}
{Vidotto}, A.~A., {Jardine}, M., {Morin}, J., {et~al.} 2014{\natexlab{b}},
  MNRAS, 438, 1162

\bibitem[{{Vidotto} {et~al.}(2018){Vidotto}, {Lehmann}, {Jardine}, \&
  {Pevtsov}}]{Vidotto2018}
{Vidotto}, A.~A., {Lehmann}, L.~T., {Jardine}, M., \& {Pevtsov}, A.~A. 2018,
  \mnras, 480, 477

\bibitem[{{Virtanen} {et~al.}(2020){Virtanen}, {Gommers}, {Burovski},
  {Oliphant}, {Weckesser}, {Cournapeau}, {Alexbrc}, {Peterson}, {Reddy},
  {Wilson}, {Haberland}, {Mayorov}, {Endolith}, {Nelson}, {Der Van Walt},
  {Laxalde}, {Brett}, {Polat}, {Larson}, {Millman}, {Lars}, {Van Mulbregt},
  {Eric-Jones}, {Carey}, {Moore}, {Kern}, {Leslie}, {Perktold}, {Striega}, \&
  {Feng}}]{Virtanen2020}
{Virtanen}, P., {Gommers}, R., {Burovski}, E., {et~al.} 2020, {scipy/scipy:
  SciPy 1.5.3}

\bibitem[{{Wenger} {et~al.}(2000){Wenger}, {Ochsenbein}, {Egret}, {Dubois},
  {Bonnarel}, {Borde}, {Genova}, {Jasniewicz}, {Lalo{\"e}}, {Lesteven}, \&
  {Monier}}]{Wenger2000}
{Wenger}, M., {Ochsenbein}, F., {Egret}, D., {et~al.} 2000, \aaps, 143, 9

\bibitem[{{Willamo} {et~al.}(2022){Willamo}, {Lehtinen}, {Hackman},
  {K{\"a}pyl{\"a}}, {Kochukhov}, {Jeffers}, {Korhonen}, \&
  {Marsden}}]{Willamo2022}
{Willamo}, T., {Lehtinen}, J.~J., {Hackman}, T., {et~al.} 2022, \aap, 659, A71

\bibitem[{{Wilson}(1968)}]{Wilson1968}
{Wilson}, O.~C. 1968, ApJ, 153, 221

\bibitem[{{Wood}(2004)}]{Wood2004}
{Wood}, B.~E. 2004, Living Reviews in Solar Physics, 1, 2

\bibitem[{{Wood}(2006)}]{Wood2006}
{Wood}, B.~E. 2006, \ssr, 126, 3

\bibitem[{{Wright} {et~al.}(2011){Wright}, {Drake}, {Mamajek}, \&
  {Henry}}]{Wright2011}
{Wright}, N.~J., {Drake}, J.~J., {Mamajek}, E.~E., \& {Henry}, G.~W. 2011, ApJ,
  743, 48

\bibitem[{{Zeeman}(1897)}]{Zeeman1897}
{Zeeman}, P. 1897, Nature, 55, 347

\bibitem[{{Zejda} {et~al.}(2012){Zejda}, {Paunzen}, {Baumann},
  {Mikul{\'a}{\v{s}}ek}, \& {Li{\v{s}}ka}}]{Zejda2012}
{Zejda}, M., {Paunzen}, E., {Baumann}, B., {Mikul{\'a}{\v{s}}ek}, Z., \&
  {Li{\v{s}}ka}, J. 2012, \aap, 548, A97

\end{thebibliography}

\begin{appendix}

\onecolumn
\section{Journal of observations}\label{app:log}

\begin{table*}[h]
\caption{List of median activity indices and standard deviations for our stars.} 
\label{tab:log}     
\centering                       
\begin{tabular}{lcrccccccr}    
\hline\hline
Date & UT & $n_\mathrm{cyc}$ & $t_{exp}$ & S/N & $\sigma_\mathrm{LSD}$ & $\log R'_\mathrm{HK}$ & H$\alpha$ & \ion{Ca}{II} IRT & B$_l$\\
 & [hh:mm:ss] & & [s] & & [$10^{-5}I_c$] & & & & [G]\\
\hline
\multicolumn{10}{c}{2018}\\
\hline
Aug 19 & 01:58:56 & 0.00 & 4x600.0 & 450 & 4.3 & $-4.406\pm0.022$ & $0.324\pm0.001$ & $0.867\pm0.003$ & $-4.6\pm1.0$ \\
Aug 20 & 01:06:47 & 0.13 & 4x600.0 & 534 & 3.2 & $-4.413\pm0.018$ & $0.323\pm0.001$ & $0.860\pm0.002$ & $-3.7\pm0.8$ \\
Aug 21 & 01:35:49 & 0.26 & 4x600.0 & 453 & 4.2 & $-4.410\pm0.024$ & $0.323\pm0.001$ & $0.867\pm0.002$ & $1.0\pm1.0$ \\
Aug 22 & 01:27:16 & 0.39 & 4x600.0 & 597 & 2.9 & $-4.370\pm0.015$ & $0.324\pm0.001$ & $0.871\pm0.002$ & $0.8\pm0.7$ \\
Aug 23 & 01:52:38 & 0.52 & 4x600.0 & 461 & 4.2 & $-4.350\pm0.020$ & $0.326\pm0.001$ & $0.883\pm0.002$ & $1.7\pm0.9$ \\
Aug 24 & 00:26:20 & 0.64 & 4x600.0 & 439 & 4.6 & $-4.354\pm0.024$ & $0.328\pm0.001$ & $0.890\pm0.003$ & $1.0\pm1.0$ \\
Aug 27 & 01:53:18 & 1.04 & 4x600.0 & 509 & 3.4 & $-4.413\pm0.022$ & $0.323\pm0.001$ & $0.864\pm0.002$ & $-2.1\pm0.9$ \\
Aug 28 & 02:46:09 & 1.18 & 4x600.0 & 534 & 3.2 & $-4.414\pm0.017$ & $0.321\pm0.001$ & $0.855\pm0.002$ & $-3.2\pm0.8$ \\
Sep 22 & 23:41:26 & 4.55 & 4x600.0 & 440 & 4.1 & $-4.352\pm0.022$ & $0.326\pm0.001$ & $0.876\pm0.003$ & $2.0\pm1.0$ \\
Sep 24 & 23:16:43 & 4.81 & 4x600.0 & 367 & 5.2 & $-4.364\pm0.027$ & $0.327\pm0.002$ & $0.870\pm0.003$ & $-3.2\pm1.2$ \\
Sep 25 & 24:04:11 & 4.94 & 4x600.0 & 473 & 3.5 & $-4.379\pm0.019$ & $0.325\pm0.001$ & $0.866\pm0.002$ & $-5.7\pm0.9$ \\
Sep 27 & 00:30:02 & 5.07 & 4x600.0 & 418 & 3.9 & $-4.392\pm0.024$ & $0.327\pm0.001$ & $0.864\pm0.003$ & $-3.3\pm1.1$ \\
Sep 28 & 00:25:54 & 5.20 & 4x600.0 & 605 & 2.9 & $-4.386\pm0.013$ & $0.327\pm0.001$ & $0.857\pm0.002$ & $0.3\pm0.7$ \\
Oct 03 & 23:21:19 & 5.98 & 4x600.0 & 367 & 5.0 & $-4.383\pm0.029$ & $0.328\pm0.002$ & $0.863\pm0.003$ & $-4.2\pm1.2$ \\
Oct 05 & 23:34:60 & 6.24 & 4x600.0 & 616 & 3.0 & $-4.401\pm0.014$ & $0.326\pm0.001$ & $0.853\pm0.002$ & $-2.3\pm0.7$ \\
Oct 08 & 21:31:13 & 6.62 & 4x600.0 & 416 & 4.5 & $-4.338\pm0.022$ & $0.328\pm0.001$ & $0.885\pm0.003$ & $0.6\pm1.1$ \\
\hline
\end{tabular}
\tablefoot{Columns (1 and 2) Date and universal time of the observations; (3) Rotational cycle of the observations found using Eq.~\ref{eq:ephemeris}; (4) Exposure time of a polarimetric sequence, (5) S/N at 650 nm per polarimetric sequence; and (6) RMS noise level of Stokes $V$ signal in units of an unpolarised continuum.}
\end{table*}

\begin{longtable}{lcrccccccr}
\caption{Same as Table~\ref{tab:log} for HD\,28205} \\
\hline\hline
Date & UT & $n_\mathrm{cyc}$ & $t_{exp}$ & S/N & $\sigma_\mathrm{LSD}$ & $\log R'_\mathrm{HK}$ & H$\alpha$ & \ion{Ca}{II} IRT & B$_l$\\
 & [hh:mm:ss] & & [s] & & [$10^{-5}I_c$] & & & & [G]\\
\hline
\multicolumn{10}{c}{2018}\\
\hline
Nov 17 & 01:50:24 & 0.00 & 4x800.0 & 435 & 3.8 & $-4.443\pm0.023$ & $0.313\pm0.001$ & $0.847\pm0.003$ & $-0.1\pm1.2$ \\
Nov 28 & 23:50:36 & 2.03 & 4x800.0 & 282 & 6.5 & $-4.438\pm0.041$ & $0.313\pm0.002$ & $0.848\pm0.004$ & $-1.0\pm1.9$ \\
Dec 04 & 23:23:05 & 3.05 & 4x800.0 & 288 & 5.9 & $-4.437\pm0.040$ & $0.313\pm0.002$ & $0.849\pm0.004$ & $-2.9\pm1.9$ \\
Dec 06 & 24:28:10 & 3.40 & 4x800.0 & 323 & 6.1 & $-4.453\pm0.036$ & $0.313\pm0.002$ & $0.843\pm0.003$ & $0.8\pm1.7$ \\
Dec 10 & 22:51:31 & 4.07 & 4x800.0 & 406 & 4.4 & $-4.424\pm0.023$ & $0.314\pm0.002$ & $0.848\pm0.003$ & $-2.0\pm1.3$ \\
\hline
%\footnote{}
\end{longtable}

\begin{longtable}{lcrccccccr}
\caption{Same as Table~\ref{tab:log} for HD\,30495} \\
\hline\hline
Date & UT & $n_\mathrm{cyc}$ & $t_{exp}$ & S/N & $\sigma_\mathrm{LSD}$ & $\log R'_\mathrm{HK}$ & H$\alpha$ & \ion{Ca}{II} IRT & B$_l$\\
 & [hh:mm:ss] & & [s] & & [$10^{-5}I_c$] & & & & [G]\\
\hline
\multicolumn{10}{c}{2018}\\
\hline
Dec 05 & 00:46:09 & 0.00 & 4x400.0 & 445 & 3.7 & $-4.495\pm0.030$ & $0.323\pm0.001$ & $0.856\pm0.002$ & $0.4\pm1.1$ \\
Dec 06 & 23:38:31 & 0.17 & 4x400.0 & 418 & 4.2 & $-4.490\pm0.030$ & $0.323\pm0.001$ & $0.854\pm0.002$ & $1.6\pm1.2$ \\
Dec 10 & 24:10:40 & 0.53 & 4x400.0 & 654 & 2.6 & $-4.481\pm0.020$ & $0.323\pm0.001$ & $0.854\pm0.002$ & $3.0\pm0.7$ \\
\hline
%\footnote{}
\end{longtable}

\begin{longtable}{lcrccccccr}
\caption{Same as Table~\ref{tab:log} for HD\,43162} \\
\hline\hline
Date & UT & $n_\mathrm{cyc}$ & $t_{exp}$ & S/N & $\sigma_\mathrm{LSD}$ & $\log R'_\mathrm{HK}$ & H$\alpha$ & \ion{Ca}{II} IRT & B$_l$\\
 & [hh:mm:ss] & & [s] & & [$10^{-5}I_c$] & & & & [G]\\
\hline
\multicolumn{10}{c}{2018}\\
\hline
Dec 18 & 00:25:56 & 0.00 & 4x400.0 & 275 & 7.6 & $-4.348\pm0.047$ & $0.339\pm0.002$ & $0.902\pm0.004$ & $1.5\pm1.8$ \\
Dec 18 & 01:51:37 & 0.01 & 4x400.0 & 315 & 7.2 & $-4.353\pm0.039$ & $0.338\pm0.002$ & $0.902\pm0.003$ & $-0.8\pm1.5$ \\
Dec 19 & 23:06:57 & 0.23 & 4x400.0 & 136 & 17.4 & $-4.371\pm0.187$ & $0.336\pm0.004$ & $0.891\pm0.009$ & $4.4\pm3.7$ \\
\hline
\multicolumn{10}{c}{2019}\\
\hline
Jan 03 & 23:02:49 & 1.99 & 4x400.0 & 377 & 6.0 & $-4.397\pm0.059$ & $0.338\pm0.002$ & $0.898\pm0.003$ & $1.6\pm1.2$ \\
Jan 04 & 00:48:02 & 2.00 & 4x400.0 & 321 & 7.3 & $-4.398\pm0.095$ & $0.338\pm0.002$ & $0.898\pm0.004$ & $2.4\pm1.5$ \\
Jan 04 & 22:32:52 & 2.11 & 4x400.0 & 351 & 6.3 & $-4.364\pm0.046$ & $0.340\pm0.002$ & $0.905\pm0.004$ & $4.4\pm1.3$ \\
Jan 05 & 10:08:20 & 2.12 & 4x400.0 & 239 & 9.5 & $-4.368\pm0.104$ & $0.340\pm0.003$ & $0.904\pm0.005$ & $7.2\pm2.0$ \\
Jan 05 & 22:22:25 & 2.23 & 4x400.0 & 329 & 6.7 & $-4.343\pm0.034$ & $0.341\pm0.002$ & $0.909\pm0.003$ & $4.3\pm1.4$ \\
Jan 06 & 00:41:12 & 2.24 & 4x400.0 & 296 & 7.6 & $-4.332\pm0.041$ & $0.341\pm0.002$ & $0.911\pm0.004$ & $1.6\pm1.6$ \\
Jan 06 & 22:44:04 & 2.34 & 4x400.0 & 320 & 7.1 & $-4.323\pm0.034$ & $0.342\pm0.002$ & $0.918\pm0.003$ & $5.8\pm1.5$ \\
Jan 07 & 10:00:37 & 2.36 & 4x400.0 & 314 & 7.1 & $-4.315\pm0.037$ & $0.342\pm0.002$ & $0.919\pm0.003$ & $5.8\pm1.5$ \\
Mar 20 & 19:24:42 & 10.92 & 4x600.0 & 342 & 6.3 & $-4.377\pm0.044$ & $0.339\pm0.002$ & $0.903\pm0.003$ & $2.4\pm1.4$ \\
Mar 21 & 19:24:26 & 11.03 & 4x600.0 & 378 & 5.4 & $-4.379\pm0.030$ & $0.337\pm0.002$ & $0.896\pm0.003$ & $4.2\pm1.2$ \\
Mar 22 & 19:29:24 & 11.15 & 4x600.0 & 462 & 4.5 & $-4.370\pm0.025$ & $0.338\pm0.001$ & $0.898\pm0.002$ & $2.4\pm1.0$ \\
Mar 23 & 19:34:30 & 11.27 & 4x600.0 & 324 & 6.9 & $-4.391\pm0.049$ & $0.338\pm0.002$ & $0.896\pm0.003$ & $-2.7\pm1.4$ \\
Mar 25 & 19:31:48 & 11.50 & 4x600.0 & 232 & 10.4 & $-4.347\pm0.046$ & $0.339\pm0.002$ & $0.908\pm0.004$ & $-4.9\pm2.2$ \\
Mar 26 & 19:23:28 & 11.62 & 4x600.0 & 323 & 7.1 & $-4.342\pm0.035$ & $0.341\pm0.002$ & $0.911\pm0.003$ & $-4.3\pm1.4$ \\
Mar 27 & 19:21:34 & 11.74 & 4x600.0 & 325 & 6.9 & $-4.353\pm0.033$ & $0.341\pm0.002$ & $0.908\pm0.003$ & $4.1\pm1.5$ \\
Mar 28 & 19:43:50 & 11.86 & 4x600.0 & 483 & 4.4 & $-4.371\pm0.026$ & $0.339\pm0.001$ & $0.896\pm0.002$ & $4.3\pm1.0$ \\
\hline
%\footnote{}
\end{longtable}

\begin{longtable}{lcrccccccr}
\caption{Same as Table~\ref{tab:log} for HD\,82443} \\
\hline\hline
Date & UT & $n_\mathrm{cyc}$ & $t_{exp}$ & S/N & $\sigma_\mathrm{LSD}$ & $\log R'_\mathrm{HK}$ & H$\alpha$ & \ion{Ca}{II} IRT & B$_l$\\
 & [hh:mm:ss] & & [s] & & [$10^{-5}I_c$] & & & & [G]\\
\hline
\multicolumn{10}{c}{2019}\\
\hline
Mar 08 & 21:40:16 & 0.00 & 4x900.0 & 374 & 5.4 & $-4.170\pm0.017$ & $0.377\pm0.002$ & $0.987\pm0.003$ & $6.6\pm2.6$ \\
Mar 11 & 21:59:23 & 0.56 & 4x900.0 & 450 & 3.6 & $-4.130\pm0.011$ & $0.381\pm0.001$ & $0.995\pm0.002$ & $-10.9\pm1.3$ \\
Mar 20 & 23:10:32 & 2.24 & 4x900.0 & 425 & 4.6 & $-4.143\pm0.012$ & $0.380\pm0.001$ & $0.984\pm0.002$ & $15.6\pm1.3$ \\
Mar 21 & 22:36:06 & 2.42 & 4x900.0 & 484 & 3.6 & $-4.133\pm0.011$ & $0.380\pm0.001$ & $0.986\pm0.002$ & $-6.7\pm1.2$ \\
Mar 22 & 22:38:54 & 2.61 & 4x900.0 & 518 & 3.8 & $-4.159\pm0.010$ & $0.375\pm0.001$ & $0.974\pm0.002$ & $-5.5\pm1.1$ \\
Mar 23 & 22:45:25 & 2.80 & 4x900.0 & 488 & 3.8 & $-4.161\pm0.011$ & $0.375\pm0.001$ & $0.982\pm0.002$ & $-4.1\pm1.2$ \\
Mar 24 & 22:36:58 & 2.98 & 4x900.0 & 556 & 3.1 & $-4.147\pm0.009$ & $0.377\pm0.001$ & $0.983\pm0.002$ & $-7.1\pm1.0$ \\
Mar 25 & 23:23:19 & 3.17 & 4x900.0 & 297 & 7.7 & $-4.147\pm0.022$ & $0.380\pm0.002$ & $0.987\pm0.004$ & $8.7\pm2.2$ \\
Mar 26 & 22:47:25 & 3.36 & 4x900.0 & 382 & 5.4 & $-4.115\pm0.015$ & $0.383\pm0.002$ & $0.993\pm0.003$ & $3.1\pm1.6$ \\
\hline
%\footnote{}
\end{longtable}

\begin{longtable}{lcrccccccr}
\caption{Same as Table~\ref{tab:log} for HD\,114710} \\
\hline\hline
Date & UT & $n_\mathrm{cyc}$ & $t_{exp}$ & S/N & $\sigma_\mathrm{LSD}$ & $\log R'_\mathrm{HK}$ & H$\alpha$ & \ion{Ca}{II} IRT & B$_l$\\
 & [hh:mm:ss] & & [s] & & [$10^{-5}I_c$] & & & & [G]\\
\hline
\multicolumn{10}{c}{2019}\\
\hline
Mar 12 & 01:33:25 & 0.00 & 4x400.0 & 1132 & 1.4 & $-4.624\pm0.010$ & $0.313\pm0.001$ & $0.795\pm0.001$ & $0.0\pm0.5$ \\
Mar 21 & 00:34:41 & 0.73 & 4x400.0 & 1113 & 1.5 & $-4.609\pm0.009$ & $0.314\pm0.001$ & $0.797\pm0.001$ & $0.6\pm0.5$ \\
Mar 21 & 24:02:23 & 0.80 & 4x400.0 & 854 & 1.8 & $-4.608\pm0.013$ & $0.314\pm0.001$ & $0.797\pm0.001$ & $0.6\pm0.7$ \\
Mar 22 & 24:04:39 & 0.89 & 4x400.0 & 1117 & 1.4 & $-4.610\pm0.010$ & $0.312\pm0.001$ & $0.793\pm0.001$ & $-0.6\pm0.5$ \\
Mar 23 & 24:11:45 & 0.97 & 4x400.0 & 971 & 1.8 & $-4.614\pm0.012$ & $0.314\pm0.001$ & $0.799\pm0.001$ & $-0.3\pm0.6$ \\
Mar 25 & 01:23:38 & 1.05 & 4x400.0 & 1220 & 1.4 & $-4.610\pm0.009$ & $0.313\pm0.000$ & $0.796\pm0.001$ & $-1.3\pm0.5$ \\
Mar 26 & 00:47:24 & 1.13 & 4x400.0 & 859 & 1.9 & $-4.594\pm0.013$ & $0.315\pm0.001$ & $0.800\pm0.001$ & $-2.1\pm0.7$ \\
Mar 27 & 01:22:59 & 1.21 & 4x400.0 & 784 & 2.3 & $-4.591\pm0.015$ & $0.314\pm0.001$ & $0.804\pm0.001$ & $0.2\pm0.8$ \\
Mar 28 & 01:11:13 & 1.29 & 4x400.0 & 1196 & 1.3 & $-4.600\pm0.009$ & $0.313\pm0.000$ & $0.799\pm0.001$ & $0.7\pm0.5$ \\
Mar 29 & 01:29:45 & 1.38 & 4x400.0 & 1050 & 1.6 & $-4.609\pm0.010$ & $0.313\pm0.001$ & $0.797\pm0.001$ & $-0.4\pm0.5$ \\
Apr 19 & 22:56:18 & 3.15 & 4x400.0 & 887 & 1.9 & $-4.606\pm0.013$ & $0.313\pm0.001$ & $0.813\pm0.001$ & $-2.7\pm0.7$ \\
Apr 27 & 23:15:53 & 3.80 & 4x400.0 & 1154 & 1.4 & $-4.633\pm0.010$ & $0.312\pm0.001$ & $0.810\pm0.001$ & $-0.8\pm0.5$ \\
\hline
%\footnote{}
\end{longtable}

\begin{longtable}{lcrccccccr}
\caption{Same as Table~\ref{tab:log} for HD\,149026} \\
\hline\hline
Date & UT & $n_\mathrm{cyc}$ & $t_{exp}$ & S/N & $\sigma_\mathrm{LSD}$ & $\log R'_\mathrm{HK}$ & H$\alpha$ & \ion{Ca}{II} IRT & B$_l$\\
 & [hh:mm:ss] & & [s] & & [$10^{-5}I_c$] & & & & [G]\\
\hline
\multicolumn{10}{c}{2018}\\
\hline
Aug 20 & 20:51:13 & 0.00 & 4x900.0 & 327 & 6.9 & $-4.903\pm0.091$ & $0.308\pm0.002$ & $0.768\pm0.003$ & $0.3\pm2.1$ \\
Aug 21 & 20:39:25 & 0.05 & 4x900.0 & 342 & 6.7 & $-4.886\pm0.080$ & $0.308\pm0.002$ & $0.763\pm0.003$ & $3.4\pm2.0$ \\
Aug 22 & 20:52:59 & 0.11 & 4x900.0 & 306 & 7.0 & $-4.902\pm0.099$ & $0.309\pm0.002$ & $0.774\pm0.004$ & $2.2\pm2.3$ \\
\hline
%\footnote{}
\end{longtable}

\begin{longtable}{lcrccccccr}
\caption{Same as Table~\ref{tab:log} for HD\,189733} \\
\hline\hline
Date & UT & $n_\mathrm{cyc}$ & $t_{exp}$ & S/N & $\sigma_\mathrm{LSD}$ & $\log R'_\mathrm{HK}$ & H$\alpha$ & \ion{Ca}{II} IRT & B$_l$\\
 & [hh:mm:ss] & & [s] & & [$10^{-5}I_c$] & & & & [G]\\
\hline
\multicolumn{10}{c}{2018}\\
\hline
Nov 14 & 18:06:32 & 0.00 & 4x800.0 & 367 & 5.0 & $-4.471\pm0.023$ & $0.364\pm0.002$ & $0.867\pm0.003$ & $1.8\pm1.1$ \\
Nov 15 & 18:06:59 & 0.08 & 4x800.0 & 409 & 4.3 & $-4.479\pm0.021$ & $0.362\pm0.002$ & $0.867\pm0.002$ & $1.5\pm0.9$ \\
Nov 16 & 20:22:51 & 0.18 & 4x800.0 & 366 & 5.4 & $-4.485\pm0.036$ & $0.363\pm0.002$ & $0.866\pm0.003$ & $-1.7\pm1.0$ \\
Nov 17 & 18:00:16 & 0.25 & 4x800.0 & 410 & 4.4 & $-4.484\pm0.020$ & $0.363\pm0.001$ & $0.862\pm0.002$ & $0.9\pm0.9$ \\
Nov 28 & 18:00:01 & 1.17 & 4x800.0 & 206 & 9.5 & $-4.475\pm0.059$ & $0.363\pm0.003$ & $0.868\pm0.005$ & $-0.7\pm1.9$ \\
Dec 04 & 18:11:16 & 1.68 & 4x800.0 & 310 & 6.7 & $-4.449\pm0.029$ & $0.364\pm0.002$ & $0.880\pm0.003$ & $-2.5\pm1.3$ \\
Dec 10 & 17:55:26 & 2.18 & 4x800.0 & 401 & 5.0 & $-4.469\pm0.019$ & $0.363\pm0.002$ & $0.867\pm0.003$ & $0.0\pm0.9$ \\
Dec 11 & 18:00:29 & 2.26 & 4x800.0 & 332 & 5.5 & $-4.472\pm0.035$ & $0.364\pm0.002$ & $0.870\pm0.003$ & $-4.3\pm1.2$ \\
\hline
%\footnote{}
\end{longtable}

\begin{longtable}{lcrccccccr}
\caption{Same as Table~\ref{tab:log} for HD\,190406} \\
\hline\hline
Date & UT & $n_\mathrm{cyc}$ & $t_{exp}$ & S/N & $\sigma_\mathrm{LSD}$ & $\log R'_\mathrm{HK}$ & H$\alpha$ & \ion{Ca}{II} IRT & B$_l$\\
 & [hh:mm:ss] & & [s] & & [$10^{-5}I_c$] & & & & [G]\\
\hline
\multicolumn{10}{c}{2018}\\
\hline
Jul 09 & 24:07:36 & 0.00 & 4x400.0 & 562 & 3.0 & $-4.704\pm0.031$ & $0.313\pm0.001$ & $0.795\pm0.002$ & $-1.2\pm0.8$ \\
Jul 18 & 00:47:09 & 0.58 & 4x400.0 & 621 & 2.7 & $-4.714\pm0.028$ & $0.313\pm0.001$ & $0.789\pm0.002$ & $2.3\pm0.8$ \\
Jul 19 & 01:14:59 & 0.65 & 4x400.0 & 536 & 3.4 & $-4.719\pm0.036$ & $0.312\pm0.001$ & $0.790\pm0.002$ & $0.9\pm0.9$ \\
Jul 22 & 24:13:41 & 0.93 & 4x400.0 & 599 & 3.0 & $-4.698\pm0.029$ & $0.313\pm0.001$ & $0.795\pm0.002$ & $0.1\pm0.8$ \\
Jul 23 & 23:04:34 & 1.00 & 4x400.0 & 423 & 4.6 & $-4.695\pm0.044$ & $0.314\pm0.001$ & $0.796\pm0.002$ & $2.1\pm1.2$ \\
Jul 25 & 23:55:40 & 1.15 & 4x400.0 & 382 & 4.7 & $-4.694\pm0.049$ & $0.313\pm0.002$ & $0.794\pm0.003$ & $-0.1\pm1.3$ \\
Jul 28 & 00:46:38 & 1.29 & 4x400.0 & 339 & 5.3 & $-4.666\pm0.054$ & $0.312\pm0.002$ & $0.800\pm0.003$ & $1.7\pm1.5$ \\
Jul 29 & 00:38:55 & 1.36 & 4x400.0 & 516 & 3.4 & $-4.663\pm0.032$ & $0.313\pm0.001$ & $0.799\pm0.002$ & $1.2\pm0.9$ \\
\hline
%\footnote{}
\end{longtable}

\begin{longtable}{lcrccccccr}
\caption{Same as Table~\ref{tab:log} for HD\,206860} \\
\hline\hline
Date & UT & $n_\mathrm{cyc}$ & $t_{exp}$ & S/N & $\sigma_\mathrm{LSD}$ & $\log R'_\mathrm{HK}$ & H$\alpha$ & \ion{Ca}{II} IRT & B$_l$\\
 & [hh:mm:ss] & & [s] & & [$10^{-5}I_c$] & & & & [G]\\
\hline
\multicolumn{10}{c}{2018}\\
\hline
Jul 18 & 03:22:53 & 0.00 & 4x300.0 & 487 & 3.4 & $-4.294\pm0.015$ & $0.330\pm0.001$ & $0.926\pm0.003$ & $8.7\pm1.3$ \\
Jul 19 & 02:20:40 & 0.21 & 4x300.0 & 492 & 3.9 & $-4.320\pm0.015$ & $0.329\pm0.001$ & $0.919\pm0.002$ & $3.2\pm1.3$ \\
Jul 23 & 03:30:05 & 1.10 & 4x300.0 & 449 & 3.6 & $-4.295\pm0.016$ & $0.331\pm0.001$ & $0.929\pm0.003$ & $7.6\pm1.4$ \\
Jul 24 & 01:25:49 & 1.30 & 4x300.0 & 348 & 5.2 & $-4.332\pm0.026$ & $0.328\pm0.002$ & $0.915\pm0.003$ & $5.9\pm1.9$ \\
Jul 24 & 03:14:51 & 1.32 & 4x500.0 & 457 & 3.8 & $-4.331\pm0.018$ & $0.328\pm0.001$ & $0.912\pm0.003$ & $7.6\pm1.4$ \\
Jul 26 & 02:20:24 & 1.75 & 4x300.0 & 356 & 4.8 & $-4.338\pm0.026$ & $0.327\pm0.002$ & $0.908\pm0.003$ & $6.0\pm1.9$ \\
Jul 26 & 02:56:22 & 1.75 & 4x500.0 & 442 & 4.0 & $-4.335\pm0.019$ & $0.327\pm0.001$ & $0.910\pm0.003$ & $4.0\pm1.5$ \\
%*Jul 26 & 22:15:10 & 1.93 & 4x500.0 & 110 & ... & ... & ... & ... & ... \\
Jul 27 & 21:55:06 & 2.15 & 4x500.0 & 357 & 5.0 & $-4.307\pm0.027$ & $0.330\pm0.002$ & $0.920\pm0.003$ & $4.8\pm1.9$ \\
%*Jul 28 & 02:59:50 & 2.19 & 4x500.0 & 337 & ... & ... & ... & ... & ... \\
Jul 29 & 01:53:50 & 2.40 & 4x300.0 & 437 & 4.2 & $-4.336\pm0.018$ & $0.327\pm0.001$ & $0.909\pm0.003$ & $10.4\pm1.5$ \\
Jul 29 & 02:29:02 & 2.41 & 4x500.0 & 539 & 3.1 & $-4.339\pm0.014$ & $0.327\pm0.001$ & $0.910\pm0.002$ & $9.9\pm1.2$ \\
Sep 11 & 00:12:31 & 12.06 & 4x300.0 & 442 & 3.8 & $-4.323\pm0.017$ & $0.327\pm0.001$ & $0.910\pm0.003$ & $2.6\pm1.4$ \\
Sep 19 & 23:50:46 & 14.03 & 4x300.0 & 373 & 4.5 & $-4.338\pm0.024$ & $0.327\pm0.002$ & $0.912\pm0.003$ & $7.5\pm1.7$ \\
Sep 20 & 21:28:42 & 14.23 & 4x300.0 & 373 & 4.4 & $-4.339\pm0.022$ & $0.326\pm0.002$ & $0.910\pm0.003$ & $4.2\pm1.7$ \\
Sep 22 & 21:56:32 & 14.68 & 4x300.0 & 413 & 3.8 & $-4.309\pm0.018$ & $0.328\pm0.001$ & $0.921\pm0.003$ & $-2.1\pm1.5$ \\
Sep 26 & 23:48:02 & 15.57 & 4x300.0 & 325 & 5.2 & $-4.310\pm0.026$ & $0.330\pm0.002$ & $0.914\pm0.004$ & $1.2\pm2.0$ \\
\hline
\multicolumn{6}{c}{2019}\\
\hline
Jun 22 & 01:50:36 & 74.49 & 4x600.0 & 555 & 3.0 & $-4.348\pm0.013$ & $0.328\pm0.001$ & $0.908\pm0.002$ & $9.2\pm1.1$ \\
%*Jun 22 & 24:08:40 & 74.70 & 4x600.0 & 130 & ... & ... & ... & ... & ... \\
Jul 12 & 01:43:13 & 78.89 & 4x600.0 & 667 & 2.4 & $-4.348\pm0.010$ & $0.327\pm0.001$ & $0.900\pm0.002$ & $-2.5\pm0.9$ \\
Jul 16 & 02:41:07 & 79.77 & 4x600.0 & 681 & 2.5 & $-4.356\pm0.010$ & $0.327\pm0.001$ & $0.901\pm0.002$ & $4.9\pm0.9$ \\
Jul 21 & 02:46:06 & 80.87 & 4x600.0 & 690 & 2.5 & $-4.349\pm0.011$ & $0.327\pm0.001$ & $0.896\pm0.002$ & $1.3\pm0.9$ \\
Jul 22 & 00:54:03 & 81.08 & 4x600.0 & 554 & 3.0 & $-4.331\pm0.013$ & $0.328\pm0.001$ & $0.909\pm0.002$ & $4.9\pm1.1$ \\
Jul 23 & 00:48:36 & 81.30 & 4x600.0 & 622 & 2.5 & $-4.326\pm0.011$ & $0.327\pm0.001$ & $0.909\pm0.002$ & $0.6\pm0.9$ \\
Jul 25 & 02:04:19 & 81.75 & 4x600.0 & 619 & 2.7 & $-4.356\pm0.011$ & $0.326\pm0.001$ & $0.895\pm0.002$ & $6.0\pm1.0$ \\
Jul 29 & 02:09:57 & 82.63 & 4x600.0 & 557 & 2.8 & $-4.338\pm0.012$ & $0.328\pm0.001$ & $0.906\pm0.002$ & $6.3\pm1.0$ \\
Aug 02 & 02:20:10 & 83.51 & 4x600.0 & 516 & 3.6 & $-4.322\pm0.014$ & $0.329\pm0.001$ & $0.914\pm0.002$ & $5.5\pm1.2$ \\
\hline
%\footnote{}
\end{longtable}

\begin{longtable}{lcrccccccr}
\caption{Same as Table~\ref{tab:log} for HD\,219828} \\
\hline\hline
Date & UT & $n_\mathrm{cyc}$ & $t_{exp}$ & S/N & $\sigma_\mathrm{LSD}$ & $\log R'_\mathrm{HK}$ & H$\alpha$ & \ion{Ca}{II} IRT & B$_l$\\
 & [hh:mm:ss] & & [s] & & [$10^{-5}I_c$] & & & & [G]\\
\hline
\multicolumn{10}{c}{2018}\\
\hline
Sep 19 & 23:04:02 & 0.00 & 4x800.0 & 257 & 9.1 & $-5.000\pm0.147$ & $0.317\pm0.002$ & $0.783\pm0.004$ & $0.8\pm1.9$ \\
Sep 22 & 22:43:05 & 0.10 & 4x800.0 & 260 & 7.6 & $-4.994\pm0.135$ & $0.317\pm0.002$ & $0.765\pm0.004$ & $1.5\pm1.8$ \\
Sep 24 & 22:20:31 & 0.17 & 4x800.0 & 184 & 13.8 & $-4.984\pm0.257$ & $0.316\pm0.003$ & $0.764\pm0.006$ & $1.2\pm2.7$ \\
Sep 26 & 01:00:51 & 0.21 & 4x800.0 & 283 & 7.8 & $-4.998\pm0.116$ & $0.316\pm0.002$ & $0.761\pm0.004$ & $0.5\pm1.6$ \\
Sep 28 & 01:22:59 & 0.28 & 4x800.0 & 344 & 6.6 & $-4.987\pm0.087$ & $0.318\pm0.002$ & $0.760\pm0.003$ & $-0.1\pm1.3$ \\
Oct 03 & 24:17:46 & 0.49 & 4x800.0 & 242 & 9.9 & $-4.966\pm0.153$ & $0.317\pm0.003$ & $0.757\pm0.004$ & $3.6\pm2.0$ \\
Oct 06 & 00:31:03 & 0.56 & 4x800.0 & 353 & 6.3 & $-4.985\pm0.084$ & $0.318\pm0.002$ & $0.752\pm0.003$ & $0.5\pm1.3$ \\
Oct 08 & 22:34:57 & 0.66 & 4x800.0 & 257 & 9.1 & $-4.987\pm0.136$ & $0.317\pm0.002$ & $0.747\pm0.004$ & $1.0\pm1.8$ \\
Oct 22 & 22:29:16 & 1.15 & 4x800.0 & 282 & 8.0 & $-4.986\pm0.116$ & $0.317\pm0.002$ & $0.755\pm0.004$ & $0.8\pm1.7$ \\
Oct 24 & 22:07:20 & 1.22 & 4x800.0 & 268 & 9.3 & $-4.966\pm0.137$ & $0.317\pm0.002$ & $0.749\pm0.004$ & $0.7\pm1.8$ \\
\hline
%\footnote{}
\end{longtable}

\section{Stokes LSD profiles of our stars}\label{app:stokesV}

\begin{figure}[!b]
    \centering
    \includegraphics[width=0.7\textwidth]{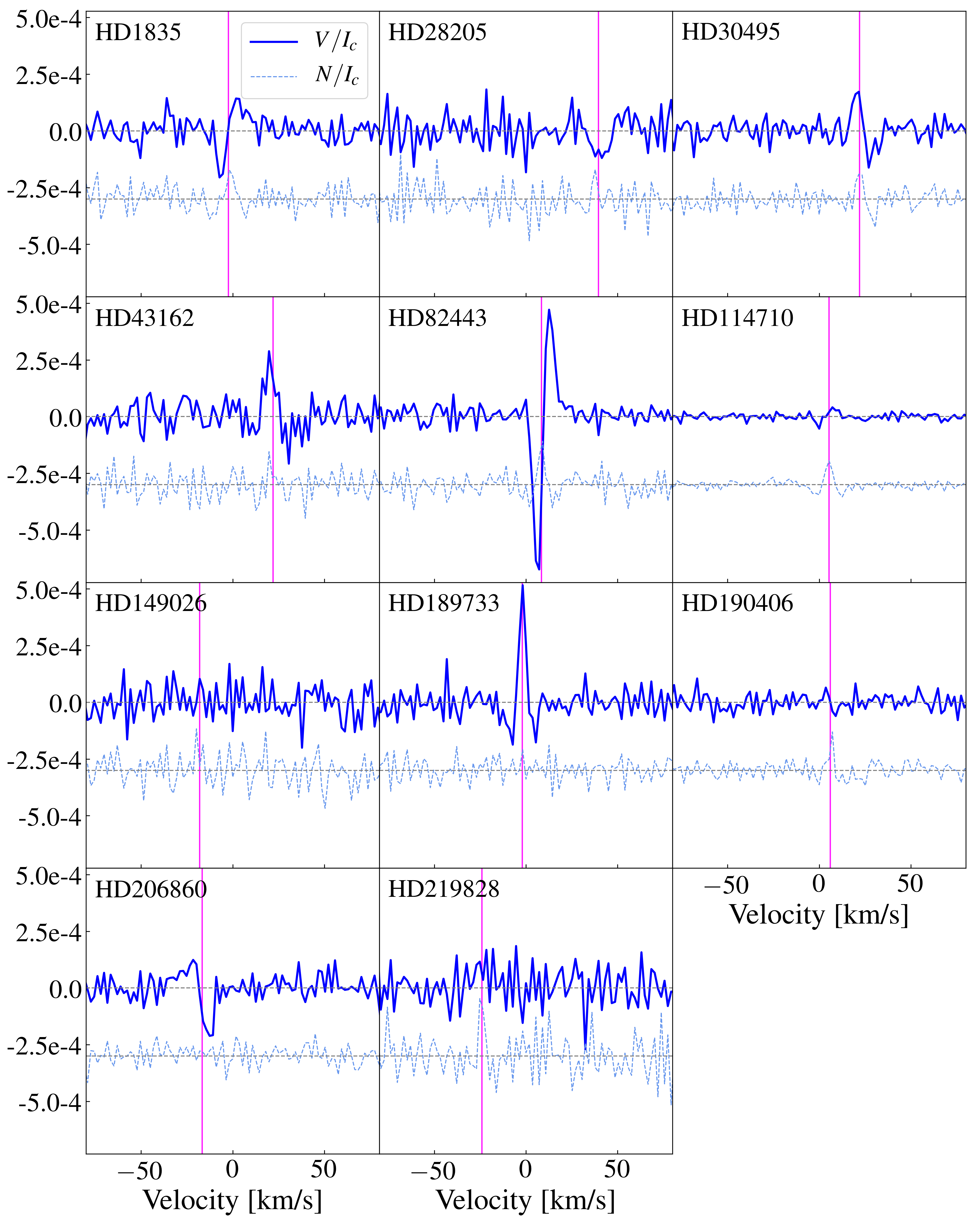}
    \caption{Least-squares deconvolution profiles for the young, Sun-like stars examined in this work. Each panel corresponds to a different star and contains the Stokes~$V$ (solid blue line) and Stokes~$N$ (dashed blue line) profiles. The vertical dotted line indicates the radial velocity of the star. We observe that some stars have a clearly detected Zeeman signature (e.g. HD\,1835, HD\,82443), while others present a flat Stokes~$V$ profile (HD~149026, HD~219828), hence they were excluded from tomographic inversion.
        }\label{fig:stokes_VN}
\end{figure}

This appendix contains examples of LSD profiles for the stars in our sample (see Fig.~\ref{fig:stokes_VN}). The vertical dotted line in the plots indicates the radial velocity of the star, computed by fitting the Stokes~$I$ profile with a Voigt kernel and a linear component to account for residuals of continuum normalisation. The radial velocity is obtained as the centroid of the fit. The properties of the line lists adopted to perform LSD are listed in Table~\ref{tab:lsd_masks}. The number of lines reported in Table~\ref{tab:lsd_masks} accounts for the removal of the wavelength intervals affected by telluric lines or containing the H$\alpha$ line, following \citet{Bellotti2022}. These intervals are: [627,632], [655.5,657], [686,697], [716,734], [759,770], [813,835], and [895,986] nm.

The \texttt{LIBRE-ESPRIT} pipeline provides the null spectrum (Stokes~$N$) as well, obtained by dividing polarimetric sub-exposures with the same polarisation state \citep[see][for more details]{Donati1997}. Least-squares deconvolution was used on the null spectrum to retrieve the corresponding average profile, which is a useful check for the presence of spurious polarisation signatures and the overall noise level in the LSD output \citep{Donati1997,Bagnulo2009}. As shown in Fig.~\ref{fig:stokes_VN}, except for HD\,43162, HD\,189733, and HD\,206860, the Stokes~$N$ LSD profile for all other stars in our study exhibits a systematically positive signature at the radial velocity of the star. Surprisingly, the Stokes~$N$ signal is persistent even when subsets of the line lists are used, more precisely when magnetically sensitive (g$_\mathrm{eff}>1.2$) or red ($\lambda>500$\,nm) lines are used. The latter subset was indeed used by \citet{Folsom2016} and \citet{Bellotti2023a} to mitigate the Stokes~$N$ signal attributed to an imperfect background subtraction during data reduction. Moreover, we note that the Stokes~$N$ signal in the data sets of a certain star may be present for the same night when observations of HD\,206860 were collected and for which this type of signal was not reported. Following \citet{Mathias2018}, we searched for (nearly) simultaneous ESPaDOnS observations of our stars to check whether the Stokes~$V$ signal was present for both instruments; whereas the Stokes~$N$ signal appeared only for Narval, possibly pointing at some instrumental effect. However, no ESPaDOnS observations are present for our stars or close enough in time to our Narval observations. These  combined aspects make the origin of the signal more difficult to decipher. 

Although the source of the Stokes~$N$ signal is not constrained, we noticed that the amplitude of Stokes~$V$ is not significantly affected by the presence of the Stokes~$N$ signal (see Fig.~\ref{fig:stokes_VN}). In practice, we observed that i) for some stars, the amplitude of Stokes~$V$ did not vary appreciably throughout the time series, while the Stokes~$N$ amplitude oscillated visibly in strength, ii) vice versa, for some other stars the Stokes~$V$ LSD profile varied while Stokes~$N$ did not, and iii) the Stokes~$N$ signal was observed even for stars whose Stokes~$V$ signature was not detected. Given the lack of correlation between the Stokes~$V$ and $N$ signals, we proceeded to perform the magnetic characterisation of the stars.

\begin{table}[h]
\caption{Synthetic line lists associated to our stars and used with LSD.} 
\label{tab:lsd_masks}     
\centering                       
\begin{tabular}{l c c c r c}    
\toprule
Name & Mask & N$_\mathrm{lines}$ & N$_\mathrm{obs}$ & $\langle\mathrm{S/N}\rangle_\mathrm{V}$ & $v_\mathrm{range}$\\
& & & & & [km\,s$^{-1}$]\\
\midrule
HD\,1835   & 6000\_4.5 & 6960 & 16    & 479  & $\pm20$\\ 
HD\,28205  & 6000\_4.5 & 6980 & 5     & 346 & ...\\ 
HD\,30495  & 6000\_4.5 & 7270 & 3     & 505  & $\pm20$\\ 
HD\,43162  & 5500\_4.5 & 5340 & 11    & 298 & $\pm16$\\ 
           & 5500\_4.5 & 5340 & 8     & 359 & $\pm16$\\ 
HD\,82443  & 5250\_4.5 & 6180 & 9     & 441  & $\pm20$\\
HD\,114710 & 6000\_4.5 & 7250 & 12    & 1028 & $\pm22$\\ 
HD\,149026 & 6000\_4.0 & 4430 & 3     & 325  & ...\\ 
HD\,189733 & 5000\_4.5 & 6615 & 8     & 350 & $\pm17$\\
HD\,190406 & 6000\_4.5 & 6960 & 8     & 497 & ...\\ 
HD\,206860 & 6000\_4.5 & 7260 & 17    & 396 & $\pm22$\\ 
           & 6000\_4.5 & 7260 & 10    & 560 & $\pm22$\\ 
HD\,219828 & 6000\_4.0 & 4430 & 10    & 273  & ...\\ 
\bottomrule 
\end{tabular}
\tablefoot{The columns indicate: identifier of the star, properties of the line list in the form temperature[K]\_logarithm surface gravity[cm\,s$^{-1}$], number of lines used in LSD, number of observations, mean signal-to-noise ratio of the Stokes~$V$ LSD profile, velocity range centred on the radial velocity of the star encompassing both the absorption of the Stokes~$I$ LSD profiles and the lobes of Stokes~$V$ LSD profiles. When the Stokes~$V$ LSD profiles does not exhibit a clear Zeeman signature, the velocity range is not shown. For two stars, HD\,43162 and HD\,206860, we have two epochs so we list the values for each.}
\end{table}

\section{Activity indices measurements}

In this Appendix, we provide the list of measurements for the following activity indices: $\log R'_\mathrm{HK}$, H$\alpha$, \ion{Ca}{II} IRT, and B$_l$. In Table~\ref{tab:index_results}, we provide the average value for each index, together as the uncertainty computed as the standard deviation of the measurements.

\begin{table*}[!t]
\caption{List of median activity indices and standard deviations for our stars.} 
\label{tab:index_results}     
\centering                       
\begin{tabular}{l c c c r}    
\toprule
Name & $\log R'_\mathrm{HK}$ & H$\alpha$ & \ion{Ca}{II} IRT & B$_l$ [G]\\
\midrule
HD\,1835   & $-4.384\pm0.025$ & $0.326\pm0.002$ & $0.866\pm0.010$ & $-2.2\pm2.5$\\
HD\,28205  & $-4.438\pm0.009$ & $0.313\pm0.001$ & $0.848\pm0.002$ & $-1.0\pm1.3$\\ 
HD\,30495  & $-4.490\pm0.006$ & $0.323\pm0.001$ & $0.855\pm0.001$ & $1.6\pm1.1$\\ 
HD\,43162  & $-4.364\pm0.023$ & $0.339\pm0.002$ & $0.903\pm0.007$ & $2.4\pm3.3$\\ 
HD\,82443  & $-4.147\pm0.016$ & $0.380\pm0.003$ & $0.986\pm0.006$ & $-4.1\pm8.4$\\ 
HD\,114710 & $-4.609\pm0.016$ & $0.313\pm0.001$ & $0.798\pm0.006$ & $-0.3\pm1.0$\\ 
HD\,149026 & $-4.900\pm0.010$ & $0.308\pm0.002$ & $0.768\pm0.004$ & $<6$ \\ 
HD\,189733 & $-4.473\pm0.011$ & $0.363\pm0.001$ & $0.867\pm0.005$ & $-0.3\pm2.0$ \\ 
HD\,190406 & $-4.697\pm0.019$ & $0.317\pm0.002$ & $0.795\pm0.003$ & $<3$ \\ 
HD\,206860 & $-4.334\pm0.017$ & $0.327\pm0.001$ & $0.910\pm0.008$ & $5.2\pm3.4$\\ 
HD\,219828 & $-4.986\pm0.011$ & $0.317\pm0.002$ & $0.758\pm0.010$ & $<5$ \\ 
\bottomrule 
\end{tabular}
\end{table*}

\section{ZDI models of Stokes~$V$ profiles}\label{app:stokesV_models}

This appendix contains the modelled time series of Stokes~$V$ LSD profiles via ZDI. For each star, we show the time series of observed Stokes~$V$ profiles, their ZDI models and the rotational cycle computed with Eq.~\ref{eq:ephemeris}. Each profile is shifted vertically for visualisation. In Table~\ref{tab:wfa_stokesI}, we also include the values of depth, Gaussian width, and Lorentzian width that describe the local Stokes~$I$ LSD line profiles. These parameters were optimised via $\chi^2$ minimisation between the observed and modelled Stokes~$I$ LSD profiles (see Sect.~\ref{sec:zdi}).

\begin{table}[!h]
\caption{Optimised Stokes~$I$ parameters for ZDI reconstruction.} 
\label{tab:wfa_stokesI}     
\centering                       
\begin{tabular}{l c c c r c}    
\toprule
Name & Depth & Gaussian width & Lorentzian width \\
& & [km\,s$^{-1}$] & [km\,s$^{-1}$]\\
\midrule
HD\,1835   & 1.43 & 0.90 & 0.36\\ 
HD\,43162  & 1.18 & 0.88 & 0.44\\ 
HD\,82443  & 1.95 & 0.55 & 0.17\\
HD\,114710 & 0.87 & 1.30 & 0.76\\
HD\,189733 & 0.85 & 1.30 & 0.68\\
HD\,206860 & 1.06 & 1.02 & 0.64\\
\bottomrule 
\end{tabular}
\tablefoot{The depth values are relative to the local continuum.}
\end{table}

\begin{figure}
\centering
    \includegraphics[width=0.88\textwidth]{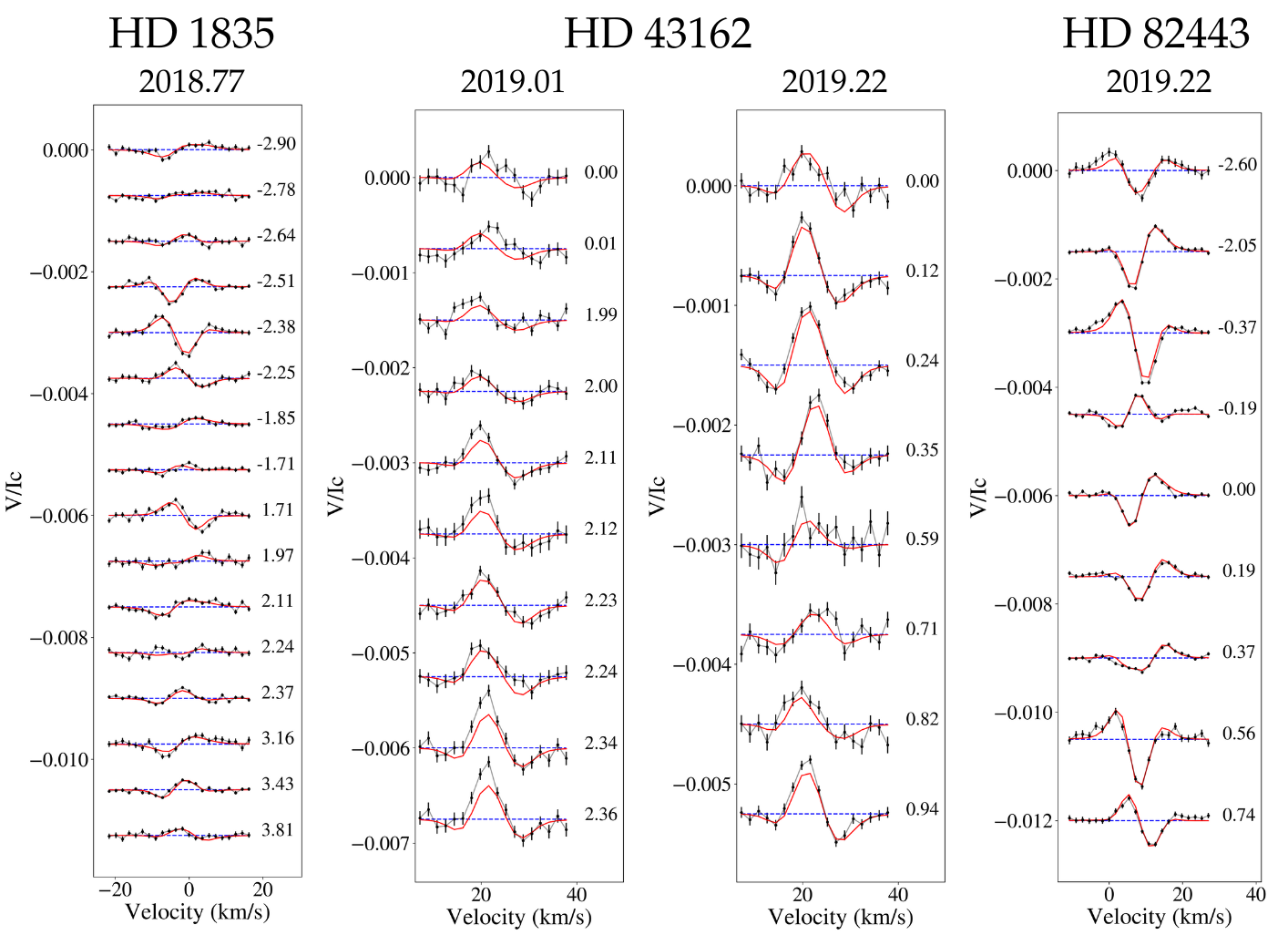}\\
    \includegraphics[width=0.88\textwidth]{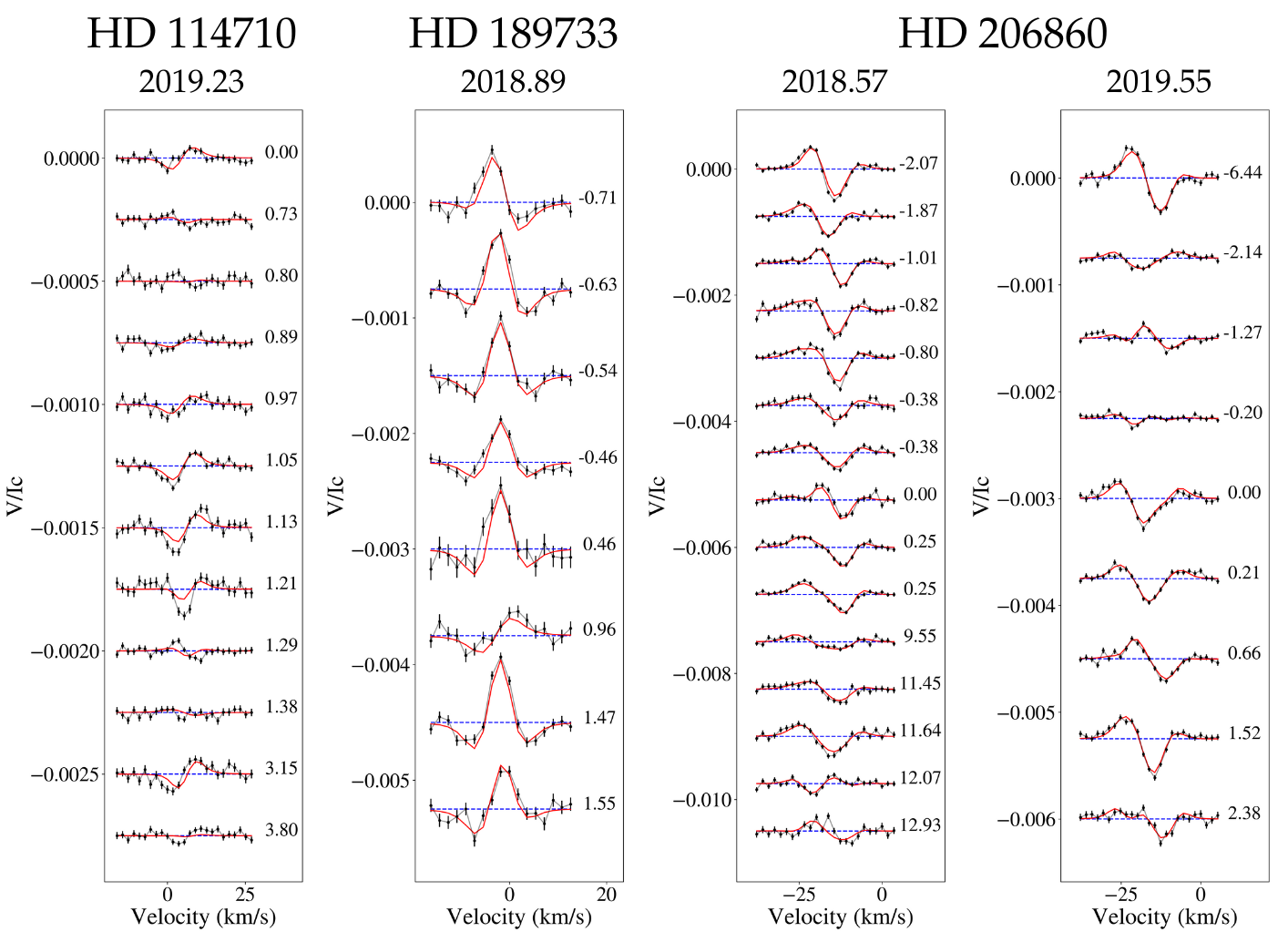}
    \caption{Stokes~$V$ profiles for all stars examined in this work. The cycle number is computed using Eq.~\ref{eq:ephemeris}. For HD\,43162 and HD\,206860, two columns are shown and correspond to two different epochs.}
    \label{fig:stokesV_A}
\end{figure}

\end{appendix}

\end{document}